%
%
%
%
%
%
%
%
\def\standardrisposta{s }\def\reducedrisposta{r }
\def\mplarisposta{mpla }\def\zerorisposta{z }
\def\doublerisposta{d }\def\cartarisposta{e }\def\amsrisposta{y }
\newcount\ingrandimento \newcount\sinnota \newcount\dimnota
\newcount\unoduecol \newdimen\collhsize \newdimen\tothsize
\newdimen\fullhsize \newcount\controllorisposta \sinnota=1
\newskip\infralinea  \global\controllorisposta=0
\immediate\write16 { ********  Welcome to PANDA macros (Plain TeX,
AP, 1991) ******** }
%
%
%
%
%
%
\def\risposta{s } 
\def\srisposta{e } 
\def\arisposta{y }
\ifx\risposta\standardrisposta \ingrandimento=1200
\message {>> This will come out UNREDUCED << }
\dimnota=2 \unoduecol=1 \global\controllorisposta=1 \fi
\ifx\risposta\reducedrisposta \ingrandimento=1095 \dimnota=1
\unoduecol=1  \global\controllorisposta=1
\message {>> This will come out REDUCED << } \fi
\ifx\risposta\doublerisposta \ingrandimento=1000 \dimnota=2
\unoduecol=2

\message {>> You must print this in
LANDSCAPE orientation << } \global\controllorisposta=1 \fi
\ifx\risposta\mplarisposta \ingrandimento=1000 \dimnota=1
\message {>> Mod. Phys. Lett. A format << }
\unoduecol=1 \global\controllorisposta=1 \fi
\ifx\risposta\zerorisposta \ingrandimento=1000 \dimnota=2
\message {>> Zero Magnification format << }
\unoduecol=1 \global\controllorisposta=1 \fi
\ifnum\controllorisposta=0  \ingrandimento=1200
\message {>>> ERROR IN INPUT, I ASSUME STANDARD
UNREDUCED FORMAT <<< }  \dimnota=2 \unoduecol=1 \fi
\magnification=\ingrandimento
%
%
%
%
\newdimen\eucolumnsize \newdimen\eudoublehsize \newdimen\eudoublevsize
\newdimen\uscolumnsize \newdimen\usdoublehsize \newdimen\usdoublevsize
\newdimen\eusinglehsize \newdimen\eusinglevsize \newdimen\ussinglehsize
\newskip\standardbaselineskip \newdimen\ussinglevsize
\newskip\reducedbaselineskip \newskip\doublebaselineskip
\eucolumnsize=12.0truecm    
\eudoublehsize=25.5truecm   
\eudoublevsize=6.7truein    
\uscolumnsize=4.4truein     
\usdoublehsize=9.4truein    
\usdoublevsize=6.8truein    
\eusinglehsize=6.5truein    
\eusinglevsize=24truecm     
\ussinglehsize=6.5truein    
\ussinglevsize=8.9truein    
\standardbaselineskip=16pt plus.2pt  
\reducedbaselineskip=14pt plus.2pt   
\doublebaselineskip=12pt plus.2pt    
%
%
\def\Portoffset{}
\def\Landoffset{\voffset=-.2truein}
\ifx\risposta\mplarisposta \def\Portoffset{\hoffset=1.8truecm} \fi
%
%
\def\Landspec{}
\tolerance=10000
\parskip=0pt plus2pt  \leftskip=0pt \rightskip=0pt
%
%
\ifx\risposta\standardrisposta \infralinea=\standardbaselineskip \fi
\ifx\risposta\reducedrisposta  \infralinea=\reducedbaselineskip \fi
\ifx\risposta\doublerisposta   \infralinea=\doublebaselineskip \fi
\ifx\risposta\mplarisposta     \infralinea=13pt \fi
\ifx\risposta\zerorisposta     \infralinea=12pt plus.2pt\fi
\ifnum\controllorisposta=0    \infralinea=\standardbaselineskip \fi
\ifx\risposta\doublerisposta   \Landoffset \else \Portoffset \fi
\ifx\risposta\doublerisposta \ifx\srisposta\cartarisposta
\tothsize=\eudoublehsize \collhsize=\eucolumnsize
\vsize=\eudoublevsize  \else  \tothsize=\usdoublehsize
\collhsize=\uscolumnsize \vsize=\usdoublevsize \fi \else
\ifx\srisposta\cartarisposta \tothsize=\eusinglehsize
\vsize=\eusinglevsize \else  \tothsize=\ussinglehsize
\vsize=\ussinglevsize \fi \collhsize=4.4truein \fi
\ifx\risposta\mplarisposta \tothsize=5.0truein
\vsize=7.8truein \collhsize=4.4truein \fi
%
%
%
%
\newcount\contaeuler \newcount\contacyrill \newcount\contaams
\font\ninerm=cmr9  \font\eightrm=cmr8  \font\sixrm=cmr6
\font\ninei=cmmi9  \font\eighti=cmmi8  \font\sixi=cmmi6
\font\ninesy=cmsy9  \font\eightsy=cmsy8  \font\sixsy=cmsy6
\font\ninebf=cmbx9  \font\eightbf=cmbx8  \font\sixbf=cmbx6
\font\ninett=cmtt9  \font\eighttt=cmtt8  \font\nineit=cmti9
\font\eightit=cmti8 \font\ninesl=cmsl9  \font\eightsl=cmsl8
\skewchar\ninei='177 \skewchar\eighti='177 \skewchar\sixi='177
\skewchar\ninesy='60 \skewchar\eightsy='60 \skewchar\sixsy='60
\hyphenchar\ninett=-1 \hyphenchar\eighttt=-1 \hyphenchar\tentt=-1
%
\font\tencmmib=cmmib10  \newfam\cmmibfam  \skewchar\tencmmib='177
\font\tencmbsy=cmbsy10  \newfam\cmbsyfam  \skewchar\tencmbsy='60
\def\scaps{\cmcsc}                 
\font\tencmcsc=cmcsc10  \newfam\cmcscfam
\ifnum\ingrandimento=1095

\font\capsone=cmcsc10 at 10.95pt \font\capstwo=cmcsc10 at 13.145pt

\else

\font\capsone=cmcsc10 at 12pt \font\capstwo=cmcsc10 at 14.4pt
\fi

\def\ttaarr{\bf}		
\def\ppaarr{\sl}		

%
%
%
\newfam\eufmfam \newfam\msamfam \newfam\msbmfam \newfam\eufbfam
\def\Loadeulerfonts{\global\contaeuler=1 \ifx\arisposta\amsrisposta
\font\teneufm=eufm10              
\font\eighteufm=eufm8 \font\nineeufm=eufm9 \font\sixeufm=eufm6
\font\seveneufm=eufm7  \font\fiveeufm=eufm5
\font\teneufb=eufb10              
\font\eighteufb=eufb8 \font\nineeufb=eufb9 \font\sixeufb=eufb6
\font\seveneufb=eufb7  \font\fiveeufb=eufb5
\font\teneurm=eurm10              
\font\eighteurm=eurm8 \font\nineeurm=eurm9
\font\teneurb=eurb10              
\font\eighteurb=eurb8 \font\nineeurb=eurb9
\font\teneusm=eusm10              
\font\eighteusm=eusm8 \font\nineeusm=eusm9
\font\teneusb=eusb10              
\font\eighteusb=eusb8 \font\nineeusb=eusb9
\else \def\eufm{\tt} \def\eufb{\tt} \def\eurm{\tt} \def\eurb{\tt}
\def\eusm{\tt} \def\eusb{\tt}    \fi}
\def\loadeuler{\Loadeulerfonts\tenpoint}
\def\loadamsmath{\global\contaams=1 \ifx\arisposta\amsrisposta
\font\tenmsam=msam10 \font\ninemsam=msam9 \font\eightmsam=msam8
\font\sevenmsam=msam7 \font\sixmsam=msam6 \font\fivemsam=msam5
\font\tenmsbm=msbm10 \font\ninemsbm=msbm9 \font\eightmsbm=msbm8
\font\sevenmsbm=msbm7 \font\sixmsbm=msbm6 \font\fivemsbm=msbm5
\else \def\msbm{\bf} \fi \def\Bbb{\msbm} \def\symbl{\msam} \tenpoint}
\def\loadcyrill{\global\contacyrill=1 \ifx\arisposta\amsrisposta
\font\tenwncyr=wncyr10 \font\ninewncyr=wncyr9 \font\eightwncyr=wncyr8
\font\tenwncyb=wncyr10 \font\ninewncyb=wncyr9 \font\eightwncyb=wncyr8
\font\tenwncyi=wncyr10 \font\ninewncyi=wncyr9 \font\eightwncyi=wncyr8
\else \def\cyrill{\sl} \def\cyrilb{\sl} \def\cyrili{\sl} \fi\tenpoint}
\ifx\arisposta\amsrisposta
\font\sevenex=cmex7               
\font\eightex=cmex8  \font\nineex=cmex9
\font\ninecmmib=cmmib9   \font\eightcmmib=cmmib8
\font\sevencmmib=cmmib7 \font\sixcmmib=cmmib6
\font\fivecmmib=cmmib5   \skewchar\ninecmmib='177
\skewchar\eightcmmib='177  \skewchar\sevencmmib='177
\skewchar\sixcmmib='177   \skewchar\fivecmmib='177
\font\ninecmbsy=cmbsy9    \font\eightcmbsy=cmbsy8
\font\sevencmbsy=cmbsy7  \font\sixcmbsy=cmbsy6
\font\fivecmbsy=cmbsy5   \skewchar\ninecmbsy='60
\skewchar\eightcmbsy='60  \skewchar\sevencmbsy='60
\skewchar\sixcmbsy='60    \skewchar\fivecmbsy='60
\font\ninecmcsc=cmcsc9    \font\eightcmcsc=cmcsc8     \else
\def\cmmib{\fam\cmmibfam\tencmmib}\textfont\cmmibfam=\tencmmib
\scriptfont\cmmibfam=\tencmmib \scriptscriptfont\cmmibfam=\tencmmib
\def\cmbsy{\fam\cmbsyfam\tencmbsy} \textfont\cmbsyfam=\tencmbsy
\scriptfont\cmbsyfam=\tencmbsy \scriptscriptfont\cmbsyfam=\tencmbsy
\scriptfont\cmcscfam=\tencmcsc \scriptscriptfont\cmcscfam=\tencmcsc
\def\cmcsc{\fam\cmcscfam\tencmcsc} \textfont\cmcscfam=\tencmcsc \fi
\catcode`@=11
\newskip\ttglue
\gdef\tenpoint{\def\rm{\fam0\tenrm}
  \textfont0=\tenrm \scriptfont0=\sevenrm \scriptscriptfont0=\fiverm
  \textfont1=\teni \scriptfont1=\seveni \scriptscriptfont1=\fivei
  \textfont2=\tensy \scriptfont2=\sevensy \scriptscriptfont2=\fivesy
  \textfont3=\tenex \scriptfont3=\tenex \scriptscriptfont3=\tenex
  \def\mcal{\fam2 \tensy}  \def\mmit{\fam1 \teni}
  \textfont\itfam=\tenit \def\it{\fam\itfam\tenit}
  \textfont\slfam=\tensl \def\sl{\fam\slfam\tensl}
  \textfont\ttfam=\tentt \scriptfont\ttfam=\eighttt
  \scriptscriptfont\ttfam=\eighttt  \def\tt{\fam\ttfam\tentt}
  \textfont\bffam=\tenbf \scriptfont\bffam=\sevenbf
  \scriptscriptfont\bffam=\fivebf \def\bf{\fam\bffam\tenbf}
     \ifx\arisposta\amsrisposta    \ifnum\contaeuler=1
  \textfont\eufmfam=\teneufm \scriptfont\eufmfam=\seveneufm
  \scriptscriptfont\eufmfam=\fiveeufm \def\eufm{\fam\eufmfam\teneufm}
  \textfont\eufbfam=\teneufb \scriptfont\eufbfam=\seveneufb
  \scriptscriptfont\eufbfam=\fiveeufb \def\eufb{\fam\eufbfam\teneufb}
  \def\eurm{\teneurm} \def\eurb{\teneurb} \def\eusm{\teneusm}
  \def\eusb{\teneusb}    \fi    \ifnum\contaams=1
  \textfont\msamfam=\tenmsam \scriptfont\msamfam=\sevenmsam
  \scriptscriptfont\msamfam=\fivemsam \def\msam{\fam\msamfam\tenmsam}
  \textfont\msbmfam=\tenmsbm \scriptfont\msbmfam=\sevenmsbm
  \scriptscriptfont\msbmfam=\fivemsbm \def\msbm{\fam\msbmfam\tenmsbm}
     \fi      \ifnum\contacyrill=1     \def\cyrill{\tenwncyr}
  \def\cyrilb{\tenwncyb}  \def\cyrili{\tenwncyi}         \fi
  \textfont3=\tenex \scriptfont3=\sevenex \scriptscriptfont3=\sevenex
  \def\cmmib{\fam\cmmibfam\tencmmib} \scriptfont\cmmibfam=\sevencmmib
  \textfont\cmmibfam=\tencmmib  \scriptscriptfont\cmmibfam=\fivecmmib
  \def\cmbsy{\fam\cmbsyfam\tencmbsy} \scriptfont\cmbsyfam=\sevencmbsy
  \textfont\cmbsyfam=\tencmbsy  \scriptscriptfont\cmbsyfam=\fivecmbsy
  \def\cmcsc{\fam\cmcscfam\tencmcsc} \scriptfont\cmcscfam=\eightcmcsc
  \textfont\cmcscfam=\tencmcsc \scriptscriptfont\cmcscfam=\eightcmcsc
     \fi            \tt \ttglue=.5em plus.25em minus.15em
  \normalbaselineskip=12pt
  \setbox\strutbox=\hbox{\vrule height8.5pt depth3.5pt width0pt}
  \let\sc=\eightrm \let\big=\tenbig   \normalbaselines
  \baselineskip=\infralinea  \rm}
\gdef\ninepoint{\def\rm{\fam0\ninerm}
  \textfont0=\ninerm \scriptfont0=\sixrm \scriptscriptfont0=\fiverm
  \textfont1=\ninei \scriptfont1=\sixi \scriptscriptfont1=\fivei
  \textfont2=\ninesy \scriptfont2=\sixsy \scriptscriptfont2=\fivesy
  \textfont3=\tenex \scriptfont3=\tenex \scriptscriptfont3=\tenex
  \def\mcal{\fam2 \ninesy}  \def\mmit{\fam1 \ninei}
  \textfont\itfam=\nineit \def\it{\fam\itfam\nineit}
  \textfont\slfam=\ninesl \def\sl{\fam\slfam\ninesl}
  \textfont\ttfam=\ninett \scriptfont\ttfam=\eighttt
  \scriptscriptfont\ttfam=\eighttt \def\tt{\fam\ttfam\ninett}
  \textfont\bffam=\ninebf \scriptfont\bffam=\sixbf
  \scriptscriptfont\bffam=\fivebf \def\bf{\fam\bffam\ninebf}
     \ifx\arisposta\amsrisposta  \ifnum\contaeuler=1
  \textfont\eufmfam=\nineeufm \scriptfont\eufmfam=\sixeufm
  \scriptscriptfont\eufmfam=\fiveeufm \def\eufm{\fam\eufmfam\nineeufm}
  \textfont\eufbfam=\nineeufb \scriptfont\eufbfam=\sixeufb
  \scriptscriptfont\eufbfam=\fiveeufb \def\eufb{\fam\eufbfam\nineeufb}
  \def\eurm{\nineeurm} \def\eurb{\nineeurb} \def\eusm{\nineeusm}
  \def\eusb{\nineeusb}     \fi   \ifnum\contaams=1
  \textfont\msamfam=\ninemsam \scriptfont\msamfam=\sixmsam
  \scriptscriptfont\msamfam=\fivemsam \def\msam{\fam\msamfam\ninemsam}
  \textfont\msbmfam=\ninemsbm \scriptfont\msbmfam=\sixmsbm
  \scriptscriptfont\msbmfam=\fivemsbm \def\msbm{\fam\msbmfam\ninemsbm}
     \fi       \ifnum\contacyrill=1     \def\cyrill{\ninewncyr}
  \def\cyrilb{\ninewncyb}  \def\cyrili{\ninewncyi}         \fi
  \textfont3=\nineex \scriptfont3=\sevenex \scriptscriptfont3=\sevenex
  \def\cmmib{\fam\cmmibfam\ninecmmib}  \textfont\cmmibfam=\ninecmmib
  \scriptfont\cmmibfam=\sixcmmib \scriptscriptfont\cmmibfam=\fivecmmib
  \def\cmbsy{\fam\cmbsyfam\ninecmbsy}  \textfont\cmbsyfam=\ninecmbsy
  \scriptfont\cmbsyfam=\sixcmbsy \scriptscriptfont\cmbsyfam=\fivecmbsy
  \def\cmcsc{\fam\cmcscfam\ninecmcsc} \scriptfont\cmcscfam=\eightcmcsc
  \textfont\cmcscfam=\ninecmcsc \scriptscriptfont\cmcscfam=\eightcmcsc
     \fi            \tt \ttglue=.5em plus.25em minus.15em
  \normalbaselineskip=11pt
  \setbox\strutbox=\hbox{\vrule height8pt depth3pt width0pt}
  \let\sc=\sevenrm \let\big=\ninebig \normalbaselines\rm}
\gdef\eightpoint{\def\rm{\fam0\eightrm}
  \textfont0=\eightrm \scriptfont0=\sixrm \scriptscriptfont0=\fiverm
  \textfont1=\eighti \scriptfont1=\sixi \scriptscriptfont1=\fivei
  \textfont2=\eightsy \scriptfont2=\sixsy \scriptscriptfont2=\fivesy
  \textfont3=\tenex \scriptfont3=\tenex \scriptscriptfont3=\tenex
  \def\mcal{\fam2 \eightsy}  \def\mmit{\fam1 \eighti}
  \textfont\itfam=\eightit \def\it{\fam\itfam\eightit}
  \textfont\slfam=\eightsl \def\sl{\fam\slfam\eightsl}
  \textfont\ttfam=\eighttt \scriptfont\ttfam=\eighttt
  \scriptscriptfont\ttfam=\eighttt \def\tt{\fam\ttfam\eighttt}
  \textfont\bffam=\eightbf \scriptfont\bffam=\sixbf
  \scriptscriptfont\bffam=\fivebf \def\bf{\fam\bffam\eightbf}
     \ifx\arisposta\amsrisposta   \ifnum\contaeuler=1
  \textfont\eufmfam=\eighteufm \scriptfont\eufmfam=\sixeufm
  \scriptscriptfont\eufmfam=\fiveeufm \def\eufm{\fam\eufmfam\eighteufm}
  \textfont\eufbfam=\eighteufb \scriptfont\eufbfam=\sixeufb
  \scriptscriptfont\eufbfam=\fiveeufb \def\eufb{\fam\eufbfam\eighteufb}
  \def\eurm{\eighteurm} \def\eurb{\eighteurb} \def\eusm{\eighteusm}
  \def\eusb{\eighteusb}       \fi    \ifnum\contaams=1
  \textfont\msamfam=\eightmsam \scriptfont\msamfam=\sixmsam
  \scriptscriptfont\msamfam=\fivemsam \def\msam{\fam\msamfam\eightmsam}
  \textfont\msbmfam=\eightmsbm \scriptfont\msbmfam=\sixmsbm
  \scriptscriptfont\msbmfam=\fivemsbm \def\msbm{\fam\msbmfam\eightmsbm}
     \fi       \ifnum\contacyrill=1     \def\cyrill{\eightwncyr}
  \def\cyrilb{\eightwncyb}  \def\cyrili{\eightwncyi}         \fi
  \textfont3=\eightex \scriptfont3=\sevenex \scriptscriptfont3=\sevenex
  \def\cmmib{\fam\cmmibfam\eightcmmib}  \textfont\cmmibfam=\eightcmmib
  \scriptfont\cmmibfam=\sixcmmib \scriptscriptfont\cmmibfam=\fivecmmib
  \def\cmbsy{\fam\cmbsyfam\eightcmbsy}  \textfont\cmbsyfam=\eightcmbsy
  \scriptfont\cmbsyfam=\sixcmbsy \scriptscriptfont\cmbsyfam=\fivecmbsy
  \def\cmcsc{\fam\cmcscfam\eightcmcsc} \scriptfont\cmcscfam=\eightcmcsc
  \textfont\cmcscfam=\eightcmcsc \scriptscriptfont\cmcscfam=\eightcmcsc
     \fi             \tt \ttglue=.5em plus.25em minus.15em
  \normalbaselineskip=9pt
  \setbox\strutbox=\hbox{\vrule height7pt depth2pt width0pt}
  \let\sc=\sixrm \let\big=\eightbig \normalbaselines\rm }
\gdef\tenbig#1{{\hbox{$\left#1\vbox to8.5pt{}\right.\n@space$}}}
\gdef\ninebig#1{{\hbox{$\textfont0=\tenrm\textfont2=\tensy
   \left#1\vbox to7.25pt{}\right.\n@space$}}}
\gdef\eightbig#1{{\hbox{$\textfont0=\ninerm\textfont2=\ninesy
   \left#1\vbox to6.5pt{}\right.\n@space$}}}
\def\alternativefont#1#2{\ifx\arisposta\amsrisposta \relax \else
\xdef#1{#2} \fi}
\global\contaeuler=0 \global\contacyrill=0 \global\contaams=0
%
%
%
%
\newbox\fotlinebb \newbox\hedlinebb \newbox\leftcolumn
\gdef\makeheadline{\vbox to 0pt{\vskip-22.5pt
     \fullline{\vbox to8.5pt{}\the\headline}\vss}\nointerlineskip}
\gdef\makehedlinebb{\vbox to 0pt{\vskip-22.5pt
     \fullline{\vbox to8.5pt{}\copy\hedlinebb\hfil
     \line{\hfill\the\headline\hfill}}\vss} \nointerlineskip}
\gdef\makefootline{\baselineskip=24pt \fullline{\the\footline}}
\gdef\makefotlinebb{\baselineskip=24pt
    \fullline{\copy\fotlinebb\hfil\line{\hfill\the\footline\hfill}}}
\gdef\doubleformat{\shipout\vbox{\Landspec\makehedlinebb
     \fullline{\box\leftcolumn\hfil\columnbox}\makefotlinebb}
     \advancepageno}
\gdef\columnbox{\leftline{\pagebody}}
\gdef\line#1{\hbox to\hsize{\hskip\leftskip#1\hskip\rightskip}}
\gdef\fullline#1{\hbox to\fullhsize{\hskip\leftskip{#1}%
\hskip\rightskip}}
\gdef\footnote#1{\let\@sf=\empty
         \ifhmode\edef\#sf{\spacefactor=\the\spacefactor}\/\fi
         #1\@sf\vfootnote{#1}}
\gdef\vfootnote#1{\insert\footins\bgroup
         \ifnum\dimnota=1  \eightpoint\fi
         \ifnum\dimnota=2  \ninepoint\fi
         \ifnum\dimnota=0  \tenpoint\fi
         \interlinepenalty=\interfootnotelinepenalty
         \splittopskip=\ht\strutbox
         \splitmaxdepth=\dp\strutbox \floatingpenalty=20000
         \leftskip=\oldssposta \rightskip=\olddsposta
         \spaceskip=0pt \xspaceskip=0pt
         \ifnum\sinnota=0   \textindent{#1}\fi
         \ifnum\sinnota=1   \item{#1}\fi
         \footstrut\futurelet\next\fo@t}
\gdef\fo@t{\ifcat\bgroup\noexpand\next \let\next\f@@t
             \else\let\next\f@t\fi \next}
\gdef\f@@t{\bgroup\aftergroup\@foot\let\next}
\gdef\f@t#1{#1\@foot} \gdef\@foot{\strut\egroup}
\gdef\footstrut{\vbox to\splittopskip{}}
\skip\footins=\bigskipamount
\count\footins=1000  \dimen\footins=8in
\catcode`@=12
\tenpoint
\ifnum\unoduecol=1 \hsize=\tothsize   \fullhsize=\tothsize \fi
\ifnum\unoduecol=2 \hsize=\collhsize  \fullhsize=\tothsize \fi
\global\let\lrcol=L      \ifnum\unoduecol=1
\output{\plainoutput{\ifnum\tipbnota=2 \clearnmbnota\fi}} \fi
\ifnum\unoduecol=2 \output{\if L\lrcol
     \global\setbox\leftcolumn=\columnbox
     \global\setbox\fotlinebb=\line{\hfill\the\footline\hfill}
     \global\setbox\hedlinebb=\line{\hfill\the\headline\hfill}
     \advancepageno  \global\let\lrcol=R
     \else  \doubleformat \global\let\lrcol=L \fi
     \ifnum\outputpenalty>-20000 \else\dosupereject\fi
     \ifnum\tipbnota=2\clearnmbnota\fi }\fi
\def\ifdoublepage{\ifnum\unoduecol=2 }
\gdef\yespagenumbers{\footline={\hss\tenrm\folio\hss}}
\gdef\ciao{ \ifnum\fdefcontre=1 \endfdef\fi
     \par\vfill\supereject \ifnum\unoduecol=2
     \if R\lrcol  \headline={}\nopagenumbers\null\vfill\eject
     \fi\fi \end}

\newskip\olddsposta \newskip\oldssposta
\global\oldssposta=\leftskip \global\olddsposta=\rightskip

\def\filldots{\leaders\hbox to 1em{\hss.\hss}\hfill}
\def\inquadrb#1 {\vbox {\hrule  \hbox{\vrule \vbox {\vskip .2cm
    \hbox {\ #1\ } \vskip .2cm } \vrule  }  \hrule} }
 \def\newline{\hfil\break}
\def\jump{\vskip\baselineskip} \newskip\iinnffrr
\def\sjump{\iinnffrr=\baselineskip
          \divide\iinnffrr by 2 \vskip\iinnffrr}
\def\bjump{\vskip\baselineskip \vskip\baselineskip}
\newcount\nmbnota  \def\clearnmbnota{\global\nmbnota=0}
\newcount\tipbnota \def\letterfootnote{\global\tipbnota=1}

\def\note#1{\global\advance\nmbnota by 1 \ifnum\tipbnota=1
    \footnote{$^{\rm\nttlett}$}{#1} \else {\ifnum\tipbnota=2
    \footnote{$^{\nttsymb}$}{#1}
    \else\footnote{$^{\the\nmbnota}$}{#1}\fi}\fi}
\def\nttlett{\ifcase\nmbnota \or a\or b\or c\or d\or e\or f\or
g\or h\or i\or j\or k\or l\or m\or n\or o\or p\or q\or r\or
s\or t\or u\or v\or w\or y\or x\or z\fi}
\def\nttsymb{\ifcase\nmbnota \or\dag\or\sharp\or\ddag\or\star\or
\natural\or\flat\or\clubsuit\or\diamondsuit\or\heartsuit
\or\spadesuit\fi}   \clearnmbnota
\def\numberfootnote{\global\tipbnota=0} \numberfootnote
\def\setnote#1{\expandafter\xdef\csname#1\endcsname{
\ifnum\tipbnota=1 {\rm\nttlett} \else {\ifnum\tipbnota=2
{\nttsymb} \else \the\nmbnota\fi}\fi} }
\newcount\nbmfig  \def\clearnbmfig{\global\nbmfig=0}
\gdef\figure{\global\advance\nbmfig by 1
      {\rm fig. \the\nbmfig}}   \clearnbmfig
\def\setfig#1{\expandafter\xdef\csname#1\endcsname{fig. \the\nbmfig}}
 \def\endformula{\eqno\numero $$}
 \def\efr{\endformula}
\newcount\frmcount \def\clearfrmcount{\global\frmcount=0}
\def\numero{\global\advance\frmcount by 1   \ifnum\indappcount=0
  {\ifnum\cpcount <1 {\hbox{\rm (\the\frmcount )}}  \else
  {\hbox{\rm (\the\cpcount .\the\frmcount )}} \fi}  \else
  {\hbox{\rm (\applett .\the\frmcount )}} \fi}
\def\nameformula#1{\global\advance\frmcount by 1%
\ifnum\draftnum=0  {\ifnum\indappcount=0%
{\ifnum\cpcount<1\xdef\spzzttrra{(\the\frmcount )}%
\else\xdef\spzzttrra{(\the\cpcount .\the\frmcount )}\fi}%
\else\xdef\spzzttrra{(\applett .\the\frmcount )}\fi}%
\else\xdef\spzzttrra{(#1)}\fi%
\expandafter\xdef\csname#1\endcsname{\spzzttrra}
\eqno \hbox{\rm\spzzttrra} $$}
\def\nfr{\nameformula}    \def\numali{\numero}
\def\nameali#1{\global\advance\frmcount by 1%
\ifnum\draftnum=0  {\ifnum\indappcount=0%
{\ifnum\cpcount<1\xdef\spzzttrra{(\the\frmcount )}%
\else\xdef\spzzttrra{(\the\cpcount .\the\frmcount )}\fi}%
\else\xdef\spzzttrra{(\applett .\the\frmcount )}\fi}%
\else\xdef\spzzttrra{(#1)}\fi%
\expandafter\xdef\csname#1\endcsname{\spzzttrra}
  \hbox{\rm\spzzttrra} }      \clearfrmcount
\newcount\cpcount \def\clearcpcount{\global\cpcount=0}
\newcount\subcpcount \def\clearsubcpcount{\global\subcpcount=0}
\newcount\appcount \def\clearappcount{\global\appcount=0}
\newcount\indappcount \def\clearindappcount{\indappcount=0}
\newcount\sottoparcount 

\def\applett{\ifcase\appcount  \or {A}\or {B}\or {C}\or
{D}\or {E}\or {F}\or {G}\or {H}\or {I}\or {J}\or {K}\or {L}\or
{M}\or {N}\or {O}\or {P}\or {Q}\or {R}\or {S}\or {T}\or {U}\or
{V}\or {W}\or {X}\or {Y}\or {Z}\fi    \ifnum\appcount<0
\immediate\write16 {Panda ERROR - Appendix: counter "appcount"
out of range}\fi  \ifnum\appcount>26  \immediate\write16 {Panda
ERROR - Appendix: counter "appcount" out of range}\fi}
\clearappcount  \clearindappcount \newcount\connttrre
\def\clearconnttrre{\global\connttrre=0} \newcount\countref
\def\clearcountref{\global\countref=0} \clearcountref
\def\chapter#1{\global\advance\cpcount by 1 \clearfrmcount
                 \goodbreak\null\vbox{\jump\nobreak
                 \clearsubcpcount\clearindappcount
                 \itemitem{\ttaarr\the\cpcount .\qquad}{\ttaarr #1}
                 \par\nobreak\jump\sjump}\nobreak}
\def\section#1{\global\advance\subcpcount by 1 \goodbreak\null
               \vbox{\sjump\nobreak\ifnum\indappcount=0
                 {\ifnum\cpcount=0 {\itemitem{\ppaarr
               .\the\subcpcount\quad\enskip\ }{\ppaarr #1}\par} \else
                 {\itemitem{\ppaarr\the\cpcount .\the\subcpcount\quad
                  \enskip\ }{\ppaarr #1} \par}  \fi}
                \else{\itemitem{\ppaarr\applett .\the\subcpcount\quad
                 \enskip\ }{\ppaarr #1}\par}\fi\nobreak\jump}\nobreak}
\clearsubcpcount
\def\appendix#1{\global\advance\appcount by 1 \clearfrmcount
                  \goodbreak\null\vbox{\jump\nobreak
                  \global\advance\indappcount by 1 \clearsubcpcount
          \itemitem{ }{\hskip-40pt\ttaarr #1}
             \nobreak\jump\sjump}\nobreak}
\clearappcount \clearindappcount
\def\references{\goodbreak\null\vbox{\jump\nobreak
   \noindent{\ttaarr References} \nobreak\jump\sjump}\nobreak}

\clearcpcount\clearcountref

\def\setchap#1{\ifnum\indappcount=0{\ifnum\subcpcount=0%
\xdef\spzzttrra{\the\cpcount}%
\else\xdef\spzzttrra{\the\cpcount .\the\subcpcount}\fi}
\else{\ifnum\subcpcount=0 \xdef\spzzttrra{\applett}%
\else\xdef\spzzttrra{\applett .\the\subcpcount}\fi}\fi
\expandafter\xdef\csname#1\endcsname{\spzzttrra}}
\newcount\draftnum \newcount\ppora   \newcount\ppminuti
\global\ppora=\time   \global\ppminuti=\time
\global\divide\ppora by 60  \draftnum=\ppora
\multiply\draftnum by 60    \global\advance\ppminuti by -\draftnum
\def\droggi{\number\day /\number\month /\number\year\ \the\ppora
:\the\ppminuti}     \global\draftnum=0
\def\draftcomment#1{\ifnum\draftnum=0 \relax \else
{\ {\bf ***}\ #1\ {\bf ***}\ }\fi} 
%
%
\catcode`@=11
\gdef\Ref#1{\expandafter\ifx\csname @rrxx@#1\endcsname\relax%
{\global\advance\countref by 1    \ifnum\countref>200
\immediate\write16 {Panda ERROR - Ref: maximum number of references
exceeded}  \expandafter\xdef\csname @rrxx@#1\endcsname{0}\else
\expandafter\xdef\csname @rrxx@#1\endcsname{\the\countref}\fi}\fi
\ifnum\draftnum=0 \csname @rrxx@#1\endcsname \else#1\fi}
\gdef\beginref{\ifnum\draftnum=0  \gdef\Rref{\fairef}
\gdef\endref{\scriviref} \else\relax\fi
\ifx\risposta\mplarisposta \ninepoint \fi
\parskip 2pt plus.2pt \baselineskip=12pt}
\def\Reflab#1{[#1]} \gdef\Rref#1#2{\item{\Reflab{#1}}{#2}}
\gdef\endref{\relax}  \newcount\conttemp
\gdef\fairef#1#2{\expandafter\ifx\csname @rrxx@#1\endcsname\relax
{\global\conttemp=0 \immediate\write16 {Panda ERROR - Ref: reference
[#1] undefined}} \else
{\global\conttemp=\csname @rrxx@#1\endcsname } \fi
\global\advance\conttemp by 50  \global\setbox\conttemp=\hbox{#2} }
\gdef\scriviref{\clearconnttrre\conttemp=50
\loop\ifnum\connttrre<\countref \advance\conttemp by 1
\advance\connttrre by 1
\item{\Reflab{\the\connttrre}}{\unhcopy\conttemp} \repeat}
\clearcountref \clearconnttrre
\catcode`@=12
\ifx\risposta\mplarisposta \def\Reflab#1{#1.} \letterfootnote \fi

\def\slashchar#1{\setbox0=\hbox{$#1$} \dimen0=\wd0
     \setbox1=\hbox{/} \dimen1=\wd1 \ifdim\dimen0>\dimen1
      \rlap{\hbox to \dimen0{\hfil/\hfil}} #1 \else
      \rlap{\hbox to \dimen1{\hfil$#1$\hfil}} / \fi}
\ifx\oldchi\undefined \let\oldchi=\chi
  \def\cchi{{\raise 1pt\hbox{$\oldchi$}}} \let\chi=\cchi \fi

\def\frac#1#2{{\textstyle{#1 \over #2}}}

\def\half{\ifinner {\scriptstyle {1 \over 2}}\else {1 \over 2} \fi}
\def\bra#1{\langle#1\vert}  \def\ket#1{\vert#1\rangle}

\def\simge{\rlap{\raise 2pt \hbox{$>$}}{\lower 2pt \hbox{$\sim$}}}
\def\simle{\rlap{\raise 2pt \hbox{$<$}}{\lower 2pt \hbox{$\sim$}}}

\def\buildchar#1#2#3{{\null\!\mathop{#1}\limits^{#2}_{#3}\!\null}}
\def\overcirc#1{\buildchar{#1}{\circ}{}}

\def\vbig#1#2{{\vbigd@men=#2\divide\vbigd@men by 2%
\hbox{$\left#1\vbox to \vbigd@men{}\right.\n@space$}}}

%
%
\newcount\fdefcontre \newcount\fdefcount \newcount\indcount
\newread\filefdef  \newread\fileftmp  \newwrite\filefdef
\newwrite\fileftmp     \def\strip#1*.A {#1}
\def\futuredef#1{\beginfdef
\expandafter\ifx\csname#1\endcsname\relax%
{\immediate\write\fileftmp {#1*.A}
\immediate\write16 {Panda Warning - fdef: macro "#1" on page
\the\pageno \space undefined}
\ifnum\draftnum=0 \expandafter\xdef\csname#1\endcsname{(?)}
\else \expandafter\xdef\csname#1\endcsname{(#1)} \fi
\global\advance\fdefcount by 1}\fi   \csname#1\endcsname}

\def\beginfdef{\ifnum\fdefcontre=0
\immediate\openin\filefdef \jobname.fdef
\immediate\openout\fileftmp \jobname.ftmp
\global\fdefcontre=1  \ifeof\filefdef \immediate\write16 {Panda
WARNING - fdef: file \jobname.fdef not found, run TeX again}
\else \immediate\read\filefdef to\spzzttrra
\global\advance\fdefcount by \spzzttrra
\indcount=0      \loop\ifnum\indcount<\fdefcount
\advance\indcount by 1   \immediate\read\filefdef to\spezttrra
\immediate\read\filefdef to\sppzttrra
\edef\spzzttrra{\expandafter\strip\spezttrra}
\immediate\write\fileftmp {\spzzttrra *.A}
\expandafter\xdef\csname\spzzttrra\endcsname{\sppzttrra}
\repeat \fi \immediate\closein\filefdef \fi}
\def\endfdef{\immediate\closeout\fileftmp   \ifnum\fdefcount>0
\immediate\openin\fileftmp \jobname.ftmp
\immediate\openout\filefdef \jobname.fdef
\immediate\write\filefdef {\the\fdefcount}   \indcount=0
\loop\ifnum\indcount<\fdefcount    \advance\indcount by 1
\immediate\read\fileftmp to\spezttrra
\edef\spzzttrra{\expandafter\strip\spezttrra}
\immediate\write\filefdef{\spzzttrra *.A}
\edef\spezttrra{\string{\csname\spzzttrra\endcsname\string}}
\iwritel\filefdef{\spezttrra}
\repeat  \immediate\closein\fileftmp \immediate\closeout\filefdef
\immediate\write16 {Panda Warning - fdef: Label(s) may have changed,
re-run TeX to get them right}\fi}
\def\iwritel#1#2{\newlinechar=-1
{\newlinechar=`\ \immediate\write#1{#2}}\newlinechar=-1}
\global\fdefcontre=0 \global\fdefcount=0 \global\indcount=0
%
%
\null
%
%
%
%
%
%
\loadamsmath
\loadeuler
%
%
%
%
\def\hkm{\widetilde{\eufm h}}
\def\hh{\widetilde{\eufm h}}
\def\nn{\widetilde{\eufm n}}
\def\gkm{\widetilde{\eufm g}}
\def\g{{\eufm g}}
\def\h{{\eufm h}}
\def\ss{\widehat{\eufm s}}
\def\gg#1{\widetilde{\eufm g}_{#1}'}
\def\GG{{\>\widetilde{G}}}
\def\ie{{\it i.e.\/}}
\def\eg{{\it e.g.\/}}
\def\s{{\bf s}}
\def\sh{{\bf s}_{\rm h}}
\def\m{{\bf m}}
\def\HSA{{\cal H}}
\def\LAX{{\cal L}}
\def\ad{{\rm ad\>}}
\def\Ker{{\rm Ker\/}}
\def\Im{{\rm Im\/}}
\def\pa{\partial}
\def\Pos{{\rm P}_{\geq0[\s]}}
\def\Neg{{\rm P}_{<0[\s]}}
\def\z{{\eufm Z}}
\def\bil#1#2{\bigl(#1 \> \vert \> #2 \bigr)}
%
%
%
%
\pageno=0
\nopagenumbers{\baselineskip=12pt
\line{\hfill US-FT/4-98}
\line{\hfill\tt hep-th/9809052}
\line{\hfill September 1998}\jump
\ifdoublepage \bjump\bjump\bjump\else\jump\vfill\fi
\centerline{\capstwo Tau-Functions generating the Conservation Laws}
\sjump
\centerline{\capstwo for Generalized Integrable Hierarchies}
\sjump
\centerline{\capstwo of KdV and Affine-Toda type} 
\bjump\jump
\centerline{{\scaps J. Luis Miramontes}}
\jump\jump
\centerline{\sl Departamento de F\'\i sica de Part\'\i
culas,}
\centerline{\sl Facultad de F\'\i sica,}
\centerline{\sl Universidad de Santiago,}
\centerline{\sl E-15706 Santiago de Compostela, Spain}
\sjump
\centerline{\tt e-mail: miramont@fpaxp1.usc.es} 
\bjump\bjump
\ifdoublepage
\vfill
{\noindent
\line{September 1998\hfill}}
\eject\null\vfill\fi
\centerline{\capsone ABSTRACT}\jump

\noindent
For a class of
generalized integrable hierarchies associated with affine (twisted or
untwisted) Kac-Moody algebras, an explicit representation of their local
conserved densities by means of a single scalar tau-function is deduced.
This tau-function acts as a partition function for the conserved
densities, which fits its potential interpretation as the effective
action of some quantum system. The class consists of multi-component
generalizations of the Drinfel'd-Sokolov and the two-dimensional affine
Toda lattice hierarchies. The relationship between the former and the
approach of Feigin, Frenkel and Enriquez to  soliton equations of KdV and
mKdV type is also discussed. These results considerably simplify the
calculation of the conserved charges carried by the soliton solutions to
the equations of the hierarchy, which is important to establish their
interpretation as particles. By way of illustration, we calculate the
charges carried by a set of constrained KP solitons recently
constructed.
\jump\sjump
\sjump
\sjump
\vfill
\ifdoublepage \else
\noindent
\line{September 1998\hfill}\fi
\eject}
%
%

\yespagenumbers\pageno=1
\footline={\hss\tenrm-- \folio\ --\hss}

\chapter{Introduction}

The purpose of this article is to show that the local conserved
densities of a large class of integrable hierarchies of zero-curvature
equations are given by the second order partial derivatives of a single
(complex) scalar tau-function. In other words, this tau-function
provides a generating function for the infinite conserved densities of
the hierarchy, which matches the interpretation of tau-functions as
the effective actions of certain quantum systems where the conserved
densities become correlation functions~[\Ref{TAUQ},\Ref{DIJK},\Ref{SW}].
Among the hierarchies included in this class, there are
(multi-component) generalizations of the Korteweg-de Vries (KdV), the
modified Korteweg-de Vries (mKdV), the non-linear Schr\"odinger (NLS),
the constrained Kadomtsev-Petviashvili (cKP), and the affine Toda
equations associated with affine (twisted or untwisted) Kac-Moody
algebras. Our results generalize the well known relation between
conserved quantities and tau-functions in the context of the
Kadomtsev-Petviashvili (KP) hierarchy and its reductions of KdV type,
whose important role in the matrix model approach to two-dimensional
quantum and topological gravity is well known~[\Ref{DIJK}].

Recall the definition of the KP hierarchy by means of a
pseudo-differential scalar Lax operator $L=\pa + u_0 \pa^{-1} + u_1
\pa^{-2} + \cdots$. Its conserved densities $J_k = {\rm res\/}(
L^k)$ are related to the single tau-function of the hierarchy
by means of $J_k = \pa^2 \ln \tau/\pa x_1 \pa x_k$~[\Ref{DICK}]. 
The extension of this relationship to other integrable hierarchies
is far from obvious. Actually, in contrast with the KP hierarchy, the
number of tau-functions required to reconstruct the variables of the
zero-curvature formalism is often larger that one, and it is not {\it a
priori} clear if any of them plays a more fundamental role than the
others. This already occurs in the hierarchies associated with
$sl(2,{\Bbb C})$. For example, although the KdV hierarchy can be
described by a single tau-function related to the original variable by
the celebrated formula $u=2\pa_{x}^2 \ln \tau$, the mKdV hierarchy
already requires two tau-functions, and the change of variables is $q=
\pa \ln \tau_0/\tau_1$. Similarly, the complexified version
of the NLS equation, known as the AKNS equations, involves three
tau-functions. Moreover, the number of tau-functions usually increases
with the size of the algebra the hierarchy is associated with. A well
known example of this is provided by the two-dimensional affine Toda
lattice associated with a Lie algebra $\g$ of rank $r_\g$, which needs
$r_{\g}+1$ tau-functions attached to the vertices of the extended 
Dynkin diagram of~$\g$.
 
Out of the different methods used to investigate integrable
hierarchies of equations, the tau-function approach has proved to be
particularly useful concerning their applications in particle physics
and string theory. This method started with the work of Hirota as a
change of variables designed to simplify the direct construction of
various types of solutions to the equations of the hierarchy; especially,
the multi-soliton solutions~[\Ref{HIR}]. Far from being just a new
solution method, the tau-function approach uncovered a deep underlying
structure of the integrable hierarchies where affine Kac-Moody algebras
(central extensions of loop algebras) play a central role. This
connection was already apparent in the pioneering work of the Kyoto
school~[\Ref{KYOTO}], and made even clearer by many authors like Segal
and Wilson~[\Ref{SEGWIL}], Wilson~[\Ref{WILTAU}], and Kac and
Wakimoto~[\Ref{KW}]. According to their work, the tau-function formalism
concerns a particular class of solutions specified by the elements of
an infinite manifold associated with a Kac-Moody group (centrally
extended loop group). For example, in the particular case of the KdV
equation, the manifold is a Grassmannian and the class includes the
multi-soliton solutions and the algebro-geometric solutions
of Krichever~[\Ref{SEGWIL}]. A systematic description of the relation
between the tau-functions and the variables of the zero-curvature
formalism is provided by the work of Wilson~[\Ref{WILTAU}]. There, the
Kac-Moody group is used to induce a group of transformations, known as
`dressing transformations', acting on the space of solutions. Then, the
relevant class of solutions is generated by the action of this group on
some trivial `vacua'. Wilson's ideas have been recently
used in~[\Ref{JMP}] to develop the tau-function formalism for a large
class of integrable hierarchies of zero-curvature equations
characterized by means of their vacuum solutions . Actually, the results
of this last reference apply to the hierarchies that will be considered
in this article.

The tau-function or Hirota method has been used to construct
the soliton solutions to a large number of equations and, in
particular, to the generalizations of the sine-Gordon
equation: the equations-of-motion of the abelian and non-abelian affine
Toda field theories~[\Ref{SOLTAU},\Ref{CAT}]. They are relevant to
particle physics mainly because of two reasons. First, as
two-dimensional models where the spectrum contains not only perturbative
particles but also solitons (\eg, kinks, monopoles, dyons,~$\ldots$).
Second, because some of them describe massive integrable perturbations
of two-dimensional conformal field theories.

Our results show that all the information about the
soliton solutions is encoded in a single (complex) scalar tau-function
and, in particular, that the value of the conserved charges they carry
(\eg, their energy and momentum) is given by the asymptotic
behaviour of this function through expressions like
$$
Q_{n}^{\pm}\> =\> \int_{-\infty}^{+\infty}\> dx_\pm \>
J_{\pm}^{(\pm1)}[\pm n, \m] \> =\> -{N_{\s'}\over n N_\m}\> \pa_{\pm n}
\ln
\tau_{\overline{\m}}^{\rm T}(h) \Bigm|_{x_\pm = -\infty}^{x_\pm =
+\infty} 
$$
(see eq.~(6.24)). This is a generalization of the well
known properties of the energy-momentum tensor of the affine Toda
equations restricted to the, so called, `solitonic specialization' of
their general solution originally proposed by Olive {\it et
al.\/}~[\Ref{EMSPEC},\Ref{SOLSPEC}]. This class of solutions is
conjectured to include the multi-soliton solutions, which
has been extensively checked for the $A_n$~theories~[\Ref{ANSOL}]. In
the abelian affine Toda theories, this conjecture is supported by the
explicit evaluation of the energy-momentum tensor for the general
solution. It splits into two parts: a `radiation' piece that vanishes
for solutions describing solitons and an `improvement' which, after
integration, yields the energy and momentum carried by the solitons as
surface terms~[\Ref{EMSPEC}]. In the non-abelian affine Toda theories,
the explicit evaluation of the energy-momentum tensor has not been
actually carried out, and the split is usually justified by means of
their relation with an extension of the theory that exhibits conformal
symmetry~[\Ref{CAT},\Ref{LUIZ}]. Restricted to the (abelian and
non-abelian) affine Toda equations, the class of solutions that we will
consider coincides with the solitonic specialization. Hence, our results
provide an alternative proof that the energy-momentum tensor carried by
these solutions is actually given by an improvement term and, moreover,
they also show that the same is true for all the other conserved
densities.

The construction of soliton solutions in two-dimensional field theories
is not the only way in which tau-functions arise in the context of
particle physics. Nowadays, there is an increasing number of quantum
systems whose effective action turns out to be the tau-function of some
classical integrable system~[\Ref{TAUQ}]. A renowned example of this is
provided by two-dimensional quantum and topological gravity (and its
coupling to topological matter), whose partition function is a
tau-function of the KdV hierarchy (and its generalizations)~[\Ref{DIJK}].
Another important example is given by the exact solution of $N=2$
supersymmetric Yang-Mills theory in four dimensions constructed by
Seiberg and Witten. There, the pre-potential of the low-energy effective
action is found to be related with the tau-functions of the
one-dimensional periodic Toda lattice hierarchy~[\Ref{SW}], which
also play an important role in the topologically twisted version of these
theories: namely, Donalson theory~[\Ref{MARCOS}]. It is worth noticing
that the latter example concerns a four-dimensional quantum field
theory, which illustrates that even though the relevant classical
integrable hierarchies of equations are formulated in one or two
dimensions, their applications are not necessarily restricted to
low-dimensional quantum systems. In this respect, our results identify a
subset of tau-functions that operate as generating functions for the
conserved densities of the hierarchy in the sense that the latter are
given by the partial derivatives of the former:
$$
J^{(j)}[n;\m]\> =\> -\> {N_{\s'}\over n\> N_{\bf m}}\> \pa_{j}\> \pa_{n}
\ln \tau_{\overline{\m}}^{\rm DS}(h)
$$
(see eq.~(4.14) and~(5.6)). This looks like the classical
counterpart of their interpretation as the partition function of some
quantum system whose coupling constants would be provided by the times of
the hierarchy.

The article is organized as follows. In Section~2, we describe the two
types of integrable hierarchies of zero-curvature equations
that will be considered. The first corresponds to the generalized
Drinfel'd-Sokolov hierarchies constructed
in~[\Ref{GEN},\Ref{GEN2},\Ref{GENTAU}], which, in particular, provide
multi-component generalizations of the KdV, mKdV, NLS, and cKP equations.
The second is a new non-abelian generalization of the two-dimensional
affine Toda lattice hierarchy that can be viewed as the result of
coupling two generalized mKdV hierarchies: the resulting hierarchies
always include a non-abelian affine Toda (NAAT) equation. They exhibit
an infinite number of local conserved densities whose precise definition
by means of the Drinfel'd-Sokolov procedure is discussed in Section~3.
The core of the article is Section~4 where we specify the class of
solutions and calculate the corresponding conserved densities. The
solutions are singled out by requiring that all the conjugations that
arise in the Drinfel'd-Sokolov approach make sense in the Kac-Moody
group. They turn out to be generated by the action of the group of
dressing transformations, in the manner described by Wilson, on the
vacuum solutions of the hierarchies. Moreover, they can also be viewed
as a generalization of both the class of solutions to the KdV equation
considered by Segal and Wilson, and the solitonic specialization of the
general solution to affine Toda equations proposed by Olive {\it et
al\/}. In the particular case of the generalized Drinfel'd-Sokolov
hierarchies, we also show that their restriction to these solutions fits
into the approach of Feigin, Frenkel and Enriquez to soliton equations
of KdV and mKdV type~[\Ref{FRENKEL}]. In Section~5, we discuss the
relations between different integrable hierarchies implied by the
results of the previous section. They generalize the well known
connection between the conserved quantities of the KdV, mKdV, and
sine-Gordon equations originally pointed out by Chodos~[\Ref{CHODOS}].
For the purpose of illustration, in Section~6 we specialize our results
to the original Drinfel'd-Sokolov KdV hierarchies, to the non-abelian
affine Toda equations, and to the calculation of the charges carried by
the solitons of the cKP hierarchy recently constructed by Aratyn {\it et
al.}~[\Ref{LUIZKP}]. Our conventions and some relevant properties of
Kac-Moody algebras are collected in the appendix.

\chapter{The zero-curvature hierarchies}

We will consider two types of integrable systems. The first one 
corresponds to the generalized Drinfel'd-Sokolov hierarchies constructed
in~[\Ref{GEN},\Ref{GEN2},\Ref{GENTAU}], which includes most of the
generalizations of the original Drinfel'd-Sokolov
construction~[\Ref{DS}] available in the literature. In order to
illustrate the extent of this class, let us recall their interpretation
by means of pseudo-differential operators, which has been investigated
independently by Feh\'er {\it et al.\/}~[\Ref{FHM}] and Aratyn {\it et
al.\/}~[\Ref{ARA}]. Their results show that some of these hierarchies
provide matrix generalizations of both the standard Gelfand-Dickey
hierarchies and the constrained KP hierarchy. Moreover, the construction
of~[\Ref{GEN},\Ref{GEN2}] can be generalized once more in a quite
natural way to interpret certain non-standard Gelfand-Dickey
hierarchies~[\Ref{DELD}]. Finally, the link between the hierarchies
of~[\Ref{GEN},\Ref{GEN2}] and the Adler-Kostant-Symes (AKS)
construction has been explained in~[\Ref{FHM},\Ref{AKS}],
where it is shown that the former corresponds to the `nice reductions'
of the AKS system that exhibit local monodromy invariants.  

The second type of integrable systems to be considered is, as far as we 
know, new. It consists of hierarchies that include a generalized
non-abelian affine Toda equation together with a generalized mKdV
equation. We will call them Toda-mKdV hierarchies, and they can be
viewed as the result of coupling two different mKdV hierarchies.
Alternatively, they can also be described as non-abelian generalizations
of the usual two-dimensional Toda lattice hierarchy~[\Ref{TL}]. 

All these equations admit a unified group theoretical description where 
they are associated with the different Heisenberg subalgebras
$\HSA[\s']$ of a (twisted or untwisted) affine Kac-Moody algebra $\gkm =
\g^{(r)}$ (our conventions concerning Kac-Moody algebras are collected
in the appendix). For a particular hierarchy, there will be a flow
associated with each element $b_j$ in a properly defined subalgebra of
$\HSA[\s']$. These flows will be defined by Lax operators of the form
$$
\LAX_j\> =\> \pa_j\> - \> A_j
\nfr{LaxOp}
where $\pa_j = \pa/ \pa t_j$. Then, the integrable hierarchy of equations
consists of a set of zero-curvature equations 
$$
[\LAX_j \>, \> \LAX_k]\> =\> 0\>,
\nfr{ZeroCurv}
which can also be viewed as the integrability conditions of an associated
linear problem
$$
\bigl( \pa_j \> - \> A_j \bigr) \> \Psi \> =\> 0
\nfr{LinProb}
where $\Psi$ is the, so called, `Baker-Akhiezer' or wave function. 

An important common feature of all these hierarchies is that they admit 
trivial `vacuum solutions' of the form~[\Ref{JMP}]
$$
A_{j}^{({\rm vac})}\> =\> b_j  \> +\> {1\over2}\> (\vert j\vert -
j)\> t_{- j}\> K 
\nfr{VacLax}
or, equivalently,
$$ 
\Psi^{({\rm vac})}\>  =\> \exp\Bigl[ \>\sum_{j\geq0} t_j \> b_j\> 
\Bigr]\> \exp\Bigl[ \>\sum_{j>0} t_{-j} \> b_{-j}\> \Bigr] \>. 
\nfr{VacLin}
Actually, as will be explained in Section~4, most of our results 
will be restricted to the special class of solutions generated from these
trivial ones through the action of the group of dressing
transformations~[\Ref{JMP}]. As it is well known, it is in 
this context where a systematic description of $\tau$-functions
can be given~[\Ref{WILTAU},\Ref{KW}]. 

\section{Generalized Drinfel'd-Sokolov hierarchies}

In this first class, a particular hierarchy will be specified by a 
constant element $\Lambda  = b_i \in \HSA[\s']$ of positive grade $i>0$,
and an additional gradation $\s$ such that $\s\preceq \s'$ with respect
to the  partial ordering defined in the appendix (Section~A.1). This
class contains generalizations of the mKdV and KdV hierarchies, which
are recovered when 
$\s$ is maximal ($\s\simeq \s'$) and minimal ($s_j\not=0$ for a single
$j$), respectively. Otherwise, the auxiliary gradation $\s$ sets the
`degree of modification' and the hierarchy is said to be a partially
modified KdV (pmKdV) hierarchy. In~[\Ref{GEN},\Ref{GEN2}], the
construction was undertaken in the loop algebra or Kac-Moody algebra
with zero centre, while in~[\Ref{GENTAU},\Ref{JMP}] it was presented in
a representation independent way in the full Kac-Moody algebra including
the centre. However, in~[\Ref{GENTAU}] it was implicitly assumed that
the hierarchy was of `type-I' ($\Lambda$ regular and, hence, $\Ker(\ad
\Lambda) =\HSA[\s']$), while the results of ref.~[\Ref{JMP}] are
restricted to generalized mKdV hierarchies ($\s\simeq \s'$). In
contrast, in this paper we will be concerned with the most general case.
Therefore, the choice of $\s$ and $\Lambda$ will be constrained only by
the condition
$$
\Ker(\ad \Lambda) \cap P \> = \emptyset \>,
\nfr{Degen}
where $P$ will be defined in eq.~(2.18), which is needed to ensure
that the equations of the hierarchy are of polynomial form (see
eq.~(2.19)).

Since we will need it to develop the Toda-mKdV hierarchies, we will
briefly summarized the construction and main features of these
hierarchies in the general case following
refs.~[\Ref{GEN},\Ref{GEN2},\Ref{GENTAU},\Ref{JMP}]. Consider the
Lax operator
$$ 
L \> =\>  \pa_x \> -\> \Lambda \>-\> q \>,
\nfr{LaxDS}
where $\pa_x = {\pa/ \pa x}$ and $q$ is a function taking values on the
subspace $\gg{\geq0}(\s) \cap \gg{<i}(\s')$.  The construction of the
hierarchy is based on the existence of a  transformation that `abelianizes'
the Lax operator {\it \`a la\/} Drinfel'd-Sokolov, \ie,    
$$ 
\Phi^{-1}\> L\> \Phi\> =\> \pa_x \> -\>
\Lambda \> -\> h^\Phi\>, 
\nfr{Abelianize}
where $\Phi = {\rm e}^y$ is a function with $y \in \gg{<0}(\s')$, and 
$h^\Phi $ takes values on $\Ker(\ad \Lambda) \cap \gg{<i}(\s')$.\note{By
$\Phi^{-1} L \Phi$ we mean the exponentiated adjoint action defined the
formal power series expansion
$$
\Phi^{-1} L \Phi = \exp(-\ad y) (L) = L - [y,L] + {1\over 2}[y,[y,L]] -
{1\over 3!}[y,[y,[y,L]]] +
\cdots\>.
$$
}
This transformation can be constructed because $\Lambda$ is semisimple
and, hence, $\gg{}$ admits the decomposition $ \gg{} = \Ker(\ad \Lambda) 
+ \Im(\ad \Lambda)$~[\Ref{DS},\Ref{WILSON},\Ref{GEN}]. Recall that the
subalgebra $\Ker(\ad \Lambda)$ is not abelian in general, and $\Ker(\ad
\Lambda)\cap \Im(\ad \Lambda)={\Bbb C}K$ (see the appendix, Section~A.2).
At this point, it is important to  notice that $q$ always has a
component along the centre, say $q_c\> K$. Then, it is straightforward
to realize that neither $\Phi$ nor $h^\Phi - q_c K$ depend on $q_c$.  

Let $ \z_\Lambda = {\rm Cent\/}(\Ker(\ad \Lambda))$ be the subalgebra of
$\HSA[\s']$ generated by all the $b_{j}$'s that commute with all the 
elements in $\Ker(\ad \Lambda)$ up to the centre, \ie, 
$$ 
\z_\Lambda \> =\>  {\Bbb C}\> K
\>\oplus \> \sum_{j\in E_\Lambda} {\Bbb C} \> b_j \quad{\rm and} \quad 
[b_j\>,\> \Ker(\ad \Lambda)] \in {\Bbb C}\> K
\efr
for all $j\in E_\Lambda$. The flows of the generalized
Drinfel'd-Sokolov hierarchy corresponding to the data $\{ \gkm =\g^{(r)},
\HSA[\s'],\Lambda, \s \}$  are associated with the non-central graded
elements of the  subalgebra
$$
\ss_{{\rm DS}}\> =\> \z_\Lambda \cap \gg{\geq0}(\s')\>. 
\nfr{Shat}
The flow corresponding to $b_j\in \ss_{{\rm DS}}$ is defined by means of a Lax
operator $\LAX_j = \pa_j -A_j$ where 
$$
A_j\> =\> \Pos(\Phi\> b_{j}\> \Phi^{-1})\> + \> h^{\Phi}_{c}(j)\> K\>,
\nfr{LaxPos}
and $h^{\Phi}_{c}(j)$ is an, {\it a priori}, arbitrary complex function. 

Recall eq.~\Abelianize, and consider
$$
\Phi^{-1}\> \LAX_j \> \Phi \> =\> \partial_j\> - \> b_j \> - \>  h^\Phi
(j)\> -\> h^{\Phi}_{c}(j)\> K\>,
\nfr{Abelj}
where $h^\Phi (j) \in \gg{<0}(\s')$ as a straightforward consequence
of~\LaxPos. Then, it is not difficult to check that $[L, \LAX_j]=0$
implies that 
$$
h^\Phi(j) \in \Ker(\ad \Lambda) \cap \gg{<0}(\s')
\efr
and, moreover,
$$
[\partial_x \> -\> h^\Phi\>, \> \partial_j\> - \> h^\Phi (j)\> -\>
h^{\Phi}_{c}(j)\> K]\> =\> {r\over N_{\s'}}\> \Bigl( j\bil{b_j}{h^\Phi}\> -\>
i\bil{\Lambda}{h^\Phi (j)} \Bigr)\> K \>.
\nfr{Centres}
In particular, this shows that 
$$
\partial_j\> \bil{b_n}{h^\Phi}\> =\> \partial_x \> \bil{b_n}{h^\Phi(j)}
\quad {\rm for\;\;  each}\quad b_n\in \z_\Lambda \cap \gg{>0}(\s')\>,
\nfr{DensOr}
which means that the components of ${\rm P}_{<0[\s']}(h^\Phi)$ along
$\z_\Lambda$ provide the infinite conserved densities of the hierarchy.

It will be emphasized in the next section that the choice of $\Phi$ is not 
unique. Actually, since they do not change the form of~\Abelianize, it is
defined modulo the transformations $\Phi \mapsto
\Phi e^\eta$, where $\eta$ is an arbitrary function that takes values on
$\Ker(\ad \Lambda) \cap \gg{<0}(\s')$. This justifies the notation
like $h^\Phi$ and $h^\Phi (j)$. However, since $b_j \in \z_\Lambda$, 
$$
(\Phi e^\eta)\> b_j \> (\Phi e^\eta)^{-1} \> =\> \Phi\> b_{j}\> \Phi^{-1}\>
-\> j\> {r\over N_{\s'}}\>\bil{b_j}{\eta}\> K
\nfr{Indep}
and, hence, different choices of $\Phi$ just lead to different values of 
the arbitrary complex functions $h^{\Phi}_{c}(j)$, which will not modify
the  hierarchy of integrable equations. In any case, the abelianization 
can always be chosen such that both $\Phi={\rm e}^y$ and $h^\Phi - q_c\>
K$ are `local' functionals of the components of $\widetilde{q}= q - q_c\>
K$, simply  by enforcing the condition $y\in \Im(\ad
\Lambda)$~[\Ref{GEN}]. At the end of the day, this is the reason why
these hierarchies exhibit local monodromy invariants provided by
the components of ${\rm P}_{<0[\s']}(h^\Phi)$ along $\z_\Lambda$ when
$\Phi$ is chosen as indicated.  

In this construction, the function $q$ plays the role of the
phase space coordinate. However, there exist symmetries corresponding to
the gauge transformations 
$$
L \mapsto U\> L \> U^{-1}\>,
\nfr{Gauge}
in the sense that they do not change the form of the Lax operator. 
In~\Gauge, $U$ is a function that takes values on the group
generated by the nilpotent finite dimensional subalgebra
$$
P\> =\> \gg{0}(\s) \cap \gg{<0}(\s') \>.
\nfr{GaugeGen}
Therefore, the actual phase space of the hierarchy is the set
of gauge equivalent classes, and the equations are to be thought of as a
set of partial differential equations on a consistent gauge slice of $q$,
denoted $q^{\rm GS}$~[\Ref{DS},\Ref{GEN}]. Since $\Lambda$ satisfies
eq.~\Degen, this gauge slice can be chosen following the standard
Drinfel'd-Sokolov procedure~[\Ref{ANALS}]. Then, if we split the component
along the centre, $q^{\rm GS} =
\widetilde{q\/}^{\rm GS} + q_{c}\> K$, the equations of the hierarchy can be
put in the form
$$ 
{\pa \widetilde{q}^{\> \rm GS} \over \pa t_{j} } \> =\> F_j\Bigl( 
\widetilde{q}^{\> \rm GS},  {\pa \widetilde{q}^{\> \rm GS}\over \pa x},  
{\pa \widetilde{q}^{\> \rm GS}\over \pa x}\>  ,\>  {\pa^2 
\widetilde{q}^{\> \rm GS}\over  \pa x^2}, \ldots\Bigr)\>, 
\nfr{PdeA} 
for some polynomial functions $F_j$, \ie, differential polynomials in the
components of $\widetilde{q}^{\> \rm GS}$. Notice that 
the evolution of $\widetilde{q}^{\> \rm GS}$ is
independent of $q_c$, which explains why these hierarchies can be
associated with both Kac-Moody algebras or loop algebras.  In
contrast, the evolution of $q_c$ is of the form
$$
{\pa q_c \over \pa t_{j} } \> =\> G\Bigl( 
\widetilde{q}^{\> \rm GS},  {\pa \widetilde{q}^{\> \rm GS}\over \pa x},  
{\pa \widetilde{q}^{\> \rm GS}\over \pa x}\>  ,\>  {\pa^2 
\widetilde{q}^{\> \rm GS}\over  \pa x^2}, \ldots\Bigr)\>, 
\nfr{PdeB}
which shows that $q_c$ is completely determined once a solution
of~\PdeA\ is provided. Moreover, the precise form of~\PdeB\ depends on 
the value of the arbitrary functions $h^{\Phi}_{c}(j)$ and, in
particular, they can be chosen such that $G$ is also polynomial.
Consequently, $q_c$ has to be regarded not as an actual degree of
freedom of the hierarchy but just as an auxiliary function.  

Although $\Lambda = b_{i}$, $x$ cannot be identified in general with 
$t_i$. The precise connection between both variables is provided
by~\Abelianize, which implies that  
$$ 
L\> =\> \pa_x \> -\> \Lambda\> - \> q\> =\> {\rm P}_{\geq0[\s]}\Bigl(
\Phi 
\bigl(\pa_x \> -\> h^\Phi \bigr)\Phi^{-1} \Bigr)\> -\> \Pos(\Phi \>
\Lambda \> \Phi^{-1}) \>. 
\nfr{Symm}
In this paper, we will restrict ourselves to hierarchies where
$\pa_x =\pa_i$, which requires to constrain the form of $q$ such that
$$
{\rm P}_{\geq0[\s']}(h^\Phi)\> \in \> {\Bbb C}\> K \>. 
\nfr{ConstDS} 
Notice that eq.~\Centres\ implies that 
$$
\partial_j\> {\rm P}_{\geq0[\s']}(h^\Phi)\> +\> {\rm
P}_{\geq0[\s']}\Bigl([{\rm P}_{>0[\s']}(h^\Phi)\>, \> h^\Phi(j)]\Bigr) \>
\in \> {\Bbb C}\> K \>,
\efr
which shows that the constraints~\ConstDS\ are preserved by all the flows.
Then, following~[\Ref{GENTAU}], the gauge slice will be chosen as follows.
Let us split the component  of $h^\Phi$ along the centre, $h^\Phi =
\widetilde{h}^\Phi + h^{\Phi}_c\> K$, and introduce a (non-local) 
functional of $q$, say $\chi^\Phi= e^\rho$, such that  
$$
\partial_x \> -\> \widetilde{h}^\Phi \>= \> \chi^\Phi \>\pa_x \>
{\chi^\Phi}^{ -1}
\quad {\rm and} \quad  \rho \in \Ker(\ad \Lambda) \cap \gg{<0} (\s') \>.
\nfr{ChiA}
This allows one to write the first term on the right-hand-side of~\Symm\ 
as ${\rm P}_{0[\s]}[( \Phi\chi^\Phi) \>{\pa_x} ( \Phi\chi^\Phi)^{-1}] - 
h_{c}^\Phi\> K $. Then, if $\Phi \chi^\Phi = {\rm e\> }^{v}$, it has
been shown in~[\Ref{GENTAU}] that there always exists a consistent gauge
slice where ${\rm P}_{0[\s]}(v)=0$ and, from now on, we will assume that it
has been chosen. In other words, taking into account~\ConstDS\ and the
non-uniqueness of $\Phi$ (see eq.~\Indep), our gauge slice will be
specified by the constraints 
$$
y \in \gg{<0}[\s]\quad {\rm and} \quad {\rm P}_{\geq0[\s]}(h^\Phi)\> \in
\> {\Bbb C}\> K \>. 
\nfr{ConstPlus} 
where $\Phi = {\rm e}^y$. Therefore, eq.~\Symm\ becomes 
$$
\Lambda \> +\> q^{\rm GS}\> =\> \Pos(\Phi\> \Lambda  \> \Phi^{-1})\> + \>
h^{\Phi}_c \> K \>,   
\efr
which, upon comparison with eq.~\LaxPos, exhibits that $\pa_x=\pa_i$
and, consequently, $h^\Phi =h^\Phi(i) + h_{c}^{\Phi}(i)\> K$.
This implies that all the quantities in the hierarchy depend on $x$ and
$t_i$ through the combination $x+t_i$; however, we will often indicate
only the dependence on either $x$ or $t_i$.

Finally, let us recall that these hierarchies exhibit a
scale invariance, \ie, under the transformation $x\mapsto \lambda x$, for
constant $\lambda$, each quantity in the equations can be assigned a
scaling dimension such that the equations are invariant. Let us split $q$
into components such that $q=\sum_{j<i} \> q^j\> e_j$ and $e_j \in
\gg{j}(\s')$. Then, the hierarchy is invariant under the
transformations~[\Ref{GEN2}]
$$
x\>\mapsto\> \lambda\> x\>, \qquad t_j \>\mapsto\> \lambda^{{j/ i}}\> t_j\>, \quad
{\rm and} \quad q^j \>\mapsto\> \lambda^{{j/ i}-1}\> q^j\>,
\nfr{Scaling}
Notice that, under~\Scaling,\note{In the following expressions, we mean 
$$
\lambda^{{d_{\s'}/ i}}\> L \> \lambda^{-{d_{\s'}/ i}}\>
=\> \exp\bigl( \ln\lambda\> (\ad d_{\s'}/i)\bigr) (L)\>.
$$
}
$$
L\> =\> \partial_x \> -\> \Lambda \> -\> q \longmapsto \lambda^{-1}\>
\lambda^{d_{\s'}/i}\> L \> \lambda^{-d_{\s'}/i}\>,
\efr
which induces the following changes in~\Abelianize:
$$
\Phi \>\mapsto\> \lambda^{d_{\s'}/i}\> \Phi \> \lambda^{-d_{\s'}/i}\>,
\qquad (h^\Phi)^j \>\mapsto\> \lambda^{{j/ i}-1} \> (h^\Phi)^j\>,
\nfr{ScaleH}
where $(h^\Phi)^j$ is the component of $h^\Phi$ with $\s'$-grade~$j$.
Therefore, taking into account~\LaxPos, the transformation~\Scaling\
implies that
$$
\LAX_j \longmapsto \lambda^{-{j/ i}}\> \lambda^{d_{\s'}/i}\> \LAX_j
\> \lambda^{-d_{\s'}/i}\>,
\efr
which do not change neither the form of the Lax operators nor the
zero-curvature equations and, hence, exhibits that~\Scaling\ is an
actual symmetry of these hierarchies.  

\section{Toda-mKdV hierarchies}

Each Toda-mKdV hierarchy will be specified by two
constant elements $\Lambda_+ =b_i \in \HSA[\s']$ and $\Lambda_- =b_{-i} \in 
\HSA[\s']$, with $i>0$, constrained by the condition $\Ker(\ad \Lambda_+) =
\Ker(\ad \Lambda_-)$. Then, consider the two Lax operators
$$
L_\pm = \partial_\pm - \Lambda_\pm - q_\pm
\efr
where $\partial_\pm = \partial/\partial x_\pm$, 
$$
q_+\>= \> \sum_{1\leq j< i} \> F_{j}^+ \> -\> B^{-1}\partial_+ B\>,
\quad {\rm and}\quad
q_-\>= \> \sum_{1\leq j< i} \> F_{j}^- \> +\> \partial_-
B B^{-1} \>.
\nfr{TwoLax}
Here, $F_{j}^\pm$ and $B$ are functions taking values on $\gg{\pm
j}(\s')$ and on the finite Lie group $\GG_0(\s')$ corresponding to the
subalgebra $\gg{0}(\s')$, respectively. They provide the phase space
coordinates of the hierarchy and, in this case, there will not be any
gauge symmetry.

First of all, notice that the zero-curvature condition 
$$
[ L_{+} \> , \> B^{-1}\> L_{-}\> B]\> =\>0
\nfr{NAAT}
provides the equations-of-motion of a generalized non-abelian affine
Toda theory where $\pa_{\pm}$ are the derivatives with respect to the
light-cone variables. The usual abelian affine Toda
equations are recovered with the principal Heisenberg subalgebra and
$\Lambda_\pm = b_{\pm1}$~[\Ref{ABTOD},\Ref{DS}], while its non-abelian
versions involve generic Heisenberg subalgebras together with the
constraints $F_{m}^\pm =0$~[\Ref{LS},\Ref{LUIZ}]. The most
general case with $F_{m}^\pm \not=0$ corresponds to the coupling of the
latter systems with (spinor) matter fields~[\Ref{SAVGERV}].

On the other hand, $L_+$ and $L_-$ can be viewed as the Lax operators of two
different mKdV-type hierarchies within the generalized
Drinfel'd-Sokolov construction, where, in the case of $L_-$, positive grades
have been changed into negative ones and {\it vice versa}. Therefore, they
can be abelianized as in eq.~\Abelianize, which means that there exist two
functions $\Phi_\pm = {\rm e}^{y_\pm}$ with $y_+\in \gg{<0}(\s')$ and
$y_-\in \gg{>0}(\s')$ such that
$$
\Phi_{\pm}^{-1} \> L_\pm \> \Phi_{\pm} \> = \> \partial_\pm \> -\> 
\Lambda_\pm \> -\> h_{\pm}^{\Phi_\pm}\>,
\nfr{AbelTwo}
where $h_{+}^{\Phi_+}$ and $h_{-}^{\Phi_-}$ take values on 
$\Ker(\ad \Lambda_+) \cap \gg{<i}(\s')$ and $\Ker(\ad \Lambda_-) \cap
\gg{>-i}(\s')$, respectively. Again, the choice of $\Phi_\pm$ is not unique,
but they can always be chosen such that both $\Phi_\pm$ and
$h_{\pm}^{\Phi_\pm} - q_{\pm c}K$ are local functionals of the components
of $\widetilde{q}_{\pm}= q_\pm - q_{\pm c}\> K$, simply by taking $y_\pm
\in \Im(\ad \Lambda_\pm)$. In these equations, $q_{\pm c}\> K$ stand for 
the central components of $q_\pm$.

The flows of the Toda-mKdV hierarchy corresponding to the data $\{\gkm
=\g^{(r)}, \HSA[\s'], \Lambda_+,\Lambda_-\}$ are associated with the
(non-central) elements of the subalgebra of $\HSA[\s']$
$$
\eqalign{
\ss_{{\rm T}} \>  =\> \z_{\Lambda_\pm}\> & =\> {\rm Cent\/}(\Ker(\ad
\Lambda_+))\> =\> {\rm Cent\/}(\Ker(\ad \Lambda_-)) \cr
& = \> {\Bbb C}\> K \>\oplus \> \sum_{j\in E_{\Lambda_\pm}} {\Bbb C} \>
b_j\>. \cr}
\nfr{ShatTod}
The flow corresponding to $b_j\in \ss_{{\rm T}}$ is defined by means of the
following Lax operators: 
$$
\LAX_j\> =\> \partial_j\> -\> {\rm
P}_{\geq0[\s']}(\Phi_+\> b_{j}\>
\Phi_{+}^{-1})\> - \> h^{\Phi_+}_{c}(j)\> K \>,
\nfr{LaxNegA}
if $j\in E_{\Lambda_\pm}\geq0$, and
$$
\LAX_{j}\> =\>B^{-1}\> \Bigl[ \partial_{j}\> -\> {\rm
P}_{\leq0[\s']}(\Phi_-\> b_{{j}}\>
\Phi_{-}^{-1})\> - \> h^{\Phi_-}_{c}(j)\> K \Bigr]\> B\>, 
\nfr{LaxNegB} 
if $j\in E_{\Lambda_\pm}<0$, where $h^{\Phi_\pm}_{c}(j)$ are arbitrary
complex functions. Moreover, the evolution of $B$ will be constrained such
that
$$
B^{-1}\> \partial_j B\> = \> -\> {\rm P}_{0[\s']}(\Phi_+\>
b_{j}\>
\Phi_{+}^{-1}) - \> h^{\Phi_+}_{c}(j)\> K \>, 
\nfr{FlowBa}
if $j\geq0$, and
$$
\partial_{j} B\> B^{-1}\> =\> {\rm P}_{0[\s']}(\Phi_-\> b_{{j}}\>
\Phi_{-}^{-1})\> + \> h^{\Phi_-}_{c}(j)\> K\>, 
\nfr{FlowBb} 
if $j<0$. Then, if 
$$
\LAX_j \>=\> \pa_j \> -\> A_j \qquad {\rm and}\qquad B\> \LAX_j\> B^{-1}
\>=\> \pa_j \> -\> \widehat{A}_j \>,
\nfr{Names} 
it is straightforward to check that $A_{j} \in \gg{<0}(\s')$ if $j<0$, and
$\widehat{A}_k \in \gg{>0}(\s')$ for $k\geq0$. 

Recall eq.~\AbelTwo\ and consider
$$
\Phi_{+}^{-1}\> \LAX_j \Phi_{+}\> =\> \partial_j \> -\> b_j\> -\>
h_{+}^{\Phi_+}(j)\> -\> h_{c}^{\Phi_+}(j) K\>.
\nfr{AbeljTp}
As a consequence of the form of the Lax operators, it is not difficult to
check that
$h_{+}^{\Phi_+}(j) \in \gg{<0}(\s')$ for all $j$, and
that $h_{c}^{\Phi_+}(k)=0$ for $k<0$. Then, $[L_+,\LAX_j]=0$ implies that 
$$
h^{\Phi_+}_+(j)\> \in\> \Ker(\ad \Lambda_\pm)\cap \gg{<0}(\s')
\efr
for all $j$ and, moreover,
$$
\eqalignno{
& [\partial_+ \> -\> h_{+}^{\Phi_+}\>, \> \partial_j\> - \> h_{+}^{\Phi_+}
(j)\> -\> h_{c}^{\Phi_+}(j) K]\> =\> -\> i\> \delta_{i+j,0}\> K\cr 
\noalign{\vskip 0.3cm}
& \qquad \qquad +\> {r\over N_{\s'}}\> \Bigl( j\bil{b_j}{h_{+}^{\Phi_+}}\>
-\> i\bil{\Lambda_+}{h_{+}^{\Phi_+}(j)} \Bigr)\> K \>, &\nameali{CentresTa}
\cr}
$$ 
which is the analogue of eq.~\Centres. Therefore, this shows that
$$
\partial_j\> \bil{b_n}{h_{+}^{\Phi_+}}\> =\> \partial_+ \>
\bil{b_n}{h_{+}^{\Phi_+} (j)}
\nfr{DensOrA}
for each $n\in E_{\Lambda_\pm}>0$ and $ j\in E_{\Lambda_\pm}$,
which means that the components of ${\rm
P}_{<0[\s']}(h_{+}^{\Phi_+})$ along $\z_{\Lambda_\pm}$ provide an infinite
set of conserved densities with respect to $\partial_+$. 

Similarly, consider
$$
\Phi_{-}^{-1}\> B\> \LAX_j\> B^{-1}\>  \Phi_{-}\> =\> \partial_j \> -\>
b_j\> -\> h_{-}^{\Phi_-}(j)\> -\> h_{c}^{\Phi_-}(j) K\>.
\nfr{AbeljTn}
Then, $[L_-,\LAX_j]=0$ implies that 
$$
h^{\Phi_-}_-(j)\> \in\> \Ker(\ad \Lambda_-)\cap \gg{>0}(\s')
\efr
for all $j$, $h_{c}^{\Phi_-}(k)=0$ for $k\geq0$, and
$$
\eqalignno{
& [\partial_- \> -\> h_{-}^{\Phi_-}\>, \> \partial_j\> - \> h_{-}^{\Phi_-}
(j)\> -\> h_{c}^{\Phi_-}(j) K]\> =\>  i\> \delta_{i-j,0}\> K\cr 
\noalign{\vskip 0.3cm}
& \qquad \qquad +\> {r\over N_{\s'}}\> \Bigl( j\bil{b_j}{h_{-}^{\Phi_-}}\>
+\> i\bil{\Lambda_-}{h_{-}^{\Phi_-}(j)} \Bigr)\> K \>. &\nameali{CentresTb}
\cr}
$$ 
Therefore, 
$$
\partial_j\> \bil{b_n}{h_{-}^{\Phi_-}}\> =\> \partial_- \>
\bil{b_n}{h_{-}^{\Phi_-} (j)}
\nfr{DensOrB}
for each $n \in E_{\Lambda_\pm}<0$ and $j\in E_{\Lambda_\pm}$ and, hence,
the components of ${\rm P}_{>0[\s']}(h_{-}^{\Phi_-})$ along
$\z_{\Lambda_\pm}$ provide another infinite set of conserved densities; but
now  with respect to $\partial_-$.

In the following, we will restrict ourselves to the Toda-mKdV hierarchies
where $\pa_\pm=\pa_{\pm i}$. The precise connection is
provided by~\AbelTwo, which implies that
$$
\eqalignno{ 
& L_+ \> =\> \pa_+ \>  -\> {\rm P}_{\geq0[\s']}(\Phi_+
\> \Lambda_+ \> \Phi_{+}^{-1}) \> -\>  {\rm
P}_{\geq0[\s']}\left(\Phi_+ \> h_{+}^{\Phi_+} \> \Phi_{+}^{-1} \right)\>,
\cr & L_- \> =\> \pa_- \>  -\> {\rm P}_{\geq0[\s']}(\Phi_{-}
\> \Lambda_- \> \Phi_{-}^{-1}) \> -\>  {\rm
P}_{\leq0[\s']}\left(\Phi_{-} \> h_{-}^{\Phi_-} \> \Phi_{-}^{-1}
\right) \>. &\nameali{SymmT} \cr}
$$
Therefore, the identification $\pa_\pm = \pa_{\pm i}$ requires to
constrain the functions $F_{j}^\pm$ and $B$ by means of
$$
{\rm
P}_{\geq0[\s']}\left( h_{+}^{\Phi_+}\right) ,\>\> 
{\rm
P}_{\leq0[\s']}\left( h_{-}^{\Phi_-}\right) \in {\Bbb C}\>K \>,
\nfr{ConstT}
which implies that 
$$
L_+ \> =\>\LAX_i \quad{\rm and} \quad L_- \> =\>B\LAX_{-i} B^{-1}\>.
\efr

At this point, it is worth mentioning that the constraints~\ConstT\ 
have a nice physical interpretation in the context of non-abelian affine
Toda theories. Consider the non-abelian Toda equations
corresponding to $\Lambda_\pm = b_{\pm 1}$. In this case,
${\rm P}_{0[\s']}\left( h_{+}^{\Phi_+}\right)$ and
${\rm P}_{0[\s']}\left( h_{-}^{\Phi_-}\right)$ are just the components
of $B^{-1}\>\partial_+ B$ and $\partial_-
B\> B^{-1}$ along $\Ker(\ad \Lambda_\pm)$, respectively. Then, the
constraints~\ConstT\ are the necessary conditions to ensure that the
theory has a mass-gap and, hence, that it admits an $S$-matrix
description~[\Ref{NOS}].

These hierarchies also exhibit a scale invariance. In this case,
the scale transformation is
$$
\eqalign{
& x_\pm \>\mapsto\> \lambda^{\pm1}\> x_\pm\>, \qquad t_j \>\mapsto\>
\lambda^{j/i}\> t_j\>, \cr
& F_{j}^\pm \>\mapsto\> \lambda^{\pm(j/i-1)}\>, \quad {\rm and} \quad
B \>\mapsto\> B\>, \cr}
\nfr{Lorentz}
which amounts just to a conjugation of $L_\pm$:
$$
L_\pm  \longmapsto \lambda^{\mp1}\>
\lambda^{d_{\s'}/i}\> L_\pm \> \lambda^{-d_{\s'}/i}\>.
\efr
In~\AbelTwo, it induces
$$
(h_{\pm}^{\Phi_\pm})^j \>\mapsto\> \lambda^{\pm(j/ i-1)} \>
(h_{\pm}^{\Phi_\pm})^j\>.
\efr
This scale invariance has a special meaning when
$x_\pm$ are viewed as light-cone variables, \ie, $x_\pm = t\pm x$, as it
happens in the affine Toda field theories. Then, the scale
transformation~\Lorentz\ is just a Lorentz transformation in $1+1$
dimensions and, hence, scale invariance implies relativistic invariance.

\chapter{Invariant conserved charges}

The classical integrability of the zero-curvature hierarchies discussed
in the previous section is manifested in the existence of an infinite number
of local conserved charges. 

Recall eqs.~\DensOr,~\DensOrA, and~\DensOrB.
They show that, with suitable boundary conditions, the infinite
set of quantities
$$
{\cal Q}^\Phi[n]\> =\> \int dx\> \bil{b_n}{h^\Phi}\> , \quad
b_n \in \z_\Lambda \cap \gg{>0}(\s')\>,
\nfr{QconDS}
and
$$
{\cal Q}_{\pm}^{\Phi_\pm}[n]\> =\> \int dx_\pm\>
\bil{b_n}{h_{\pm}^{\Phi_\pm}}\> , 
\quad
b_{\pm n} \in \z_\Lambda \cap \gg{>0}(\s')\>,
\nfr{QconT}
are conserved by all the flows of the generalized Drinfel'd-Sokolov and
the Toda-mKdV hierarchies, respectively. Unfortunately, this definition of
the conserved quantities is ambiguous as a consequence of the fact that
neither $\Phi$ nor $\Phi_\pm$ are uniquely defined. Indeed, as explained
in Section~2.1 around eq.~\Indep, the form of~\Abelianize\ is preserved
if $ \Phi \mapsto \Phi \> {\rm e\/}^{\eta}$, where $\eta \in \Ker(\ad
\Lambda) \cap \gg{<0}(\s')$. This transformation implies that 
$$
h^\Phi \mapsto
{\rm e\/}^{- \eta}\> h^\Phi\> {\rm e\/}^{\eta}\> -\>  {\rm e\/}^{-\eta}
(\pa_x {\rm e\/}^{\eta} )\>,
\efr
and, since $b_{n} \in \z_\Lambda$,   
$$ 
\bil{b_n}{h^\Phi} \mapsto \bil{b_n}{h^\Phi} \> - \> \pa_x
\bil{b_{n}}{\eta}\>. 
\efr 
Therefore, the conserved densities are ambiguous and, consequently, the
quantities ${\cal Q}^\Phi[n]$ depend on the arbitrary asymptotic values of
$\eta$. The same is true for the conserved densities ${\cal 
Q}_{\pm}^{\Phi_\pm}[n]$. 

A non-ambiguous definition of the conserved quantities can be achieved by
constraining $\Phi={\rm e\>}^y$ and $\Phi_\pm={\rm e\>}^{y_\pm}$ such that
$y\in \Im(\ad \Lambda)$ and $y_\pm \in \Im(\ad \Lambda_\pm)$. In fact,
this choice ensures that the conserved quantities are local functionals
of the components of
$\widetilde{q\/}^{\> \rm GS}$ and $q_\pm$, respectively; \ie, 
$$
\eqalignno{
& {\cal Q}^\Phi[n] \longrightarrow Q_{n} \> =\> \int dx\>\>
T_{n}\Bigl(\widetilde{q\/}^{\> \rm GS}\Bigr) \>,\cr
& {\cal Q}_{+}^{\Phi_+}[n] \longrightarrow Q_{n}^{+} \> =\> \int dx_+\>\>
T^{+}_{n}\Bigl( B^{-1}\pa_+ B \> ,\> F_{j}^+\Bigr) \>, \cr
& {\cal Q}_{-}^{\Phi_-}[n] \longrightarrow Q^{-}_{n} \> =\> \int dx_-\>\>
T^{-}_{n} \Bigl( \pa_- B B^{-1}\>, \> F_{j}^-\Bigr) \>, & \nameali{Monos}
\cr}
$$
where $T_{n}$, $T^{+}_{n}$ and $T^{-}_{n}$ are differential
polynomials in the components of their arguments. In other words, $Q_{n}$
and $Q_{n}^{\pm}$ provide an infinite set of local conserved quantities. 

However, sometimes it is not convenient to constrain $\Phi$ or
$\Phi_\pm$ to be local and, hence, an improved `invariant'
definition of the conserved quantities $Q[n]$ and $Q_\pm[n]$ is required.
This problem was originally pointed out by Wilson~[\Ref{WILSON}], but it
was Freeman who first succeeded in defining the local conserved densities
in a $\eta$-independent way in the context of the abelian affine Toda
equation~[\Ref{FREEMAN}]. In this section, we will generalize his
construction to our zero-curvature hierarchies.

We start by writing the conserved densities of the generalized
Drinfel'd-Sokolov hierarchy associated with $\{ \gkm =\g^{(r)},
\HSA[\s'],\Lambda, \s \}$ by means of commutators. Since
$b_{n} \in \z_\Lambda$ and $h^{\Phi}(j) \in \Ker(\ad
\Lambda)$,\note{Recall that $[b_n, \Ker(\ad
\Lambda)]\in {\Bbb C}K$, $[d_{\s'},b_n] = n b_n$, and
$\bil{d_{\s'}}{K}=N_{\s'}/r$ (see the appendix, Section~A.1).}
$$
[b_{n} \>, \> h^{\Phi}(j) ]\> =\> {n\> r \over N_{\s'}}\>
\bil{b_n}{h^{\Phi}(j)} \> K\>,
\efr
which, using eqs.~\LaxPos\ and~\Abelj, implies that
$$
\eqalignno{
{n\> r \over N_{\s'}}\> \bil{b_n}{h^{\Phi}(j)} \> K \> &=\> \Phi\> [b_{n}
\>, \> h^{\Phi}(j)]\> \Phi^{-1}  \cr
& =\> 
[\Phi\> b_{n}\> \Phi^{-1}\>, \> A_j]\> +\> \pa_{j}  (\Phi \> 
b_{n}\> \Phi^{-1})\>. & \numali \cr}
$$
The last term in the last line is a total
$\pa_j$-derivative that trivially satisfies eq.~\DensOr. In contrast, the
first term is explicitly $\eta$-independent (see eq.~\Indep) and can be
used to define an invariant conserved density. In order to do it, one has
to project its component along $K$ out, which can be done by means of an
auxiliary derivation~$d_\m$. Then, we
define the invariant conserved densities of a generalized
Drinfel'd-Sokolov hierarchy by 
$$
\eqalignno{
J^{(j)}[n ; \m]\> & =\> {N_{\s'}\over n\> N_\m}\> \bil{d_\m}{[\Phi\>
b_n\>
\Phi^{-1}\> ,\> A_j]}\cr
&=\> \bil{b_n}{h^{\Phi}(j)}\> -\> {N_{\s'} \over n\> N_\m}\> \pa_j
\bil{d_\m}{\Phi\> b_n\> \Phi^{-1}}\>, & \nameali{DefDS}\cr}
$$
where $j\in E_{\Lambda}\geq0$ and $n\in E_{\Lambda} >0$. In particular,
since $\Lambda =b_i$ and
$h^{\Phi} =h^{\Phi}(i)$, these densities lead to the conserved quantities
$$
Q_n\>  =\> \int dx\> J^{(i)}[n; \m]\>,
\nfr{Def}
whose value seems to depend on the arbitrary gradation~$\m$. However, 
the conserved density $J^{(i)}[n; \m]$ depends on~\m\ only through an
improvement term that is a total $\pa_x$-derivative.  Then, if
$\Phi$ is chosen to be a local functional of $\widetilde{q\/}^{\> \rm GS}$, 
this term is the total derivative of a quantity that is also local in
$\widetilde{q\/}^{\> \rm GS}$ and, hence, it does not contribute to the value 
of $Q_n$. Since the definition of $J^{(i)}[n; \m]$ is manifestly
independent of the choice of $\Phi$, this also ensures the
$\m$-independence of $Q_n$.

It is worth mentioning that $Q_n$ has definite scaling dimension
with respect to the transformations given by~\Scaling. Actually, taking
into account~\ScaleH, its transformation is 
$$
Q_n \longmapsto \lambda^{n/i} \> Q_n\>.
\efr

Let us consider the behaviour of the invariant conserved
densities defined by~\DefDS\ with respect to the group of gauge
transformations given by~\Gauge. Since they do not change the form of the
Lax operator, they induce the following changes
 in~\Abelianize:
$$
\Phi \mapsto U\> \Phi\> e^\eta\>, \qquad
h^\Phi \mapsto
{\rm e\/}^{- \eta}\> h^\Phi\> {\rm e\/}^{\eta}\> -\>  {\rm e\/}^{-\eta}
(\pa_x {\rm e\/}^{\eta} )\>,
\nfr{GaugeInd}
where we have taken into account that $\Phi$ is not uniquely
defined. Therefore, the invariant conserved density transforms
according to
$$
J^{(i)}[n;\m] \mapsto \bil{b_n}{h^{\Phi}(i)}\> -\> {N_{\s'} \over n\>
N_\m}\> \pa_x
\bil{U^{-1}\> d_\m\> U}{\Phi\> b_n\> \Phi^{-1}}\>,
\efr
\ie, it is gauge invariant up to a total $\pa_x$-derivative, which is
enough to ensure that the conserved quantities $Q_n$ are gauge invariant.
However, it is possible to choose the auxiliary gradation $\m$ such that
the conserved densities are also explicitly gauge invariant. Recall that
$U$ takes values on the subalgebra $P\subset \gg{0}(\s)$ given
by~\GaugeGen. Therefore, if
$\m$ is chosen to be $\preceq\s$, then $\gg{0}(\s)\subseteq\gg{0}(\m)$
and, hence, $U^{-1}\> d_\m\> U = d_\m$, which implies that
the density $J^{(i)}[n;\m]$ is also gauge invariant.

The dependence of the invariant conserved densities on
$t_0$ is particularly simple. Choosing the gauge slice according
with~\ConstPlus, the flow associated with $t_0$ is induced by the
Lax operator
$$
\LAX_0\> = \> \pa_0\> -\> b_0\> -\> h_{c}^\Phi(0)\>K \>.
\efr
Actually, using eq.~\Abelj\ with $j=0$, it can be shown that $\Phi$ is of
the form
$$
\Phi\> = \> {\rm e\/}^{ t_0\> b_0}\> \widetilde{\Phi} \> {\rm e\/}^{-\>
t_0\> b_0} \> {\rm e}^{\rho}\>,
\efr
where $ \widetilde{\Phi}$ is independent of $t_0$ and $\rho $
takes values on $\Ker(\ad \Lambda)\cap \gg{<0}(\s')$. Then,
using~\LaxPos, the conserved densities become
$$
J^{(j)}[n ; \m]\> =\> {N_{\s'}\over n\> N_\m}\> \bil{{\rm e\/}^{-\>  t_0\>
b_0}\> d_\m\> {\rm e\/}^{ t_0\> b_0}}{[\widetilde{\Phi}\> b_n\>
\widetilde{\Phi}^{-1}\> ,\> \Pos\bigl(\widetilde{\Phi}\> b_j\>
\widetilde{\Phi}^{-1}\bigr)]}\>. 
\efr
Therefore, if $\m\preceq \s'$  or, in particular, if $\m\preceq \s$, the
previous equation shows that all the conserved densities are explicitly
independent of $t_0$. Moreover, $J^{(0)}[n ; \m]=0$ for all $n$, and 
$$
n\> J^{(j)}[n ; \m]\> =\> j\> J^{(n)}[j ; \m] \>,
\nfr{Cross}
for all $n, j\in E_{\Lambda}>0$.

Coming back to the Toda-mKdV hierarchy associated with $\{\gkm
=\g^{(r)}, \HSA[\s'], \Lambda_+,\Lambda_-\}$, their invariant
conserved densities can be defined is a similar way:
$$
\eqalignno{
J^{(j)}_+[n ; \m]\> & =\> {N_{\s'}\over n\> N_\m}\> \bil{d_\m}{[\Phi_+\>
b_n\>
\Phi^{-1}_{+}\> ,\> A_j]} \> -\> {N_{\s'}\over r}\>
\delta_{n+j,0}\cr &=\> \bil{b_n}{h^{\Phi_+}_{+}(j)}\> -\> {N_{\s'} \over
n\> N_\m}\> \pa_j
\bil{d_\m}{\Phi_+\> b_n\> \Phi^{-1}_{+}}\>, & \nameali{DefTp}\cr}
$$
for $n\in E_{\Lambda_\pm}>0$ and any $j\in E_{\Lambda_\pm}$, and
$$
\eqalignno{
J^{(j)}_-[n ; \m]\> & =\> {N_{\s'}\over n\> N_\m}\> \bil{d_\m}{[\Phi_-\>
b_n\>
\Phi^{-1}_{-}\> ,\> \widehat{A}_j]} \> -\> {N_{\s'}\over
r}\>\delta_{n+j,0}\cr &=\> \bil{b_n}{h^{\Phi_-}_{-}(j)}\> -\> {N_{\s'}
\over n\> N_\m}\> \pa_j
\bil{d_\m}{\Phi_-\> b_n\> \Phi^{-1}_{-}}\>, & \nameali{DefTn}\cr}
$$
where now $n\in E_{\Lambda_\pm}<0$. Recall that $A_j = \LAX_j - \pa_j$
and $\widehat{A}_j = B\LAX_j B^{-1} - \pa_j$ (see eq.~\Names). 
The corresponding conserved quantities also have definite scaling
dimension with respect to the (Lorentz) transformations given
by~\Lorentz:
$$
Q_{\pm n}^\pm \>  =\> \int dx_\pm \> J^{(\pm i)}_\pm[\pm n; \m]
\longmapsto
\lambda^{\pm n/i}\> Q_{n}^\pm \>, \qquad n>0\>.
\nfr{Ex}
Finally, in this case, it is also true that choosing $\m\preceq
\s'$ makes all the conserved densities explicitly independent of
$t_0$. However, the analogue of~\Cross\ is not satisfied in general.

Once we have an unambiguous definition of the conserved densities, let us
introduce the (very non-local) choice for $\Phi$ and $\Phi_\pm$ which
provides the bridge to connect with the group of dressing
transformations, the tau-function formalism and the results
of~[\Ref{JMP}]. 

\section{Drinfel'd-Sokolov hierarchies}

Recall eq.~\Abelj\ and consider the complete set of zero-curvature 
equations $\Phi^{-1} [\LAX_j, \LAX_k] \Phi =0$, which imply that
$$
\eqalignno{
&[\pa_j \> -\> h^\Phi(j)\>, \> \pa_k \> -\> h^\Phi(k)]\> =\> 0 &
\nameali{Integra} \cr
\noalign{\vskip 0.2truecm}
&\pa_j h_{c}^{\Phi}(k)\>+ \> {k\>r \over
N_{\s'}}\> 
\bil{b_k}{h^\Phi(j)} \> =\> \pa_k h_{c}^{\Phi}(j)\> + \> {j\>r \over
N_{\s'}}\> \bil{b_j}{h^\Phi(k)}\>. & \nameali{IntegraK} \cr}
$$
Eq.~\Integra\ provides the integrability conditions for 
$$
h^\Phi (j) \> =\> (\pa_{j} \chi^{\Phi})\> \chi^{\Phi\> -1}\>,
\nfr{ChiB}
where $\chi^\Phi = {\rm e\/}^\rho$ and $\rho$ is a function that takes
values on $\Ker(\ad \Lambda) \cap \gg{<0}(\s')$;
eq.~\ChiB\ is the generalization of~\ChiA\ to all the flows. Next 
consider
$\Theta_{\rm DS} = \Phi \chi^\Phi = {\rm e\>}^{v}$, where $v\in
\gg{<0}(\s)$ in agreement with the gauge fixing prescription
specified by~\ConstPlus. It is important to stress that, even if $\Phi$ 
is chosen to be a local functional of $q$, both $\chi^\Phi$ and
$\Theta_{\rm DS}$ are always non-local. Then, it is straightforward to
demonstrate that 
$$ 
\Theta_{\rm DS}^{-1} \> \LAX_{j}\> \Theta_{\rm DS}\> =\>
\pa_{j}\> -\> b_j \> - \> h_{c}^{\Theta_{\rm DS}}(j)\> K\>,
\nfr{AbelNL}
\ie, $\Theta_{\rm DS}$ also abelianizes the Lax operators
with the characteristic feature that ${h}^{\Theta_{\rm DS}}(j) =0$ for all
$j$. Moreover,
$$
h_{c}^{\Theta_{\rm DS}}(j)\> =\> h_{c}^\Phi(j)\> +\> {j\> r\over N_{\s'}}\>
\bil{b_j}{\rho}\>,
\efr
where we have used~\ChiB\ and taken into account that $b_j\in \z_\Lambda$.
Then, eq.~\IntegraK\ becomes
$$
\pa_j h_{c}^{\Theta_{\rm DS}} (k)\> -\> \pa_k h_{c}^{\Theta_{\rm DS}} (j) \>
=\> 0\>,
\efr
which allows one to fix the, so far, arbitrary functions $h_{c}^\Phi(j)$
such that $h_{c}^{\Theta_{\rm DS}} (j)=0$ and, hence,
$$ 
\Theta_{\rm DS}^{-1} \> \LAX_{j}\> \Theta_{\rm DS}\> =\>
\pa_{j}\> -\> b_{j}\> = \> \pa_j \> -\> A_{j}^{({\rm vac})}\>, 
\nfr{DressDS}
where $A_{j}^{({\rm vac})}$ has been defined in~\VacLax\ (recall that in
this case $j\in E_{\Lambda}$ is always $\geq0$). This particular form
of the abelianization provides a one-to-one map from solutions of the (gauge
fixed) hierarchy to solutions of the associated linear problem~\LinProb\ of
the form 
$$ 
\Psi \> =\> \Theta_{\rm DS}\>
\exp\Bigl[ \> \sum_{ j\in E_{\Lambda}\geq0} t_j \> b_j \>
\Bigr]\>, 
\nfr{Form}
which is a generalization of the Theorem~2.4 of~[\Ref{GENTAU}].

Using~\DressDS, the invariant conserved densities defined by~\DefDS\ become
$$
J^{(j)}[n;\m]\> = \> - \>
{N_{\s'}\over n\> N_{\bf m}}\> \pa_{j} \bil{ d_{\bf m}}{\Theta_{\rm DS}\> 
b_{n} \> \Theta_{\rm DS}^{-1}}\>.
\nfr{DefB}
In particular, since $x=t_i$, $J^{(i)}[n;\m]$ is just a total
$\pa_x$-derivative and, hence, all the information
required to calculate the value of the conserved quantities $Q_n$ is
encoded in the boundary values of $\Theta_{\rm DS}$. 

\section{Toda-mKdV hierarchies}

In this case, the previous construction can be repeated in two different
ways, using either~\AbeljTp\ or~\AbeljTn.

Let us start with eq.~\AbeljTp\ and consider the complete set of
zero-curvature equations
$\Phi^{-1}_{+} [\LAX_j, \LAX_k] \Phi_+ =0$. They provide the integrability
conditions for 
$$
h_{+}^{\Phi_+} (j) \> =\> (\pa_{j} \chi_{+}^{\Phi_+})\> \chi_{+}^{\Phi_+\>
-1}\>,
\efr
where $\chi_{+}^{\Phi_+} = {\rm e\/}^{\rho_+}$ and $\rho_+$ is a function
that takes values on $\Ker(\ad \Lambda_\pm) \cap \gg{<0}(\s')$. Then
consider $\Theta = \Phi_+ \chi_{+}^{\Phi_+} = {\rm e\>}^{v_+}$, where
$v_+\in
\gg{<0}(\s')$. It is not difficult to prove that 
$$ 
\Theta^{-1} \> \LAX_{j}\> \Theta\> =\>
\pa_{j}\> -\> b_j \> - \> h_{c}^\Theta(j)\> K\>,
\efr
\ie, $\Theta$ abelianizes the Lax operators in the sense of eq.~\AbeljTp\ 
with the characteristic feature that ${h}_{+}^{\Theta}(j) =0$ for all
$j\in E_{\Lambda_\pm}$; moreover,
$$
h_{c}^\Theta(j)\> =\> h_{c}^{\Phi_+}(j)\> +\> {j\> r\over N_{\s'}}\>
\bil{b_j}{\rho_+}\>.
\efr
The zero-curvature equations also imply that
$$
\pa_j h_{c}^\Theta (k)\> -\> \pa_k h_{c}^\Theta (j) \> =\> j\>
\delta_{j+k,0}\>,
\efr
which allows one to fix the arbitrary functions
$h_{c}^{\Phi_+}(j)$ such that 
$$ 
\Theta^{-1} \> \LAX_{j}\> \Theta\> =\>
\pa_{j}\> -\> A_{j}^{({\rm vac})}\> +\> \pa_j \varphi\> K, 
\nfr{DressTp}
where $A_{j}^{({\rm vac})}$ is given by~\VacLax\ and\note{According
to~\AbeljTp, recall that $h_{c}^{\Theta}(j)$ has to vanish for $j<0$,
while $A_{j}^{({\rm vac})} - b_j = -\pa_j \varphi\> K \not=0$.} 
$$
\varphi \>=\> \sum_{j\in E_{\Lambda_\pm}>0}\> jt_j t_{-j}\>.
\nfr{Ristra} 
Consequently, the (positive) invariant conserved densities defined
by~\DefTp\ become
$$
J_{+}^{(j)}[n;\m]\> = \> - \>
{N_{\s'}\over n\> N_{\bf m}}\> \pa_{j} \bil{ d_{\bf m}}{\Theta\> 
b_{n} \> \Theta^{-1}}\>,
\nfr{DefTpPlus}
for all $j$ (positive and negative) such that $j \in E_{\Lambda_\pm}$ and
$n\in E_{\Lambda_\pm}>0$.

Next, let us consider eq.~\AbeljTn\ and the
zero-curvature equations
$$
\Phi^{-1}_{-} B[\LAX_j, \LAX_k] B^{-1} \Phi_- =0\>,
\efr
which provide the integrability conditions for 
$$
h_{-}^{\Phi_-} (j) \> =\> (\pa_{j} \chi_{-}^{\Phi_-})\> \chi_{-}^{\Phi_-\>
-1}\>,
\efr
where $\chi_{-}^{\Phi_-} = {\rm e\/}^{\rho_-}$ and $\rho_-$ is a function
that takes values on $\Ker(\ad \Lambda_\pm) \cap \gg{>0}(\s')$. Then
consider $\Upsilon = \Phi_- \chi_{-}^{\Phi_-} = {\rm e\>}^{v_-}$, where
$v_-\in \gg{>0}(\s')$. It is not difficult to prove that 
$$ 
\Upsilon^{-1} \> B\> \LAX_{j}\> B^{-1} \Upsilon\> =\>
\pa_{j}\> -\> b_j \> - \> h_{c}^\Upsilon(j)\> K\>,
\efr
\ie, $	\Upsilon$ abelianizes the Lax
operators in the sense of eq.~\AbeljTn\ with the characteristic feature 
that ${h}_{-}^{\Upsilon}(j) =0$ for all $j$; moreover,
$$
h_{c}^\Upsilon(j)\> =\> h_{c}^{\Phi_-}(j)\> +\> {j\> r\over N_{\s'}}\>
\bil{b_j}{\rho_-}\>.
\efr
The zero-curvature equations also imply that
$$
\pa_j h_{c}^\Upsilon (k)\> -\> \pa_k h_{c}^\Upsilon (j) \> =\> j\>
\delta_{j+k,0}\>,
\efr
which allows one to choose the arbitrary functions $h_{c}^{\Phi_-}(j)$
such that 
$$ 
\Upsilon^{-1} \> B\> \LAX_{j}\> B^{-1} \Upsilon\> =\>
\pa_{j}\> -\> A_{j}^{({\rm vac})}\>. 
\nfr{DressTn}
Therefore, the (negative) invariant conserved densities given
by~\DefTn\ become
$$
J_{-}^{(j)}[n;\m]\> = \> - \>
{N_{\s'}\over n\> N_{\bf m}}\> \pa_{j} \bil{ d_{\bf m}}{\Upsilon\> 
b_{n} \> \Upsilon^{-1}}\>,
\nfr{DefTnPlus}
for all $j \in E_{\Lambda_\pm}$ and $n\in E_{\Lambda_\pm}<0$.

\chapter{Group theoretical solutions and their conserved densities}

In the previous section, it has been shown that the Lax operators of both
the generalized Drinfel'd-Sokolov and the Toda-mKdV hierarchies can
always be conjugated to their trivial vacuum form $\LAX_j = \pa_j
-A_{j}^{({\rm vac})}$ by means of the transformations given by
eqs.~\DressDS,~\DressTp, and~\DressTn. As a consequence of the
Drinfel'd-Sokolov procedure, the appropriate conjugations are given by
the exponentiated adjoint action of certain formal series in
$\gg{<0}(\s)$ ($\Theta_{\rm DS}={\rm e}^v$), in
$\gg{<0}(\s')$  ($\Theta={\rm e}^{v_+}$), or in $\gg{>0}(\s')$
($\Upsilon={\rm e}^{v_-}$). These series are not convergent in
general. However, the construction of a particular Lax operator or
conserved density are well defined because they only need a finite
number of terms of the series.

In the following, we will restrict ourselves to the class of solutions
for which $\Theta_{\rm DS}$, $\Theta$ and $\Upsilon$ are elements of the
Kac-Moody group $\GG$. Although this constraint might seem rather
mild, it is analogous to the condition that singles out the
class of solutions of the KdV equation considered by Segal and
Wilson~[\Ref{SEGWIL}]. As it is well known, this class includes the
explicit (periodic) algebro-geometric solutions of Krichever and the
multi-soliton solutions. We will show that our class
of solutions consists precisely of those ones generated by means of the
action of the group of dressing transformations on the vacuum solutions
given by eqs.~\VacLax\ and~\VacLin~[\Ref{JMP}]. Moreover, in the
particular case of the affine Toda equations, these solutions coincide
with the `solitonic specialization' proposed by Olive {\it et
al.\/}~[\Ref{EMSPEC},\Ref{SOLSPEC}].

\section{Solutions of the generalized Drinfel'd-Sokolov hierarchies}

Consider eq.~\DressDS\ where, now, $\Theta_{\rm DS}$ is a function that
takes values on $\GG_-(\s)$, the subgroup of the Kac-Moody group $\GG$
corresponding to $\gg{<0}(\s)$. Taking into account~\LaxPos,
eq.~\DressDS\ splits into
$$
\eqalignno{
& \LAX_j\> =\> \pa_j \> -\> \Pos(\Theta_{\rm DS}\> b_{j}\> \Theta_{\rm
DS}^{-1}) \>, & \nameali{LaxTheta} \cr
\noalign{\vskip 0.3cm}
& \pa_j \Theta_{\rm DS}^{-1} \> =\> \Theta_{\rm
DS}^{-1}\>\> \Neg(\Theta_{\rm DS}\> b_{j}\> \Theta_{\rm DS}^{-1})\>.
&\nameali{FlowTheta} \cr}
$$

These equations lead to an alternative interpretation of the generalized
Drinfel'd-Sokolov hierarchies restricted to this class of solutions. They
can be viewed as a generalization of the approach of Feigin, Frenkel
and Enriquez to soliton equations of KdV and mKdV type~[\Ref{FRENKEL}],
where the hierarchies consists of flows defined on functions that
take values on the Kac Moody group. Given its relevance, let us make this
relationship more explicit. Since $\Lambda = b_i$ and $L =\LAX_i$,
eq.~\LaxTheta\ provides the following one-to-one map between the (gauge
fixed) $q^{\> \rm GS}$, restricted to this class of solutions, and
$\Theta_{\rm DS}
\in
\GG_-(\s)$:\note{Notice that, according to~\ConstPlus, the condition
$\Theta_{\rm DS} \in \GG_-(\s)$ already fixes the gauge slice.}
$$
q^{\> \rm GS} \>= \> \Pos(\Theta_{\rm DS}\> \Lambda \> \Theta_{\rm
DS}^{-1})\> -\> \Lambda\>.
\nfr{MapFrenk}
Consider the transformation $ \Theta_{\rm DS} \rightarrow \Theta_{\rm 
DS}\> {\rm e}^\rho$, where $\rho \in \Ker(\ad \Lambda) \cap
\gg{<0}(\s)$, which implies that $q^{\> \rm GS} \rightarrow q^{\> \rm GS}
\> - \> ir/N_{\s'}\>
\bil{\Lambda}{\rho}\> K$. However, it was shown in Section~2 that the
component of $q^{\> \rm GS}$ along the centre is not an actual degree of
freedom of the hierarchy (see the comments below eq.~\PdeB). Therefore,
eq.~\MapFrenk\ has to be viewed as a one-to-one map between
$\widetilde{q}^{\> \rm GS} = q^{\> \rm GS} - q^{\> \rm GS}_c K$ and $\Theta_{\rm
DS} \in
\GG_-(\s)/K^{\Lambda}_-(\s)$, where
$K^{\Lambda}_-(\s)$ is the subgroup of $\GG$ corresponding to the
subalgebra $\Ker(\ad \Lambda) \cap \gg{<0}(\s)$. On $\GG_-(\s)$, the flows
are defined by~\FlowTheta, which can be put in the form
$$
\pa_j \Theta_{\rm DS}^{-1} \> =\> b_{j}^L\cdot \Theta_{\rm DS}^{-1}\>,
\nfr{DerRara}
where $b_{j}^L$ is the derivation of $\GG_-(\s)$ corresponding to the
left infinitesimal action of $b_j$. It is defined
as follows~[\Ref{FRENKEL}]. Consider a one-parameter subgroup
$b_j(\epsilon)$ of $\GG$ such that $b_j(\epsilon)  = 1 + \epsilon b_j +
o(\epsilon^2)$, and a generic element $h \in \GG_-(\s)$. This group acts
on $h$ from the left according to
$$
b_j(\epsilon) \cdot h = h + \epsilon\> b_j h + o(\epsilon^2) =
h( 1+ \epsilon\> h^{-1} b_j h  + o(\epsilon^2))\>,
\efr
which, for infinitesimal $\epsilon$, can be factorized as an element of
$\GG_-(\s)\> \GG_0(\s) \> \GG_+(\s)$. Then, its component in $\GG_-(\s)$ 
equals  $h \> +\> \epsilon\> h\>\Neg(h^{-1} b_j h) \>+ \> o(\epsilon^2)$,
which leads to $b_j^L\cdot h = h\> \Neg(h^{-1} b_j h)$ and, hence,
proves~\DerRara.

Coming back to eqs.~\LaxTheta\ and~\FlowTheta, notice that $\Theta_{\rm
DS}^{-1}$ satisfies a system of first order differential equations.
Their general solution is
$$
\Theta_{\rm DS}^{-1}[h]\> =\> \biggl(\exp\Bigl[ \> \sum_{ j\in 
E_{\Lambda}\geq0} t_j \> b_j \> \Bigr]\> h \biggr)_-\>,
\nfr{SolDS}
where $(\cdot)_-$ indicates the projection on $\GG_{<0}(\s)$  (see the
appendix, Section~A.3), and $h$ is an arbitrary constant element in
$\GG_{<0}(\s)$ that fixes the initial conditions for~\FlowTheta, \ie, 
$$
\Theta_{\rm DS}^{-1}[h] \Bigm|_{t_j\> = \>0} \> = \> h\>.
\efr
Taking into account~\Form, eq.~\SolDS\ provides the following solution 
for the associated linear problem
$$
\Psi\> =\> \Bigl[\bigl(\Psi^{({\rm vac})}_{\rm DS}\> 
h\bigr)_-\Bigl]^{-1}\> \Psi^{({\rm vac})}_{\rm DS}\>,
\nfr{SolDSLin}
where, in this case,
$$
\Psi^{({\rm vac})}_{\rm DS}\>  =\> \exp\Bigl[ \>\sum_{j\in 
E_{\Lambda}\geq0} t_j \> b_j\> \Bigr]
\efr
(see eqs.~\VacLax\ and~\VacLin). It is worth noticing the similarity of this
formula with the Baker-Akhiezer function for the KdV equation obtained by
Segal and Wilson~[\Ref{SEGWIL}]. Moreover, it is
remarkable that~\SolDS\ and~\SolDSLin\ correspond precisely to the
solutions generated by the action of the group of dressing transformations
on the trivial vacuum solution~[\Ref{GENTAU},\Ref{JMP}]. 

At this point, it is convenient to remind
that $\GG_-(\s)$ can be identified with the quotient $\GG/P_\s$, where
$P_\s$ is the `parabolic' group corresponding to the subalgebra
$\gg{\geq0}(\s)$ generated by the $e_{i}^-$, for each $i$ such that
$s_i=0$, and the $h_i$ and $e_{i}^+$, for all $i=0,1,\ldots, l$. In the
generalized Drinfel'd-Sokolov hierarchies,
$\GG/P_\s$ plays the same role as the Grassmannian (or flag manifold)
in the approach of Segal and Wilson to the KdV (or mKdV)
equation~[\Ref{SEGWIL}]. Moreover, consider a constant element $\chi\in
K^{\Lambda}_-(\s)$. Since $[b_j, \Ker(\ad \Lambda) ]\in {\Bbb C}K$, it is
easy to check that $\Theta_{\rm DS}[\chi h] = \Theta_{\rm DS}[h] \>
\chi^{-1}$. Therefore, taking into account the previous paragraphs, we 
conclude that the non-equivalent solutions for $\widetilde{q}^{\> \rm GS}$ are
in one-to-one relation with the elements of
$$
K^{\Lambda}_-(\s)\> \backslash\> \GG_-(\s) \>= \>
K^{\Lambda}_-(\s)\> \backslash \> \GG \> /\> P_\s\>.
\nfr{Manifold}

Let us calculate the invariant conserved densities~\DefB\ corresponding to
the solutions given by~\SolDS\ and~\SolDSLin. Since $\Theta_{\rm DS}$ is
an element of the Kac-Moody group, we will calculate them by considering
a suitable integrable representation of
$\gkm$. Then, using~(A.25), eq.~\DefB\ can be written as
$$
\eqalignno{
J^{(j)}[n;\m]\> &=\> -\> {N_{\s'}\over n\> N_{\bf m}}\> \pa_{j}\> 
\bra{v_{\overline{\m}}}\> \Theta_{\rm DS}[h]\>  b_{n} \> \Theta_{\rm
DS}^{-1}[h] \> \ket{v_{\overline{\m}}} \cr
\noalign{\vskip 0.3cm}
& = \> -\> {N_{\s'}\over n\> N_{\bf m}}\> \pa_{j} 
\bra{v_{\overline{\m}}}\>  b_n \> \bigl(\Psi^{({\rm vac})}_{\rm DS}\> h 
\bigr)_- \> \ket{v_{\overline{\m}}}  \>, & \numali \cr}
$$
where we have taken into account that $\bra{v_{\overline{\m}}}$ is a
highest weight vector and, hence, $\bra{v_{\overline{\m}}}\>\gg{<0}(\s) 
=0$. Let us choose the arbitrary gradation $\m$ such that
$\m\preceq\s$ with respect to the partial ordering, and consider the
factorization
$$
\Psi^{({\rm vac})}_{\rm DS}\> h \> =\>  \bigl(\Psi^{({\rm vac})}_{\rm DS}
\> h \bigr)_-
\> \bigl(\Psi^{({\rm vac})}_{\rm DS}\> h \bigr)_0
\> \bigl(\Psi^{({\rm vac})}_{\rm DS}\> h \bigr)_+
\efr
with respect to the gradation $\s$.
Then, since $\overline{\m} \simeq \m \preceq\s$, the highest weight vector
$\ket{v_{\overline{\m}}}$ is annihilated by $\gg{>0}(\s)$ and is an
eigenvector of $\gg{0}(\s)$ (see the appendix, Section~A.3), which allows
one to write
$$
\bra{v_{\overline{\m}}}\>  b_n \> \bigl(\Psi^{({\rm vac})}_{\rm DS}\> h 
\bigr)_- \>
\ket{v_{\overline{\m}}} \> =\> 
{\bra{v_{\overline{\m}}}\>  b_n \> \Psi^{({\rm vac})}_{\rm DS}\> h  \>
\ket{v_{\overline{\m}}} \over
\bra{v_{\overline{\m}}}\>  \Psi^{({\rm vac})}_{\rm DS}\> h  \>
\ket{v_{\overline{\m}}} }\>,
\efr
Therefore, taking into account that $\pa_n \Psi^{({\rm vac})}_{\rm DS} = 
b_n\>
\Psi^{({\rm vac})}_{\rm DS} $, the conserved densities~\DefB\ are given by
$$
J^{(j)}[n;\m]\> =\> -\> {N_{\s'}\over n\> N_{\bf m}}\> \pa_{j}\> \pa_{n}
\ln \tau_{\overline{\m}}^{\rm DS}(h)\>.
\nfr{DefC}
This exhibits that all the information
about the invariant conserved densities is summarized by the single
scalar tau-function
$$
\tau_{\overline{\m}}^{\rm DS}(h)\>  =\> \bra{v_{\overline{\m}}}\>  
\Psi^{({\rm vac})}_{\rm DS}\> h \> \ket{v_{\overline{\m}}}\>.
\nfr{TauDS}
Actually, $\tau_{\overline{\m}}^{\rm DS}(h)$ is one of the several
generalized Hirota tau-functions introduced 
in~[\Ref{GENTAU},\Ref{JMP}] to parameterize $q^{\> \rm GS}$. There,
the number of required tau-functions equals the number of components
of $q^{\> \rm GS}$. In contrast, it is very remarkable that all the
conserved quantities can be written by means of a single one. This is
the first main result of the paper.

\section{Solutions of the Toda-mKdV hierarchies}

In this case, the solutions follow by considering together the
eqs.~\DressTp\ and~\DressTn. Since
$$
\pa_{j}\> -\> A_{j}^{({\rm vac})}\> =\> \Psi^{({\rm vac})}_{\rm T}\> 
\pa_j\> {\Psi^{({\rm vac})}_{\rm T}}^{-1}\>,
\efr
they imply that
$$
\pa_j \Bigl({\rm e\/}^{\varphi\> K}\> {\Psi^{({\rm vac})}_{\rm T}}^{-1}\>
\Theta^{-1}\> B^{-1}\>  \Upsilon\> \Psi^{({\rm vac})}_{\rm T} \Bigr) \> =
\>0
\efr
for all the flows of the hierarchy. Therefore, since $\Theta$,
$B$, and $\Upsilon$ are in the Kac-Moody group $\GG$, the general
solution is provided by the decomposition
$$
{\rm e\/}^{-\> \varphi\> K}\>\Psi^{({\rm vac})}_{\rm T}\> h\> {\Psi^{({\rm
vac})}_{\rm T}}^{-1} \> =\> \Theta^{-1}[h]\> B^{-1}[h]\> \Upsilon[h]\>,
\nfr{SolT}
where $h$ is a constant arbitrary element in the big cell of $\GG$.
In other words, and taking into account the results of~[\Ref{JMP}], the
class of solutions of these hierarchies  compatible with the constraint
that $\Theta$, $B$, and $\Upsilon$ are in $\GG$ is provided by the orbit
of the group of dressing transformations acting on the vacuum solutions
given, in this case, by  
$$  
\Psi^{({\rm vac})}_{\rm T}\>  =\> \exp\Bigl[ \>\sum_{j\in
E_{\Lambda_\pm}\geq0} t_j \> b_j\>
\Bigr]\> 
\exp\Bigl[ \>\sum_{j\in E_{\Lambda_\pm}>0} t_{-j} \> b_{-j}\> \Bigr]  
\efr
(see eqs.~\VacLax\ and~\VacLin).

Let us calculate the invariant conserved densities~\DefTpPlus\
and~\DefTnPlus\ corresponding to the solutions given by~\SolT.
Using~(A.25), eq.~\DefTpPlus\ can be written as
$$
\eqalignno{
J^{(j)}_+[n;\m]\> &=\> -\> {N_{\s'}\over n\> N_{\bf m}}\> \pa_{j}\> 
\bra{v_{\overline{\m}}}\> \Theta[h]\>  b_{n} \> \Theta^{-1}[h] \>
\ket{v_{\overline{\m}}} \cr
\noalign{\vskip 0.3cm}
& = \> -\> {N_{\s'}\over n\> N_{\bf m}}\> \pa_{j} \>
\bra{v_{\overline{\m}}}\>  b_n \> \bigl(\Psi^{({\rm vac})}_{\rm T}\> h 
\> {\Psi^{({\rm vac})}_{\rm T}}^{-1}\bigr)_- \> \ket{v_{\overline{\m}}}  
\>, & \numali \cr}
$$
where we have used that $\bra{v_{\overline{\m}}} \gg{<0}(\s') =0$.
Let us choose the arbitrary gradation $\m$ such that
$\m\preceq\s'$ with respect to the partial ordering, and consider the
factorization of $ \Psi^{({\rm vac})}_{\rm T}\> h \> {\Psi^{({\rm
vac})}_{\rm T}}^{-1}$ with respect to the gradation $\s'$. Then, since
$\overline{\m} \simeq \m \preceq\s'$, the highest weight vector
$\ket{v_{\overline{\m}}}$ is annihilated by $\gg{>0}(\s')$ and is an eigenvector of $\gg{0}(\s')$, which
allows one to write
$$
\bra{v_{\overline{\m}}}\>  b_n \> \bigl(\Psi^{({\rm vac})}_{\rm T}\> h \>
{\Psi^{({\rm vac})}_{\rm T}}^{-1} 
\bigr)_- \>
\ket{v_{\overline{\m}}} \> =\> 
{\bra{v_{\overline{\m}}}\>  b_n \> \Psi^{({\rm vac})}_{\rm T}\> h \>
{\Psi^{({\rm vac})}_{\rm T}}^{-1}  \>
\ket{v_{\overline{\m}}} \over
\bra{v_{\overline{\m}}}\>  \Psi^{({\rm vac})}_{\rm T}\> h \> {\Psi^{({\rm
vac})}_{\rm T}}^{-1}  \>
\ket{v_{\overline{\m}}} }\>,
\efr
Therefore, taking into account that $b_n\> \Psi^{({\rm vac})}_{\rm T}
= \pa_n \Psi^{({\rm vac})}_{\rm T} $ and $b_n \ket{v_{\overline{\m}}}=0$
for $n\in E_{\Lambda_\pm}>0$, the (positive) conserved densities~\DefTp\
are given by
$$
J^{(j)}_+[n;\m]\> =\> -\> {N_{\s'}\over n\> N_{\bf m}}\> \pa_{j}\> \pa_{n}
\ln \tau_{\overline{\m}}^{\rm T}(h)\>.
\nfr{DefTpC}
This means that all the information
about the invariant conserved densities for this class of solutions is also
summarized by the single scalar tau-function
$$
\tau_{\overline{\m}}^{\rm T}(h)\>  =\> \bra{v_{\overline{\m}}}\>  
\Psi^{({\rm vac})}_{\rm T}\> h \> {\Psi^{({\rm vac})}_{\rm T}}^{-1} \>
\ket{v_{\overline{\m}}}\>.
\nfr{TauT}

Similarly, the (negative) conserved densities~\DefTnPlus\ can be written 
as
$$
\eqalignno{
J^{(j)}_-[n;\m]\> &=\> -\> {N_{\s'}\over n\> N_{\bf m}}\> \pa_{j}\> 
\bra{v_{\overline{\m}}}\> \Upsilon [h]\>  b_{n} \> \Upsilon^{-1}[h] \>
\ket{v_{\overline{\m}}} \cr
\noalign{\vskip 0.3cm}
& = \> -\> {N_{\s'}\over n\> N_{\bf m}}\> \pa_{j} \>
{\bra{v_{\overline{\m}}}\>  \Psi^{({\rm vac})}_{\rm T}\> h \>
{\Psi^{({\rm vac})}_{\rm T}}^{-1}  \> b_n\>
\ket{v_{\overline{\m}}} \over
\bra{v_{\overline{\m}}}\>  \Psi^{({\rm vac})}_{\rm T}\> h \> {\Psi^{({\rm
vac})}_{\rm T}}^{-1}  \>
\ket{v_{\overline{\m}}} }\>, &\numali \cr}
$$
which, taking into account that (see eq.~\VacLax)
$$
{\Psi^{({\rm vac})}_{\rm T}}^{-1}\>  b_n\> =\> -\> \bigl(
\pa_n \> +\> |n|\> t_{|n|}\> K\bigr)\> {\Psi^{({\rm vac})}_{\rm T}}^{-1}
\efr
and $\bra{v_{\overline{\m}}}b_n =0$ for $n\in E_{\Lambda_\pm}<0$, become
$$
J^{(j)}_-[n;\m]\> =\> -\> {N_{\s'}\over |n|\> N_{\bf m}}\> \pa_{j}\>
\pa_{n} \ln \tau_{\overline{\m}}^{\rm T}(h)\>.
\nfr{DefTnC}

In the last equation, $\tau_{\overline{\m}}^{\rm T}(h)$ is just one of
the several generalized Hirota tau-functions introduced in~[\Ref{JMP}] to
parameterize the components of $q_\pm$, \ie, $F_{j}^\pm$ and $B$ (see
eq.~\TwoLax). Taking into account the results of~[\Ref{JMP}] about the
number of required tau-functions, it is very remarkable that all the
conserved quantities can be written by means of a single one also for
these hierarchies. This is the second main result of the paper.

\chapter{Relationships between the conserved quantities of the
generalized Drinfel'd-Sokolov and Toda-mKdV hierarchies}

The main result of the previous section can be summarized as follows.
Consider the data $\{ \gkm , \HSA[\s'], \m \}$, where $\HSA[\s']$
is a Heisenberg subalgebra of $\gkm$; \ie,
$$
\HSA[\s']\> = {\Bbb C}\> K \oplus \sum_{i\in E} {\Bbb C} \> b_i\>,
\qquad b_i \in \gg{i}(\s')\>, \qquad [b_j\>, \> b_k]\> =\> j\>
\delta_{j+k,0} \> K\>.
\efr
Then, for each element $h=h_- h_0 h_+$ in the big cell of the Kac-Moody
group corresponding to $\gkm$, construct the tau-function
$$
\tau_{\overline{\m}}(h)\> =\> \exp\bigl[{N_\m\over r} \sum_{j\in E>0} j\>
t_j t_{-j} \bigr]\> \bra{v_{\overline{\m}}}\> \exp\bigl[\sum_{j\in E\geq0}
t_j  b_j\bigr] \> h \> \exp\bigl[-\sum_{j\in E<0} t_j 
b_j\bigr] \>\ket{v_{\overline{\m}}}\>, 
\nfr{TauH}
where $K_{\overline{\m}} = N_{\m}/r$ is the level of the representation
$L(\overline{\m})$ (see eq.~(A.26) and below), and the normalization is
chosen such that $\tau_{\overline{\m}}(I)=1$. Moreover, for each element
$\Lambda$ of $\HSA[\s']$ with non-zero $\s'$-grade, consider the
subalgebra of $\HSA[\s']$
$$
\z_\Lambda \>=\> {\rm Cent\/}(\Ker(\ad \Lambda))\> =\> 
{\Bbb C}\> K \>\oplus \> \sum_{j\in E_\Lambda} \HSA[\s']\cap
\gg{j}(\s')\>,
\efr
where $E_\Lambda$ is the set of $j$'s such that $\z_\Lambda\cap
\gg{j}(\s') $ is not empty.

Then, according to the previous section, the invariant conserved
densities for the generalized Drinfel'd-Sokolov hierarchy associated
with the data $\{ \gkm =\g^{(r)}, \HSA[\s'],\Lambda, \s \}$ are given by
eq.~\DefC, where
$$
\tau_{\overline{\m}}^{\rm DS} (h)\> =\> \tau_{\overline{\m}}(h)
\Bigm|_{t_j=0 {\rm\; for\;} j\not\in E_\Lambda\geq0}\>.
\efr
Notice that 
$$ 
\tau_{\overline{\m}}^{\rm DS} (h) \>=\>
\bra{v_{\overline{\m}}}\>   h_0 \>\ket{v_{\overline{\m}}}\; 
\tau_{\overline{\m}}^{\rm DS}(h_-) \>,
\efr
but $\bra{v_{\overline{\m}}}   h_0  \ket{v_{\overline{\m}}}$ is just a
constant.

Similarly, the invariant conserved
densities for the Toda-mKdV hierarchy associated
with $\{\gkm =\g^{(r)}, \HSA[\s'], \Lambda_+,\Lambda_-\}$ are given by
eqs.~\DefTpC\ and~\DefTnC, which can be written together as
$$
J^{(j)}_\pm[\pm n;\m]\> =\> -\> {N_{\s'}\over n\> N_{\bf m}}\> \pa_{j}\>
\pa_{\pm n} \ln \tau_{\overline{\m}}^{\rm T}(h)\>,
\nfr{DefTtau}
where $n>0$, and 
$$
\tau_{\overline{\m}}^{\rm T} (h)\> =\> \tau_{\overline{\m}}(h)
\Bigm|_{t_j=0 {\rm\; for\;} j\not\in
E_{\Lambda_\pm}}\>.
\efr  
 
These equations manifest the following relationship between the conserved
densities of different integrable hierarchies. Consider some $j$
such that $s_{j}' \not=0$ and the (minimal) gradation $\s^{(j)}$
specified by $s^{(j)}_i = \delta_{ij}$, \ie, $\s^{(j)} \preceq \s'$. 
Then, since one can choose $\m= \s^{(j)}$ both in eqs.~\DefC\
and~\DefTtau, the class of group theoretical solutions for all the
generalized Drinfel'd-Sokolov and Toda-mKdV hierarchies associated to 
$\{\gkm, \HSA[\s'], \Lambda, \s\}$, with $\s^{(j)} \preceq \s \preceq
\s'$, and to $\{\gkm, \HSA[\s'], \Lambda_+=\Lambda, \Lambda_-\}$,
respectively, share the same conserved quantities. 

These results should have been expected because of the following. First,
the Lax operator $L_+$ associated with $\{\gkm, \HSA[\s'],
\Lambda_+=\Lambda, \Lambda_-\}$ is of the same form as the Lax operator 
of the generalized mKdV Drinfel'd-Sokolov hierarchy associated with
$\{\gkm, \HSA[\s'], \Lambda, \s= \s'\}$. Then, the coincidence of their
conserved charges is a particular consequence of the existence of a map
from the solutions of the Toda-mKdV hierarchy into the solutions of
the mKdV hierarchy given by the equation
$$
\eqalign{
L_+\> -\> \partial_+\> -\> \Lambda_+\> &= \> B^{-1}\partial_+ B 
-\> \sum_{1\leq j< i} \> F_{j}^+ \> \cr
& \longmapsto  -\> q^{\rm
mKdV}\> =\> L\> -\> \partial_x\> -\> \Lambda \cr}
\efr
and the identification $\pa_x=\pa_{+}$. The second reason is that all
the generalized Drinfel'd-Sokolov hierarchies associated with $\{\gkm,
\HSA[\s'], \Lambda, \s\}$ and $\s^{(j)} \preceq
\s \preceq \s'$ are related by means of generalized Miura
transformations~[\Ref{GEN},\Ref{GEN2}]. Taking into account all this, the
resulting relationships between conserved quantities are nothing else
but the generalization of the well known connection between 
conservation laws in the KdV, mKdV, and sine-Gordon systems originally
pointed out by Chodos~[\Ref{CHODOS}]. 

\chapter{Examples}

To illustrate and complement the previous sections, we will show that our
results concerning the conserved densities generalize very well
known expressions in the context of the the original Drinfel'd-Sokolov 
KdV hierarchies and the affine Toda field theories. Moreover, we will
illustrate how to use our expressions to calculate the conserved
quantities carried by solitons by applying them to the soliton
solutions for the constrained KP hierarchy recently constructed by
Aratyn {\it et al.}~[\Ref{LUIZKP}].

\section{Drinfel'd-Sokolov KdV hierarchies}

In~[\Ref{DS}], Drinfel'd and Sokolov associated a generalized hierarchy
of KdV type with each vertex of the Dynkin diagram of an affine Kac-Moody
algebra $\gkm=\g^{(r)}$. In the formalism of~[\Ref{GEN},\Ref{GEN2}],
these hierarchies are recovered by considering the data $\{\gkm,
\HSA[\s_{\rm p}], \Lambda, \s=\s^{(j)} \}$, where $\HSA[\s_{\rm p}]$ is
the principal Heisenberg subalgebra, $\s_{\rm p}=(1,1, \ldots,1)$,
$\Lambda$ is  the constant `cyclic' element of $\gkm$, $\Lambda
=\sum_{j=0}^l e_{j}^+$, of principal grade $1$, and the `minimal'
gradation $\s^{(j)}$ is specified by $s^{(j)}_i = \delta_{ij}$.
Obviously, $\s^{(j)}$ is associated to the
$j^{\rm th}$-vertex of the Dynkin diagram of $\gkm$ and, in particular,
$\s^{(0)}=\s_{\rm h}$ is the homogeneous gradation.\note{In~[\Ref{DS}],
the principal gradation and the minimal gradation $\s^{(j)}$ are
referred to as the `canonical gradation' and the `standard gradation'
corresponding to the
$j^{\rm th}$-vertex of the Dynkin diagram, respectively.} The
cyclic element $\Lambda$ is regular and, hence, the principal Heisenberg
subalgebra can be defined by means of
$$
\HSA[\s_{\rm p}]\> =\> \Ker(\ad \Lambda) \> =\> {\Bbb C}\>K
\> +\> \sum_{j\in I_{\gkm} \; {\rm mod}\; r\> h_{\gkm}} {\Bbb C}\> b_j\>,
\nfr{Expo}
where $I_{\gkm}$ is the set of ${\rm rank}(\g)$ exponents of $\gkm=\g^{(r)}$
(with multiplicities) and $h_{\gkm}=\sum_{j=0}^l a_j$ is the Coxeter
number. Notice that ${\rm rank}(\g)\geq l= {\rm rank}(\gkm)$, and that
it is equal only for untwisted affine algebras, \ie, for
$r=1$~[\Ref{KBOOK}].

Since $\s^{(j)}$ is always $ \preceq \s_{\rm p}$ with respect to
the partial ordering, the function $q$ in~\LaxDS\ takes values on 
$\gg{0}(\s^{(j)}) \cap \gg{\leq0}(\s_{\rm p})$ and, hence, 
$$
[ d_{\s^{(j)}}\>, \> \Lambda \> +\> q]\> =\> e_{j}^+\>.
\efr
This simplifies the form of the invariant conserved
densities associated with $\Lambda =b_1$ and given by~\DefDS\ for $j=1$.
In order to ensure that the invariant conserved densities
are explicitly gauge invariant, recall that in~\DefDS\ 
one has to choose the auxiliary gradation $\m$ to be $\preceq \s^{(j)}$
(see below eq.~\GaugeInd). In our case, it will be
convenient to choose it such that
$m_i = a_{j}^\vee a_{j}^{-1} \delta_{ij}$. Then, 
$$
\eqalignno{
&J^{(1)}[n; \m]\>  = \> -\> {N_{\s_{\rm p}}\over
n\> N_{\m}}\> \bil{d_{\m}}{[\Phi\> b_n\> \Phi^{-1}\>, \>
\Lambda\> +\> q]}\cr 
& \qquad\quad = \> {h_{\gkm} \over n\> a_j}\> \bil{e_{j}^+}{\Phi\>
b_n\> \Phi^{-1}}\>, \qquad n \in I_{\gkm}\> \; {\rm mod}\; r\> h_{\gkm}\> >0
\>,  &\nameali{FrenKdV} \cr}
$$
where $a_j$ is the (Kac) label of the $j^{\rm th}$-vertex of the Dynkin
diagram of $\gkm$. Then, for the class of group theoretical solutions
given by~\MapFrenk\ and~\SolDS, and according to~\DefC, these conserved
densities can be written as
$$
J^{(1)}[n; \m]\> =\> -\> {h_{\gkm} \over n\> a_{j}^\vee }\> \pa_x
\pa_n \ln \tau_{j}^{\rm DS}(h)\>,
\nfr{TopGra}
where we have taken into account that $\overline{\m}= \s^{(j)}$ and
called $\tau_{j}^{\rm DS}(h)$ the tau-function corresponding to
$\ket{\s^{(j)}}=\ket{v_j}$. 

According to the results of Section~3, the conserved densities
$J^{(1)}[n; \m]$ are gauge invariant local functionals of $q$,
\ie, gauge invariant differential polynomials in $q$. The form of the
gauge slice has been worked out in~[\Ref{DS}]. First of all, we have the
decomposition
$$
\gg{0}(\s^{(j)})\cap \gg{\leq 0}(\s_{\rm p})\> =\> [\Lambda\>, \> P]\>
\oplus V\>,
\efr
where $V\cap \gg{i}(\s_{\rm p})=\emptyset$ unless $i$ is an exponent of
$\gkm$, and $P= \gg{0}(\s^{(j)})\cap \gg{< 0}(\s_{\rm p})$ is the subalgebra
of generators of gauge transformations (see~\GaugeGen), which satisfies
$[\Lambda , P] = [\Lambda - e_{j}^+ , P] $. Therefore, the gauge slice
will be of the form
$$
q^{\> \rm GS}\> =\> \sum_{i\in I_{\gkm}} u_i\> \beta_{-i}\>,
\efr
where $\{\beta_{-i}\}$ is a basis of $V$ whose elements satisfy
$\bil{\beta_{-i}}{b_i}\not=0$. Then, if $V\cap \gg{n}(\s_{\rm
p})\not=\emptyset$, one can check that
$$
J^{(1)}[n; \s^{(j)}]\> =\> c_n\> u_n \> +\> \cdots\>,
\efr
where $c_n$ is a non-vanishing constant complex number, and the ellipses
indicate differential polynomials in $u_j$ with $j>n$. Thus, this set of
invariant conserved densities provide a set of coordinates
on the gauge slice. In other words,~\TopGra\ shows
that, in this case, the single tau-function $\tau_{j}^{\rm DS}(h)$ allows
one to reconstruct not only the invariant conserved densities, but all
the variables of the gauged-fixed hierarchy. This result is well known
and was widely used in the context of the matrix model approach to
quantum and topological two-dimensional gravity. There, the tau-function
plays the role of a partition function, the times are coupling
constants, and the conserved densities are  correlation
functions~[\Ref{DIJK}]. 

From another point of view, for each
$n$ such that $V\cap \gg{n}(\s_{\rm p})\not=\emptyset$,  eq.~\FrenKdV\
with $\Phi \rightarrow \Theta_{\rm DS}$ can be viewed as the
generalization of the set of functions used by Feigin, Frenkel and
Enriquez in their approach to the Drinfel'd-Sokolov KdV
hierarchies~[\Ref{FRENKEL}], where $J^{(1)}[n;
\s^{(j)}]$ are considered as functions taking values on the manifold 
$$
K^{\Lambda}_-(\s^{(j)})\> \backslash\> \GG_-(\s^{(j)}) \>= \>
K^{\Lambda}_-(\s^{(j)})\> \backslash \> \GG \> /\> P_{\s^{(j)}}\>.
\efr
   
\section{Non-abelian affine Toda equations}

Let us consider a hierarchy of the Toda-mKdV type associated with $\{\gkm
= \g^{(r)},\HSA[\s'], \Lambda_+, \Lambda_-\}$ where $\Lambda_\pm
=b_{\pm1}$. Then, the zero-curvature condition~\NAAT\ 
is just the non-abelian affine Toda (NAAT) equation
$$
\partial_- \bigl( B^{-1}\> \partial_+\> B\bigr)\> =\> \bigl[\Lambda_+\>,
\> B^{-1}\> \Lambda_-\> B\bigr]\>.
\nfr{NAATexp}
Consider the following decomposition of $\gg{0}(\s')$ (see
eq.~(A.13)):
$$
\gg{0}(\s')\> =\> \gg{0}(\s'; \s', q)\> \oplus\> {\Bbb C}\> K\>,
\efr
where the positive integer $q$ is chosen such that $s_{q}' \not=0$, and 
the corresponding decomposition of the Toda field
$$
B\> =\> B_0\> {\rm e\/}^{( \nu\> +\> x_+\> x_-)\> K}\>,
\nfr{BceroNu}
where $B_0$ takes values in the finite Lie group corresponding
to the subalgebra $\gg{0}(\s'; \s', q)$. Then, the NAAT equation
becomes\note{The normalization $[\Lambda_+, \Lambda_-]= K$ is equivalent
to $\bil{\Lambda_+}{\Lambda_-}= \bil{d_{\s'}}{K}=N_{\s'}/r$.}
$$
\eqalignno{
& \partial_- \bigl( B_{0}^{-1}\> \partial_+\> B_{0}\bigr)\> =\> 
\bigl[\Lambda_+\>, \> B_{0}^{-1}\> \Lambda_-\>
B_{0}\bigr]\Bigm|_{\scriptstyle \gg{0}(\s'; \s', q)}\>, & \nameali{TodaFin}
\cr & \partial_+\partial_- \nu\> =\>  
{\bil{\Lambda_+}{B_{0}^{-1}\> \Lambda_-\> B_{0}\> -\> \Lambda_-}\over 
\bil{\Lambda_+}{\Lambda_-}}\>. &\nameali{TodaCent}\cr}
$$
As it is well known, eq.~\TodaFin\ provides the classical
equations-of-motion for 
$$
S[B_0]\> =\> {1\over \beta^2}\> \Bigl[ S_{WZW} [B_{0}]\> -\> \int d^2x\> 
V(B_{0}) \Bigr]\>,
\nfr{Act}
where $S_{WZW} [B_{0}]$ is the Wess-Zumino-Witten action associated with 
the finite Lie group corresponding to $\gg{0}(\s'; \s', q)$.
The potential is given by
$$
V(B_{0})\> =\> {1\over \pi}\> \bil{\Lambda_+}{B_{0}^{-1}\>
\Lambda_-\> B_{0}\> -\> \Lambda_-}\> ,
\efr
and it is assumed that $V(B_0)$ has an absolute minimum at $B_{0}=I$, 
where it obviously vanishes. Since both $\Lambda_\pm$ and $B_{0}$ take
values in a complex algebra and group, respectively, the action~\Act\ is
a complex functional of the Toda field $B_{0}$. However, additional
restrictions on the data can be imposed such that~\Act\ gives rise to a
unitary theory~[\Ref{NOS}]. 

Restricted to the NAAT equation, and according to the results
of~[\Ref{JMP}], the class of solutions obtained in Section~4 solutions
corresponds precisely to the non-abelian version of the solitonic
specialization originally proposed by Olive {\it et al.\/} in  the
context of the abelian affine Toda (AAT)
theory~[\Ref{EMSPEC},\Ref{SOLSPEC}].  This class of solutions is
conjectured to include the multi-soliton solutions, which has been
extensively checked for the usual abelian affine Toda
theories~[\Ref{SOLTAU},\Ref{CAT},\Ref{ANSOL}]. As explained in the
introduction, in the abelian affine Toda theories this conjecture is
supported by the explicit evaluation of the energy-momentum tensor. It
splits into a radiation piece that vanishes for soliton solutions and
an improvement~[\Ref{EMSPEC}]. 

In the non-abelian case, the energy-momentum tensor for the action~\Act\
can be computed by standard means and it reads
$$
\eqalign{
T_{\mu \nu}\> & =\> {1\over \pi\> \beta^2}\> \biggl[ {1\over4}\>
\bil{\partial_\mu B_{0}}{\partial_\nu B_{0}^{-1}}\> -\> {1\over8}\> 
g_{\mu \nu}\> 
\bil{\partial_\rho B_{0}}{\partial^\rho B_{0}^{-1}}\cr &\qquad \qquad 
\qquad -\>  m^2\> g_{\mu \nu}\> \bil{\Lambda_+}{B_{0}^{-1}\> \Lambda_-\>
B_{0}\> -\> \Lambda_-}
\biggr]\>. \cr}
\nfr{EMT}
In~[\Ref{LUIZ}], it was shown that $T_{\mu \nu}$ actually splits as
follows: 
$$
T_{\mu \nu}\>  =\> \Theta_{\mu \nu}\> -  \> { 1\over 2\pi 
\beta^2}\> {N_{\s'}\over r}\>\bigl(\partial_\mu \partial_\nu\> -\> 
g_{\mu \nu}\>
\partial_\rho \partial^\rho\bigr)\> \nu\>.
\nfr{GCAT}
The first piece,  $\Theta_{\mu \nu}$, is the traceless energy-momentum of 
a generalized non-abelian `conformal' affine Toda (G-CAT) theory
associated with the affine Kac-Moody algebra $\g^{(r)}$, while the second
is just an improvement term given by the component of the Toda field 
along the central element (see eq.~\BceroNu). Now, following the
reasoning of refs.~[\Ref{CAT},\Ref{LUIZ}], suppose that we have a
multi-soliton solution that carries finite energy and momentum
characterized by the masses of the individual solitons. Then, since the
G-CAT theory is conformal invariant and has no mass scale, the
contribution of $\Theta_{\mu \nu}$  has to vanish and, hence, the value
of the energy and  momentum carried by the solitons has to be given by
the boundary behaviour of $\nu$. This is the analogue of the property
used by Olive {\it et al.} to justify the solitonic specialization, and
it is agreed as a definite criterion to characterize the multi-soliton
solutions. 

The components of the energy-momentum tensor~\EMT\ can be
identified with some of the invariant conserved densities defined in
Section~3. In order to do it, let $\Phi_\pm = \exp
\bigl(y_{\pm}^{1} + y_{\pm}^{2} +\cdots \bigr)$ and
$h^{\Phi_\pm}_\pm = h^{\Phi_\pm}_\pm(\pm1)= h_{\pm}^0 + h_{\pm}^{1} +
\cdots$ in~\Abelianize, where  $y_{\pm}^{i} $ and $h_{\pm}^{i}$ are in
$\gg{\mp i}(\s')$. Then,~\Abelianize\ provides the following equations
$$
\eqalign{
& h_{\pm}^{0}\> +\> [y_{\pm}^{1}\>, \> \Lambda_\pm]\> =\> q_\pm \cr
& h_{\pm}^{1}\> +\> [y_{\pm}^{2}\>, \> \Lambda_\pm ]\> =\> 
{1\over2}\> [y_{\pm}^{1}\>, \>[ y_{\pm}^{1}\>, \>\Lambda_\pm ]] \>- \>
[y_{\pm}^{1}\>, \> q_\pm]\> -\> \partial_\pm \> y_{\pm}^{1}\>, \cr}
\nfr{AberConcr}
where~\ConstT\ implies that $h_{\pm}^{0}\in {\Bbb C}K$. They allow one
to calculate the invariant conserved densities
$$
\eqalign{
J_{\pm}^{(\pm 1)}[\pm1; \m]\> &=\>
\bil{\Lambda_\pm}{h_{\pm}^{\Phi_\pm}}\> -\> {N_{\s'}\over N_{\m}}\>
\partial_\pm
\bil{d_{\m}}{\Phi_\pm\> \Lambda_\pm\> \Phi^{-1}_\pm}\> \cr
&= \> {1\over2}\> \bil{q_\pm }{q_\pm }\> +\> \partial_\pm \bil{d_{\s'}\>
-\> {N_{\s'} \over N_{\m}}\> d_{\m}}{q_\pm \> -\> h_{\pm}^0}\>,}
\nfr{DensOne}
where $\m$ is an arbitrary gradation $\preceq\s'$ and $d_{\m}$ the 
corresponding derivation. Eq.~(A.6) shows that the
combination $d_{\s'}\> -\> {N_{\s'} \over N_{\m}}\> d_{\m}$ does not
contain any component along $d$ and, hence, the total derivative piece is
a local functional of the non-central components of $q_\pm$. This confirms
that $J_{\pm}^{(\pm 1)}[\pm1; \m]$ actually lead to 
$\m$-independent conserved charges, as explained below eq.~\Def. The
explicit relation between the conserved densities and the
energy-momentum tensor requires the choice $\m=\s'$; then
$$
J^{(\pm1)}_\pm [\pm 1; \s']\> =\> -\> 2\pi \beta^2 \> T_{\pm \pm}\>.
\nfr{RelOne}
Moreover, using~\AbeljTp\ and~\AbeljTn, it is not difficult to show that
$$
J^{(-1)}_+ [+ 1; \s']\> =\> J^{(+1)}_- [- 1; \s']\> =\> +\> 2\pi
\beta^2\> T_{+ -}\>.
\nfr{RelTwo}

Let us now consider the specialization of these equations for the class of
group theoretical solutions given by~\SolT. In this case, eq.~\DefTtau\
implies that
$$
T_{\mu\nu}\> =\> {1\over 2\pi \beta^2}\> \bigl(\partial_\mu\partial_\nu\> 
-\> g_{\mu\nu}\> \partial_\rho \partial^\rho\bigr) \> \ln\>
\tau_{\overline{\s'}}^{\rm T}(h)\>,
\nfr{Known}
where (see eqs.~\BceroNu, \SolT and~\Ristra)
$$
\eqalignno{
\ln\> \tau_{\overline{\s'}}^{\rm T}(h)\> & =
\ln\> \>\bra{v_{\overline{\s'}}} \Psi^{(\rm vac)}\>  h\>  {\Psi^{(\rm
vac)}}^{-1} \ket{v_{\overline{\s'}}} \cr
\noalign{\vskip 0.2truecm}
& = \> -\> {N_{\s'}\over r}\>
\bigl( \nu[h]\> + \> x_+ x_- -\> \sum_{ j\in E_{\Lambda_\pm}>0}\> jt_j
t_{-j} \bigr)\>, &\numali \cr}
$$
which compared with~\GCAT, and taking into account that
$t_{\pm1}=x_\pm$, shows that~$\Theta_{\mu\nu}$ actually vanishes  for
all the solutions given by~\SolT. 

This result provides
additional support to the conjecture that the class of solutions
corresponding to the solitonic specialization provide the multi-soliton
solutions also for the non-abelian affine Toda equations. However, it is
important to recall that not all these solutions are expected to be
solitons. In fact, this usually requires to constrain the parameters
entering the group element $h$ to ensure that the solution behaves as a
soliton, \ie, that it carries finite conserved quantities, etc. In this
respect, our results allow one to deduce these constraints by analysing
the properties of a single complex tau-function that encodes all the
information about the solution.

Although the precise relationship between the conserved densities
$J_{\pm}^{(\pm1)}[\pm1, \m]$ and the components of the energy-momentum
tensor requires the choice $\m=\s'$, it is worth recalling that the
conserved quantities are $\m$-independent; \ie, taking into
account~\Ex\ and~\DefTtau,
$$
Q_{n}^{\pm}\> =\> \int_{-\infty}^{+\infty}\> dx_\pm \>
J_{\pm}^{(\pm1)}[\pm n, \m] \> =\> -{N_{\s'}\over n N_\m}\> \pa_{\pm n}
\ln
\tau_{\overline{\m}}^{\rm T}(h) \Bigm|_{x_\pm = -\infty}^{x_\pm =
+\infty} 
\efr
does not depend on $\m$ as far as $\m \preceq \s'$. A similar result
restricted to the context of the abelian affine Toda equation can
be found in~[\Ref{NIEDER}].

\section{Conserved quantities of constrained KP solitons}

So far we have been concerned with the relationship between
tau-functions and conserved densities. The purpose of our last example
is to illustrate the use of eqs.~\DefC\ and~\DefTtau\ to calculate the
conserved quantities carried by multi-soliton configurations. 

Consider a generalized Drinfel'd-Sokolov hierarchy where the
function $q$ is defined for $-L < x <L$, and it is either periodic or 
rapidly decreasing at $x\rightarrow \pm\infty$ depending on whether
$L<\infty$ or $L=\infty$, respectively. Then, the conserved quantities
carried by the solutions specified in Section~4.1 can be calculated by
means of~\Def\ and~\DefC:
$$
Q_n \>= \> \int_{-L}^{+L}\> dx \> J^{(i)}[n,\m]\> =\>
-\> {N_{\s'}\over n\> N_{\m}}\> \pa_n \ln \tau_{\overline{\m}}^{\rm DS}
\Bigm|_{x=-L}^{x=+L}\>,
\nfr{CargaA}
which is independent of the auxiliary gradation $\m\preceq \s'$.
Moreover, since we impose the constraint~\ConstDS, $\pa_x =\pa_i$
and, hence, $\tau_{\overline{\m}}^{\rm DS}$ depends on $x$ and
$t_i$ only through the combination $x+t_i$. Therefore,~\CargaA\ is
equivalent to the following equation
$$
\ln\> {\tau_{\overline{\m}}^{\rm DS} [\ldots, t_i \> +\> L, \ldots]
\over
\tau_{\overline{\m}}^{\rm DS} [\ldots, t_i \> -\> L, \ldots]}\> =\>
C_\m \> -\> {N_\m \over N_{\s'}}\> \sum_{n\in E_\Lambda>0} n\>
t_n \> Q_n \>,
\nfr{Cargas}
where $C_\m$ is a constant whose value usually diverges in the limit
$L\rightarrow \infty$. It is important to emphasize that, as far as
$\m\preceq\s'$, the constants $C_\m$ and $N_\m$ are the only quantities
that depend on $\m$ on the right-hand-side of~\Cargas.

To illustrate the use of~\Cargas, we will calculate the conserved
quantities carried by the soliton solutions for the constrained KP (cKP)
hierarchy associated with $\widehat{sl}(3)$ (${\rm
cKP}_{2,1}$ or Yaijma-Oikawa hierarchy) recently worked out by Aratyn
{\it et al.}~[\Ref{LUIZKP}]. In the notation of Section~2, this
hierarchy is associated to the untwisted affine Kac-Moody algebra
$\gkm=A_{2}^{(1)}$, the Heisenberg subalgebra
$\HSA[\s']= \Ker(\ad \Lambda)$, where $\Lambda = e_{2}^+ + e_{2}^-$ and
$\s' =(1,0,1)$, and the additional gradation $\s=\s'$. 
The $\s'$-grade of $\Lambda$ is $i=1$, and $\HSA[\s']= {\Bbb C}K \oplus
\sum_{j\in E_\Lambda} {\Bbb C}b_j$ where $E_\Lambda =\Bbb Z$.

According to~[\Ref{LUIZKP}], one needs four generalized
Hirota tau-functions to reconstruct the Lax operator
$$
L\> =\> \pa_x \> -\> \pmatrix{0 & q& 0\cr r & U_2& 1\cr
0& \lambda& -U_2\cr}\> -\> \nu\> K\>.
\efr
However, according to our results, the conserved quantities can be
obtained by means of a single one associated to any gradation
$\m\preceq \s'$. Since the relevant tau-functions are
provided by the authors of~[\Ref{LUIZKP}], we will consider
$\m_0=(1,0,0)=\s^{(0)}$ and $\m_2=(0,0,1)=\s^{(2)}$ to illustrate the
$\m$-independence of the result. The relation between their notation and
ours is provided by
$$
\tau_{\m_\sigma}^{\rm DS}\> =\>  \tau_{\sigma}^{(0)}\>, \qquad
\sigma\> =\> 0,2\>.
\efr

The multi-soliton solutions considered in~[\Ref{LUIZKP}] are given by
eq.~\SolDS\ or, equivalently,~\SolDSLin, where the constant group
element $h$ is the product of exponentials of eigenvectors of the
Heisenberg subalgebra generators
$$
h\> =\> {\rm e\/}^{F_1}\> {\rm e\/}^{F_2}\> \cdots\> {\rm e\/}^{F_n}
\>, \qquad
[b_n\>, \> F_k]\> =\> \omega_{n}^{(k)}\> F_k\>, \quad k\> =\> 1,
\ldots, n\>, \quad n \in {\Bbb Z}\>.
\nfr{Expo}
The construction of the $F_k$'s involves the following Fubini-Veneziano
operators
$$
Q_{1}(z)\> =\> i\> \sum_{n\in {\Bbb Z}}{z^{-(2n+1)}\over 2n+1}\>
b_{2n+1}\>, \qquad
Q_{2}(z)\> =\> q\> - i\> p\> \ln\> z\> +\> i\>
\sum_{n\not=0}{z^{-2n}\over 2n}\> b_{2n}\>.
\efr
Then, there are three different eigenvectors corresponding to the three
positive roots of $sl(3)$:
$$
\eqalignno{
& E_{\pm 1,\sqrt{3}}(z)\> =\> \sqrt{2z^3}\> : \exp\bigl( \pm
i\>Q_1(z)\> +\> i\sqrt{3}Q_2(z)\bigr) :\>,  \cr
& E_{-2,0}(z)\> =\> -\> {1\over2}\> : \exp\bigl( -2
i\>Q_1(z)\bigr) :\> {\rm e\/}^{i\pi p}\>. & \numali \cr}
$$

We now calculate the value of the conserved quantities carried by the
soliton solutions constructed in~[\Ref{LUIZKP}]:

\jump
\indent\indent\indent 1) $n=1$ and $F_1=
E_{-2,0}(z_1)$. This  configuration is rapidly decreasing at
$x\rightarrow \pm\infty$ if
${\rm Re}(z_1)\not=0$. The corresponding tau-functions are
$$
\tau_{\sigma}^{(0)}\> =\> 1\> -\> {(-1)^{\sigma/2}\over 2}\> {\rm
e\/}^{-2(x+t_1)z_1 \> -\> 2t_3 z_{1}^3\> - \> \cdots}\>, \quad
\sigma=0,2\>, 
\nfr{TauRep}
where only the dependence on times $t_1$, $t_2$ and $t_3$ is exhibited.
Then, it is not difficult to calculate 
$$
\lim_{L\rightarrow\infty} \ln\> {\tau_{\sigma}^{(0)} [t_1 + L,
t_2, t_3, \ldots]
\over
\tau_{\sigma}^{(0)} [t_1 - L, t_2, t_3, \ldots]}\> =\>
C_\sigma \> +\> {\rm sign}[{\rm Re}(z_{1})]\> \bigl( 2z_{1}\> t_1\> +\>
2z_{1}^3\> t_3\> +\> \cdots \bigr)\>,
\nfr{CargasUno}
where 
$$
C_\sigma = \lim_{L\rightarrow\infty} \> \ln \bigl(2^{\pm1} \>
(-1)^{1+ \sigma/2}\> {\rm e\/}^{-2z_{1} L} \bigr)\>, \quad {\rm for}\quad 
\pm {\rm Re}(z_{1})>0
\efr
is the only $\sigma$- or $\m$-dependent factor. Taking into
account~\Cargas\ and~\CargasUno, the first three conserved quantities
carried by this soliton configuration are
$$
\eqalignno{
& Q_1\> =\> -\> 4\> {\rm sign}[{\rm Re}(z_{1})]\> z_{1}\>, \qquad Q_2\>
=\>  0\>, \cr
& Q_3\> =\> -\> {4\over3}\> {\rm sign}[{\rm Re}(z_{1})]\>
z_{1}^3 \>.&\numali \cr}
$$

\jump
\indent\indent\indent 2.a) $n=2$, $F_1= E_{-2,0}(z_1)$, and $F_2=
E_{1,\sqrt{3}}(z_2)$. The condition that this configuration is rapidly
decreasing at $x\rightarrow \pm\infty$ requires that $z_1+z_2=0$ and
${\rm Re}(z_1)\not= 0$. The tau-functions $\tau_{0}^{(0)}$ and
$\tau_{2}^{(0)}$ are also given by~\TauRep. Therefore, this
configuration carries the same conserved quantities as the previous
($n=1$) one.

\jump
\indent\indent\indent 2.b) $n=2$, $F_1= E_{1,\sqrt{3}}(z_1)$, and $F_2=
E_{1,-\sqrt{3}}(z_2)$. This configuration is rapidly decreasing at
$x\rightarrow \pm\infty$ whenever ${\rm Re}(z_1)$ and ${\rm Re}(z_2)$
are non-vanishing and of equal sign, \ie, whenever ${\rm Re}(z_1)\> {\rm
Re}(z_2) > 0$. The corresponding tau-functions are 
$$
\tau_{\sigma}^{(0)}\> =\> 1\> +\> {(-1)^{\sigma/2}\> 
2\> z_{1}^{1+\sigma/2}\> z_{2}^{2-\sigma/2} \over (z_1-z_2)\>
(z_1+z_2)^2}\> {\rm e\/}^{\bigl[(x+t_1)(z_1 + z_2) \> +\> \sqrt{3} t_2 
(z_{1}^2 -z_{2}^2)\> +\>  t_3 (z_{1}^3 +z_{2}^3)\> +\> \cdots\bigr]}\>,
\efr
for $\sigma=0,2$, which lead to the following conserved quantities:
$$
\eqalignno{
& Q_1\> =\> -\> 2\> {\rm sign}[{\rm Re}(z_{1})]\> (z_{1}+ z_{2})\>,
\qquad Q_2\> =\>  -\> \sqrt{3}\> {\rm sign}[{\rm Re}(z_{1})]\> (z_{1}^2
-z_{2}^2)\>, \cr 
& Q_3\> =\> -\> {2\over3}\> {\rm sign}[{\rm Re}(z_{1})]\> (z_{1}^3
+z_{2}^3)\>. &\numali \cr}
$$

Let us mention that this solution belongs to the class of
multi-soliton solutions to the cKP hierarchy constructed by Aratyn {\it
et al\/} in~[\Ref{SOLKP}]. This class is obtained through the
Darboux-B\"acklund method and, hence, the corresponding tau-functions
admit a Wronskian determinant representation.

\jump
\indent\indent\indent 3) $n=3$, $F_1= E_{-2,0}(z_1)$, $F_2=
E_{1,\sqrt{3}}(z_2)$, and $F_3= E_{1,-\sqrt{3}}(z_3)$. This
configuration is rapidly decreasing at $x\rightarrow\pm \infty$ if
${\rm Re}(z_1)\not= 0$ and ${\rm Re}(z_2)\>{\rm Re}(z_3) > 0$. The
tau-functions are more complicated than in the previous cases:
$$
\eqalignno{
\tau_{\sigma}^{(0)}\> & =\> 1 \> +\> {(-1)^{\sigma/2}\over 2}\> {\rm
e\/}^{-2(x+t_1)z_1 \> -\> 2t_3 z_{1}^3\> +\> \cdots}\cr
\noalign{\vskip 0.3truecm}
&+\> {(-1)^{\sigma/2}\> 2\>
z_{2}^{1+\sigma/2}\> z_{3}^{2-\sigma/2}
\over (z_2-z_3)\> (z_2+z_3)^2}\> {\rm e\/}^{\bigl[(x+t_1)(z_2 + z_3) \>
+\> \sqrt{3} t_2  (z_{2}^2 -z_{3}^2)\> +\>  t_3 (z_{2}^3 +z_{3}^3)\> +\>
\cdots\bigr]}\cr
\noalign{\vskip 0.3truecm}
& +\>  {z_{2}^{1+\sigma/2}\> z_{3}^{2-\sigma/2}\> (z_1+z_2)\> (z_1+z_3)
\over (z_2-z_3)\> (z_2+z_3)^2 \> (z_1-z_2)\> (z_1-z_3)}\cr 
& \qquad\qquad\qquad {\rm
e\/}^{\bigl[(x+t_1)(z_2 + z_3-2z_1) \> +\> \sqrt{3} t_2  (z_{2}^2
-z_{3}^2)\> +\>  t_3 (z_{2}^3 +z_{3}^3-2z_{1}^3)\> +\>
\cdots\bigr]} \>. &\numali \cr}
$$
However, they also lead to $\sigma$-independent conserved charges:
$$
\eqalignno{
& Q_1\> =\> -\> 4\> {\rm sign}[{\rm Re}(z_{1})]\> z_{1}\>
-\> 2\> {\rm sign}[{\rm Re}(z_{2})]\> (z_{2}+ z_{3})\>, \cr
& Q_2\> =\>  -\> \sqrt{3}\> {\rm sign}[{\rm Re}(z_{2})]\> (z_{2}^2
-z_{3}^2)\>, \cr 
& Q_3\> =\> -\> {4\over3}\> {\rm sign}[{\rm Re}(z_{1})]\> z_{1}^3
-\> {2\over3}\> {\rm sign}[{\rm Re}(z_{2})]\> (z_{2}^3
+z_{3}^3)\>.&\numali \cr}
$$

A detailed analysis of these soliton configurations is beyond the scope
of this article. However, it is worth noticing that the calculation of
the conserved quantities they carry suggests that the three
solutions with $n=1$ and $n=2$ describe single solitons,
while the last one with $n=3$ is a two-soliton solution.  This
contrasts with the interpretation proposed in~[\Ref{LUIZKP}], where
$n$ is naively identified with the number of solitons. The
correspondence between the Wronskian determinant representation
of~[\Ref{SOLKP}], when it is available, and our tau-functions also
deserves further understanding.

\chapter{Conclusions}

For a large class of integrable hierarchies, we have shown that all
the information required to calculate their local conserved densities
is provided by a single (complex) scalar tau-function. We give
explicit expressions for calculating both the conserved densities and
charges in terms of the second order partial derivatives of the
tau-function and its asymptotic behaviour.

The class where our results apply comprises two different types of
hierarchies associated with affine (twisted or untwisted) Kac-Moody
algebras. The first corresponds to the generalized Drinfel'd-Sokolov
hierarchies constructed in~[\Ref{GEN},\Ref{GEN2},\Ref{GENTAU}]. The
second is a new non-abelian generalization of the usual
two-dimensional affine Toda lattice hierarchy, which is constructed
by coupling two different generalized Drinfel'd-Sokolov mKdV
hierarchies. The tau-function formalism for most of these hierarchies
can be found in~[\Ref{GENTAU},\Ref{JMP}], where a direct relation
between the variables of the zero-curvature formalism and the
tau-functions is established. This relation exhibits that, in general,
more that one tau-function is required, which is already the case for
the usual mKdV, NLS and affine Toda lattice hierarchies, while KP and
KdV constitute exceptions from this point of view. In contrasts, our
results show that, in the whole class of hierarchies, a single
tau-function is enough to reconstruct all the (local) conserved
densities.

According
to~[\Ref{SEGWIL},\Ref{WILTAU},\Ref{KW},\Ref{GENTAU},\Ref{JMP}], the
tau-function formalism concerns a particular class of solutions
specified by the elements of an infinite dimensional manifold
associated with a Kac-Moody group. Our results assume the restriction
to these `group theoretical' solutions. For the affine Toda 
equation, this class coincides with the, so called, solitonic
specialization originally proposed by Olive {\it et
al.\/}~[\Ref{EMSPEC},\Ref{SOLSPEC}]. In the general case, the relevant
class of solutions is provided by the orbit of the group of dressing
transformations, in the manner described by Wilson~[\Ref{WILTAU}],
acting on the `vacuum solutions' of the
hierarchy~[\Ref{GENTAU},\Ref{JMP}]. In the
particular case of the generalized Drinfel'd-Sokolov hierarchies
of~[\Ref{GEN},\Ref{GEN2},\Ref{GENTAU}], we have also shown that, once
restricted, they fit nicely in the approach of Feigin, Frenkel and
Enriquez to soliton equations of KdV and mKdV type~[\Ref{FRENKEL}].

In the context of particle physics, there are two main motivations
lying behind this work. The first one comes from the increasing
number of, even four dimensional, quantum systems whose effective
action turns out to be the tau-function of some classical integrable
system~[\Ref{TAUQ},\Ref{DIJK},\Ref{SW}]. Then, among the many
tau-functions required to reconstruct the variables of the
zero-curvature formalism, the single one that provides the local
conserved densities can be viewed as the logical candidate to describe
the effective action of some quantum system where the conserved
densities would be correlation functions. 

The second motivation comes from the construction of soliton and
multi-soliton solutions to the equations of these hierarchies. Then the
main result is that all the information about the conserved charges
carried by them (\eg, their energy and momentum) turns out to be
encoded in a single (complex) scalar tau-function. In the context of
the abelian affine Toda equations, this is a generalization to all the
conserved quantities of the well known properties of the
energy-momentum tensor, which originally led to the solitonic
specialization of its general solution~[\Ref{EMSPEC},\Ref{SOLSPEC}].
However, our results apply not only to this case but to a wider class
of integrable equations including some whose soliton solutions have
recently deserved attention. 

In particular, it includes multi-component generalizations
of the constrained KP (cKP) hierarchy~[\Ref{FHM},\Ref{ARA}], whose
soliton solutions have been investigated in~[\Ref{SOLKP}] using the
Darboux-B\"acklund method and, more recently, in~[\Ref{LUIZKP}] using
tau-functions. We have applied our results
to calculate the conserved charges carried by the cKP solitons
constructed in the latter reference, and we expect that they will be
useful to clarify the relationship between them and
those of~[\Ref{SOLKP}].
 
The class of integrable equations also includes the non-abelian
generalizations of the affine Toda equations corresponding to the
equations-of-motion of the Symmetric Space (SSSG) and Homogeneous
(HSG) sine-Gordon theories constructed in~[\Ref{NOS}]. These theories
describe massive integrable perturbations of conformal field theories
given by a (gauged) Wess-Zumino-Witten action, and their spectrum
always contains solitons. In particular, the HSG theories are quantum
integrable perturbations of parafermionic theories~[\Ref{NOSQ}] whose
spectrum is entirely given by solitons~[\Ref{NOSSOL}]. Our results
considerably simplify the study of the conserved charges carried by
these solitons, which is relevant towards the overall aim of finding
the factorizable S-matrix of these theories.

\bjump\bjump

\centerline{{\bf Acknowledgements}}

\noindent
I would like to thank H.~Aratyn, L.A.~Ferreira, J.O.~Madsen
and J.~S\'anchez Guill\'en for many useful comments. I also thank
C.R.~Fern\'andez Pousa for his early colaboration on this work. This
research is supported partially by CICYT (AEN96-1673), DGICYT
(PB96-0960), and the EC Comission via a TMR Grant (FMRX-CT96-0012).

\bjump

\appendix{\bf Appendix}

In this appendix we summarize some of the details of affine Kac-Moody 
algebras which are used along the article. In order to fix the notation,
we will follow the conventions of~[\Ref{KBOOK}] where a complete
treatment of these algebras may be found (see also~[\Ref{GO}], and
references therein).

An affine Kac-Moody algebra~$\gkm =  \gkm(A) $ of rank $l$ is defined by a
generalized Cartan matrix $A=(a_{ij})$ of order $l+1$ (and rank~$l$), and is
generated by $\{h_i, e_{i}^\pm, i=0, \ldots, l\}$ and $d$ subject to the
relations
$$
\eqalign{
[h_i\> ,\> h_j]\> =0\>, \qquad &  [h_i\> ,\>e_{j}^\pm] \> =\> \pm a_{ij}\>
e_{j}^\pm\>, \cr
[e_{i}^+\>, \> e_{j}^-] \> =\> \delta_{ij}\> h_i \>, \qquad & (\ad
e_{i}^\pm)^{1-a_{ij}} \> (e_{j}^\pm)\> =\> 0\>, \cr
[d\>, \> h_i]\> =0\>, \qquad & [d\>, \> e_{i}^\pm]\> =\> \pm
\delta_{i,0}\> e_{0}^\pm\>. \cr}
\efr
The elements $\{h_0, \ldots, h_l, d\}$ span the Cartan subalgebra $\hkm$ 
of $\gkm$. The elements $e_{i}^\pm$ generate the maximal nilpotent 
subalgebras $\nn_\pm$, respectively, and are called the Chevalley
generators of $\gkm$. They also generate the derived subalgebra $\gg{} =
[\gkm,\gkm]$, and we have
$$
\gkm\> =\> \gg{} \> \oplus \> {\Bbb C}\> d \> =\> \nn_- \> \oplus \> \hh
\> \oplus \> \nn_+\>;
\efr
correspondingly, $\hkm = \hh' \oplus {\Bbb C}\> d$.
The different generalized Cartan matrices of affine type and, hence, the 
different affine Kac-Moody algebras, are usually specified in the form
$\g^{(r)}$, where $\g$ is a complex simple finite Lie algebra and $r=1$, $2$, or
$3$ is the order of an automorphism of the Dynkin diagram of $\g$.
The generalized Cartan matrix $A$ have unique right and left null 
eigenvectors whose components are positive relatively prime integers known
as the Kac labels ($a_{i}$) and the dual Kac labels ($a_{i}^\vee$) of
$\gkm$:
$$
\sum_{i=0}^l a_{i}^\vee\> a_{ij} \> =\> \sum_{i=0}^l  a_{ji} \> a_{i} \> =\>
0\>.
\efr
In all cases $a_{0}^\vee = a_\epsilon=1$, where $\epsilon=l$ if $\gkm =
A_{2l}^{(2)}$ and~$0$ otherwise. Consequently, the algebra $\gkm$ has a
one-dimensional centre ${\Bbb C}\> K$ generated by the central element
$K=\sum_{i=0}^{l} a_{i}^\vee \> h_i$. An important property of $\gkm$ is that
it has an invariant symmetric non-degenerate bi-linear form
$\bil{\cdot}{\cdot}$, which will be normalized such that
$$
\eqalign{
& \bil{e_{i}^+}{e_{j}^-}\> =\> a_{i}\>  {a_{i}^\vee}^{ -1}\> \delta_{ij} 
\>, \qquad \bil{h_i}{h_j}\> =\> a_{j}\>  {a_{j}^\vee}^{ -1}\> a_{ij} \>, \cr
& \bil{d}{h_{i}}\> =\> a_0\> \delta_{i0}\>, \qquad \bil{d}{K}\> =\> a_0 \>, 
\qquad \bil{d}{d}\> =\> 0   \>. \cr} \efr

\section{Gradations}

The different integer gradations of $\gkm$ are labelled by sets
$\s=(s_0,s_1, \ldots, s_l)$ of non-negative integers. The corresponding
gradation is induced by a derivation $d_\s$ specified by
$$
[d_\s \>, \> e_{i}^\pm] \> =\> \pm s_i\> e_{i}^\pm
\quad {\rm and}  \quad [d_\s \>, \> h_i]\> = \> [d_\s \>, \> d] \>= \>
0\>. 
\efr
In particular, the derivation $d$ corresponds to the, so called, 
`homogeneous' gradation $\sh=(1,0,\ldots,0)$. Let us denote by
$\overcirc{\gkm}$ the subalgebra of $\gkm$ generated by the $e_{i}^\pm$
with $i\not=0$. Then, the derivation
$d_\s$ is just the following element of the Cartan subalgebra of $\gkm=
\g^{(r)}$:
$$
d_\s\> =\> {N_\s \over r}\> \Bigl[{1\over a_0}\> d\> + \> u_\s\> -\>
{1\over2}\> \bil{u_\s}{u_\s}\> K \Bigr]\>,
\efr
where
$$
N_\s\> =\> r\> \sum_{i=0}^l a_i \> s_i\>, 
\efr
and $u_\s$ is the unique element in $\overcirc{\gkm}$ such that  $
\bil{u_\s}{h_i} = rN_{\s}^{-1}\> a_{i} a_{i}^{\vee-1} s_i $ for all
$i\not=0$. This derivation satisfies 
$$
\bil{d_\s}{d_\s} \>=\> 0\>, \quad \bil{d_\s}{K}\> =\> {N_\s\over r}\>
,\quad {\rm and}\quad 
\bil{d_\s}{h_i} \>=\> a_{i}\> a_{i}^{\vee-1}\> s_i
\efr 
for all $i=0,1,\ldots, l$. Under the gradation $\s$, the  derived
subalgebra of $\gkm$ has the eigenspace decomposition
$$
\gg{}\>= \> \bigoplus_{i\in {\Bbb Z}} \gg{i} (\s)\>, \qquad
[\gg{i} (\s)\>, \> \gg{j} (\s)]\> \subset \> \gg{i+j} (\s)\>.
\efr
We shall often use the notation like $\gg{>k}(\s)=\oplus_{i>k}\>
\gg{i} (\s)$, and denote by ${\rm P}_{>k[\s]}$ the projector onto
$\gg{>k}(\s)$. Then, for any gradation $\s$, we have
$$
\gg{>0}(\s) \subseteq \nn_+ \>, \qquad \gg{<0}(\s) \subseteq \nn_- \>, \quad {\rm
and} \quad \hh' \subseteq \gg{0}(\s)\>,
\efr
where the equality corresponds to the, so called, `principal' gradation
$\s_{\rm p}=(1,1,\ldots,1)$ or, in general, to any gradation $\s$ such that
$s_i\not=0$ for all $i=0,1,\ldots, l$.

There exists a partial ordering on the set of gradations such
that $\s\preceq \s'$ if $s_{i}' \not=0$ whenever $s_i\not=0$, which implies
that
$$
\eqalignno{
\gg{>0}(\s) \subseteq \gg{>0}(\s') \>, \qquad
&\gg{<0}(\s) \subseteq \gg{<0}(\s')\>, \cr
\noalign{\vskip 0.3cm}
\gg{0}(\s') \subseteq \gg{0}(\s) \>, \qquad \gg{>0}(\s') \subset
&\> \gg{\geq 0}(\s) \>, \qquad \gg{<0}(\s') \subset \gg{\leq 0}(\s) \>. 
& \numali \cr}
$$
Moreover, we will indicate by $\s \prec \s'$ the case when $\s\preceq\s'$
but $\s'\not\preceq\s$, and by $\s \simeq \s'$ when $s_{i}'=0$ if, and only if,
$s_i =0$. In the latter case, the following identities are satisfied:
$\gg{0}(\s)=\gg{0}(\s')$, $\gg{>0}(\s)=\gg{>0}(\s')$, and
$\gg{<0}(\s)=\gg{<0}(\s')$.

There is a simple algorithm for determining the `horizontal subalgebra'
$\gg{0} (\s)$. Let $i_1, \ldots, i_p$ be all the indices for which
$s_{i_1}= \ldots = s_{i_p}=0$. Then, it is the direct sum of the
$(l-p)$-dimensional centre and a semisimple Lie algebra whose Dynkin
diagram is the subdiagram of the of the (affine) Dynkin diagram of
$\gkm=\g^{(r)}$ consisting of the vertices $i_1, \ldots, i_p$.  

$\gg{0} (\s)$ always contains the centre of $\gkm$ and, when  needed, we
will project it out by means of the following procedure. Consider an
auxiliary gradation $\m \preceq \s$, choose an integer number $q$ such
that $m_q\not=0$, and construct the subalgebra $ \gg{0} (\s; \m, q)$
generated by the $e_{i}^\pm$ with $s_i=0$, and the
$$
\widetilde{h}_j \> =\> h_j\>-\> {r\over N_\m}\> a_j\> a_{j}^{\vee\> -1}\>
m_j\> K\>
\efr
for all $j\not=q$. Then, $\gg{0} (\s)$ can be decomposed as
$$
\gg{0} (\s) \> =\> \gg{0} (\s; \m, q)\> \oplus \> {\Bbb C}\> K\>,
\nfr{SplitC}
and $\bil{d_\m}{\gg{0} (\s; \m, q)} =0$. Notice that $\gg{0} (\s; \m,
q) \simeq \gg{0} (\s)/{\Bbb C}K$ for any choice of $\m$ and $q$.
Moreover, the restriction of the bi-linear form $\bil{\cdot}{\cdot}$ to
the finite Lie algebra $\gg{0} (\s; \m, q)$ is non-degenerate.

The different gradations of $\gkm= \g^{(r)}$ are in one-to-one relation
with the finite order automorphisms $\sigma$ of the underlying finite Lie algebra $\g$
fulfilling the condition that $r$ is the least positive integer such that $\sigma^r$ is
inner. Up to conjugation, they can also be labelled by sets
$\s=(s_0, s_1, \ldots, s_l)$ of non-negative integers, $\sigma=
\sigma_{\s; r}$, where $l={\rm rank\/}(\gkm)$. Then, there exists a set of
generators of $\g$, $\{E_i, i=0, \ldots, l\}$, such that 
$$  
\sigma_{\s; r}(E_j)\> =\> {\rm e\>}^{{2\pi\>i \over N_\s}\> s_j}\> E_j\>,
\efr
which shows that $\sigma_{\s;r}$ is of order $N_\s$. 
Notice that ${\rm rank}(\g)\geq l= {\rm rank}(\gkm)$, and that
it is equal only for $r=1$~[\Ref{KBOOK}]. The automorphism
$\sigma_{\s;r}$ leads to a realization of the affine Kac-Moody algebra 
$\gkm$ by means of a centrally extended loop algebra where the
gradation $\s$ is built-in. To spell this out, let us recall that
$\sigma_{\s;r}$ provides a ${\Bbb Z}/N_\s {\Bbb Z}$-gradation of $\g$ 
$$
\g\> =\> \bigoplus_{i \in {\Bbb Z}/N_\s {\Bbb Z}}\> \g_{
i}  \>, 
\efr
and associate a subalgebra ${\cal L}(\g, \sigma_{\s;r})$ of the loop
algebra of $\g$ to the automorphism $\sigma_{\s;r}$ as follows:
$$
{\cal L}(\g, \sigma_{\s;r})\> =\> \bigoplus_{i\in{\Bbb Z}}  t^i \otimes \g_{i\;
{\rm mod\;} N_\s}\>.
\efr
Then, the affine Kac-Moody algebra $\gkm= \g^{(r)}$ is isomorphic to the
central extension of ${\cal L}(\g, \sigma_{\s;r})$ plus a derivation $d_\s$
acting on ${\cal L}(\g, \sigma_{\s;r})$ as $t\> d/dt$:
$$
\g^{(r)} \simeq {\cal L}(\g, \sigma_{\s;r}) \oplus {\Bbb C}\> K \oplus {\Bbb
C}\> d_\s\>.
\nfr{Loop}  

\section{Heisenberg subalgebras}

A central role in our construction  is played by the Heisenberg subalgebras of
$\gkm$. They can be defined as follows~[\Ref{KP}]. Let $\sigma_{\s';r}$ be a
finite order automorphism of $\g$ that fixes a Cartan subalgebra $\h\subset \g$,
which means that $\sigma_{\s';r}$ is conjugated to a Dynkin diagram automorphism
of $\g$ or order $r$. Consider the realization of $\g^{(r)}$ by means of the
central extension of ${\cal L}(\g, \sigma_{\s';r})$ and the  relation between the
infinite algebra $\gkm= \g^{(r)}$ and the finite algebra $\g$ given by the
`covering homomorphism'  $ \varphi_{\sigma} : {\cal L}(\g, \sigma) \mapsto \g$
defined by $t\mapsto 1$. Then, the subspace 
$$
\HSA[\s']\> =\> \varphi^{-1}_{\sigma_{\s';r}}\bigl( \h \bigr) \oplus {\Bbb C}\>
K 
\nfr{Hsa}
is a Heisenberg subalgebra of $\g^{(r)}$. In other words, $\HSA[\s']$ is the
pull-back of $\h$ to $\gg{}$. Notice that, by construction, $\HSA[\s']$ is graded
with respect to $\s'$ and, hence, it is possible to choose a basis such that
$\HSA[\s']= {\Bbb C}\> K \oplus \sum_{i\in E} {\Bbb C} \> b_i$, where $b_i \in
\gg{i}(\s')$, $E= I \; {\rm mod}\; N_{\s'}$, and $I$ is a set of ${\rm
rank\/}(\g)$ integers $\geq0$ and $<N_{\s'}$ (with multiplicities). The algebra is
$$
[b_j\>, \> b_k]\> =\> j\> \delta_{j+k,0} \> K\>,
\efr
where we have normalized the $b_i$'s by means of $\bil{b_i}{b_j} =
(N_{\s'}/r) \delta_{i+j,0}$. Notice that, properly speaking, only the
$b_i$ with $i\not=0$ span a Heisenberg algebra, although we will apply
this name to the whole $\HSA[\s']$. Moreover, $\HSA[\s']/{\Bbb C}K$ is a
maximal abelian subalgebra of $\gg{}/{\Bbb C}K$

Heisenberg subalgebras are central to our construction because any $b \in
\HSA[\s']$  is `semisimple', which means that we have the decomposition
$$
\gg{}\> =\> \Ker(\ad b)\>  +\> \Im(\ad b)\>, \qquad
\Ker(\ad b) \cap \Im(\ad b) \>= \> {\Bbb C}\> K\>.
\efr
Moreover, if $b$ has definite $\s'$-grade then both  $\Ker(\ad b)$ and
$\Im(\ad b)$ are graded subspaces with respect to $\s'$. In general, 
$\Ker(\ad b)$ is a non-abelian subalgebra of $\gg{}$ that always
contains $\HSA[\s']$ and satisfies
$$ 
[\Ker(\ad b)\> ,\> \Im(\ad b)]\> \subset \Im(\ad b) \>.
\efr
However, $\Ker(\ad b) \not= \HSA[\s']$ unless $b$ is `regular'. 

\section{Integrable representations}

The definition of integrable representations makes use of the following
property. An element $x\in \gkm$ is said to be `locally nilpotent' on a given
representation if for any vector $\ket{v}$ there exists a positive integer
$n_v$ such that $x^{n_v} \ket{v}=0$. Then, an integrable highest-weight 
representation of $\gkm$ is a highest-weight representation where the Chevalley
generators are locally nilpotent. It can be proven that these representations
are irreducible and that they have a unique highest-weight vector
$\ket{v_\s}$, which can be labelled by a gradation $\s=(s_0,s_1,\ldots,s_l)$
such that 
$$ 
e_{i}^+\> \ket{v_\s} \>= \> \left(e_{i}^-\right)^{s_i +1} \> \ket{v_\s}
\>=\>0 \>,  \quad \left(e_{i}^-\right)^{s_i} \> \ket{v_\s}
\>\not=\>0 \>, \quad {\rm and} \quad h_i \> \ket{v_\s} \>= \> s_i \>
\ket{v_\s}\>,  
\nfr{Repre}
for all $i=0,\ldots,l$; the corresponding representation will be indicated by 
$L(\s)$. Notice that $d_\s$ can be diagonalized on $L(\s)$, but its eigenvalue
acting on $\ket{v_\s}$ is arbitrary. The eigenvalue of the central element
$K$ on $L(\s)$ is known as the level $K_\s$ of the representation:
$$
K\> \ket{v_\s} \>= \> \sum_{i=0}^l a_{i}^{\vee}\> h_i \> \ket{v_\s} \>= 
\> \left(\sum_{i=0}^l a_{i}^{\vee}\> s_i\right) \>\ket{v_\s}
\nfr{Centre}
and, hence, $K_\s=\sum_{i=0}^l a_{i}^{\vee}\> s_i $ is a positive
integer. On $L(\s)$  there is a notion of orthogonality by means of a
(unique) positive definite Hermitian form $F$ such that $F(\ket{v_\s},
\ket{v_\s}) = \langle v_\s \ket{v_\s}= 1$. 
We will use the notation $\ket{v_i}$ for the highest-weigh vector of the
`fundamental' representation $L(i)$ where $s_j = \delta_{j,i}$. In terms
of these fundamental highest-weight vectors, $\ket{v_\s}$ can be
decomposed as
$$
\ket{v_\s} \>=\> \bigotimes_{i=0}^{r}\> \bigl\{ \ket{v_i}^{\otimes s_i}
\bigr\}\>.
\nfr{Tensor}

In Section~4 we make use of the following two properties. The first
one is the identity
$$
\bil{d_\m}{x}\> =\> \bra{v_{\overline{\m}}}\> x\> \ket{v_{\overline{\m}}},
\nfr{GradRep}
where $\ket{v_{\overline{\m}}}$ is the highest weight vector of the
representation $L(\overline{\m})$ specified by
$$
\overline{m}_i \> =\>  \bil{d_\m}{h_i}\> =\> a_i \> {a_{i}^\vee}^{ -1}\>
m_i\>.
\nfr{Grado}
Obviously, $\m \simeq \overline{\m}$ with respect to the partial ordering,
and the level of $L(\overline{\m})$ is $K_{\overline{\m}} = N_{\m}/r$. 
The second property is that, for any two gradations $\m$ and $\s$, the
highest-weight vector of $L(\m)$ is always annihilated by $\gg{>0}(\s)$.
Moreover, if $\m \preceq \s$, the highest weight vector
$\ket{v_\m}$ is an eigenvector of $\gg{0}(\s)$ with eigenvalues 
$$
h_i\> \ket{v_\m}\> =\> m_i \> \ket{v_\m}\>, \quad {\rm and} \quad e_{j}^-\>
\ket{v_\m} \>=\> 0\quad {\rm whenever}\quad s_j=0\>.
\nfr{Eigen}

$L(\s)$ can be `integrated' to a representation of the Kac-Moody group $\GG$,
which is then generated by the exponentials of the generators of
$\gkm$~[\Ref{KBOOK},\Ref{KP},\Ref{PK}]. We will denote by $\GG_+(\s)$,
$\GG_-(\s)$, and $\GG_0(\s)$ the subgroups corresponding to the subalgebras
$\gg{>0} (\s)$, $\gg{<0}(\s)$, and $\gg{0}(\s)$, respectively. An important
property of $\GG$ is that, although $\gg{}$ has the obvious direct sum
decomposition
$$
\gg{}\> =\> \gg{<0}(\s) \> \oplus \>  \gg{0}(\s) \> \oplus \> \gg{>0}(\s)\>,
\efr
it is not true that $\GG$ decomposes as the product $\GG_-(\s) \GG_0(\s)
 \GG_+(\s)$, \ie, in general, the elements of the group $\GG$ do not 
admit a generalized `Gauss decomposition'. However, $\GG_-(\s) \GG_0(\s) 
\GG_+(\s)$ forms a dense  open subset of $\GG$ called the `big
cell'~[\Ref{WILTAU}]. Then, if $g$ belongs to the big cell, we shall
write $g= g_- g_0 g_+$ for its unique factorization with
$g_\pm \in \GG_\pm(\s) $ and $g_0 \in \GG_0(\s)$.

\bjump
\references

\beginref
\Rref{SOLKP}{H.~Aratyn, E.~Nissimov and S.~Pacheva, Phys. Lett. {\bf A
201} (1995) 293; Phys. Lett. {\bf A 228} (1997) 164.}
\Rref{GEN}{M.F.~de Groot, T.J.~Hollowood, and J.L. Miramontes, Commun. 
Math. Phys.{\bf 145} (1992) 57.}
\Rref{GEN2}{N.J.~Burroughs, M.F.~de Groot, T.J.~Hollowood, and J.L.
Miramontes,  Commun. Math. Phys.{\bf 153} (1993) 187; Phys. Lett. {\bf
B~277} (1992) 89.}
\Rref{GENTAU}{T.J.~Hollowood and J.L.~Miramontes, Commun. Math.
Phys.{\bf 157} (1993) 99.}
\Rref{DICK}{L.A.~Dickey, {\sl Soliton equations and Hamiltonian
systems\/}, Advanced Series in Math. Phys. Vol. 12
(World Scientific, 1991).}
\Rref{HIR}{R.~Hirota, {\sl Direct methods in soliton theory\/}, in
{\it `Soliton'} (R.K.~Bullough and P.S.~Caudrey, eds.),
(Springer-Verlag, 1980) 157.}
\Rref{KYOTO}{M.~Jimbo and T.~Miwa, Publ. RIMS, Kyoto Univ. {\bf
19} (1983) 943;\newline
E.~Date, M.~Jimbo, M.~Kashiwara, and T.~Miwa, Publ. RIMS, Kyoto Univ. 
{\bf 18} (1982) 1077.}
\Rref{SEGWIL}{G.~Segal and G.~Wilson, {\sl Loop groups and equations of
KdV type\/}, Publ. Math. IHES {\bf 63} (1985) 1.}
\Rref{WILTAU}{G.~Wilson, Phil. Trans. R.~Soc. Lond. {\bf A~315} (1985)
383; {\sl Habillage et fonctions $\tau$\/} C.~R. Acad. Sc. Paris
{\bf 299 (I)} (1984) 587; {\sl The $\tau$-Functions of the gAKNS
Equations}, in {\it `Verdier memorial conference on integrable
systems'}   (O.~Babelon, P.~Cartier, and  Y.~Kosmann-Schwarzbach, eds.),
Birkhauser (1993) 131.}
\Rref{KW}{V.G.~Kac and M.~Wakimoto, {\sl Exceptional hierarchies
of soliton equations\/}, in {\it `Proceedings of Symposia in Pure
Mathematics'}, Vol. {\bf 49} (1989) 191.}
\Rref{JMP}{L.A.~Ferreira, J.L.~Miramontes, and J.~S\'anchez Guill\'en,
J.~Math. Phys. {\bf 38} (1997) 882.}  
\Rref{SOLTAU}{T.J.~Hollowood, Nucl. Phys. {\bf B~384} (1992) 523;
\newline
N.~McKay and W.A.~McGhee, Int.~J. Mod. Phys. {\bf A 8} (1993) 
2791; Erratum ibid. {\bf A8} (1993) 3830;\newline
Z.~Zhu and D.G.~Caldi, Nucl. Phys. {\bf B~436} (1995) 659.}
\Rref{CAT}{H.~Aratyn, C.P.~Constantinidis, L.A.~Ferreira, J.F.~Gomes,
and A.H.~Zimerman, Nucl. Phys. {\bf B~406} (1993) 727.}
\Rref{SOLSPEC}{D.~Olive, M.V.~Saveliev and J.W.R. Underwood, Phys. Lett. 
{\bf B 311} (1993) 117; \newline
D.~Olive, N.~Turok and J.W.R.~Underwood, Nucl. Phys. {\bf B 409}
(1993) 509.}
\Rref{EMSPEC}{D.~Olive, N.~Turok and J.W.R.~Underwood, Nucl. Phys. {\bf
B 401}  (1993) 663.}
\Rref{ANSOL}{O.~Babelon and D.~Bernard, Int.~J. Mod. Phys. {\bf A 8}
(1993) 507; \newline
H.~Belich and R.~Paunov, J.~Math. Phys. {\bf 38} (1997) 4108; \newline
H.~Belich, G.~Cuba and R.~Paunov, {\sl Vertex Operator Representation
of the Soliton Tau Functions in the $A_{n}^{(1)}$ Toda Models by
Dressing Transformations\/}, hep-th/9712249;\newline
E.J.~Beggs and P.R.~Johnson, Nucl. Phys. {\bf B 484} (1997) 653; {\sl
Inverse scattering and solitons in $A_{n-1}$ affine Toda field
theories II\/}, hep-th/9803248.} 
\Rref{LUIZ}{L.A.~Ferreira, J.L.~Miramontes and J.~S\'anchez Guill\'en,
Nucl. Phys. {\bf B 449} (1995) 631.}
\Rref{TAUQ}{S.~Saito, Phys. Rev. {\bf D 36} (1987) 1819; Phys. Rev.
Lett. {\bf 59} (1987) 1798;\newline
D.~Bernard and A.~Le Clair, Nucl. Phys. {\bf B 426} (1994) 534;
ERRATUM ibid. {\bf B 498} (1997) 619; \newline 
O.~Babelon, D.~Bernard and F.A.~Smirnov, Commun. Math. Phys. {\bf 182}
(1996) 319; \newline 
T.~Kojima, V.~Korepin and N.~Slavnov, Commun. Math. Phys. {\bf 189}
(1997) 709; \newline A.~Mironov, {\sl $\tau$-function within group
theory approach and its quantization\/}, q-alg/9711006.}
\Rref{DIJK}{M.R.~Douglas, Phys. Lett. {\bf B 238} (1990) 176;
\newline
E.~Witten, Diff. Geom. {\bf 1} (1991) 243;\newline
M.~Fukuma, H.~Kawai and R.~Nakayama, Int.~J. Mod. Phys. {\bf A 6}
(1991) 1385; Commun. Math. Phys. {\bf 143} (1992) 371;\newline
R.~Dijkgraaf, E.~Verlinde and H.~Verlinde, Nucl. Phys. {\bf B 438}
(1991) 435;\newline 
V.~Kac and A.~Schwarz, Phys. Lett. {\bf B 257} (1991) 329; \newline
M.~Kontsevich, Commun. Math. Phys. {\bf 147} (1992) 1.}
\Rref{SW}{T.~Nakatsu and K.~Takasaki, Mod. Phys. Lett. {\bf A 11}
(1996) 157; {\sl Integrable system and $N=2$ supersymmetric
Yang-Mills theory\/}, hep-th/9603129; \newline
A.~Gorsky, I.~Krichever, A.~Marshakov, A.~Mironov and A.~Morozov,
Phys. Lett. {\bf B 355} (1995) 466.}
\Rref{MARCOS}{M.~Mari\~no and G.~Moore, {\sl The Donaldson-Witten
function for gauge groups of rank larger than one\/},
hep-th/9802185;\newline
K.~Takasaki, {\sl Integrable Hierarchies and Contact Terms in u-plane
Integrals of Topologically Twisted Supersymmetric Gauge Theories\/},
hep-th/9803217.}
\Rref{FRENKEL}{B.~Feigin and E.~Frenkel, {\sl Integrals of motion and
Quantum Groups\/}, in Lect. Notes in Math. {\bf 1620}, Springer Verlag
(1995) 349; Invent. Math. {\bf 120} (1995) 379; \newline
B.~Enriquez and E.~Frenkel, Commun. Math. Phys. {\bf 185} (1997)
211; \newline
E.~Frenkel, {\sl Five lectures on soliton equations\/}, 
q-alg/9712005.}
\Rref{CHODOS}{A.~Chodos, Phys. Rev. {\bf D 21} (1980) 2818.}
\Rref{DS}{V.G.~Drinfel'd and V.V.~Sokolov, J.~Sov. Math. {\bf 30}
(1985) 1975.}
\Rref{FHM}{L.~Feh\'er, J.~Harnad and I.~Marshall, Commun. Math. Phys.
{\bf 154} (1993) 181;\newline
L.~Feh\'er and I.~Marshall, Commun. Math. Phys. {\bf 183} (1997)
423.}
\Rref{ARA}{H.~Aratyn, L.A.~Ferreira, J.F.~Gomes and A.H.~Zimerman,
J.~Math. Phys. {\bf 38} (1997) 1559.}
\Rref{DELD}{F.~Delduc, L.~Feh\'er and L.~Gallot, {\sl Nonstandard
Drinfeld-Sokolov reduction\/}, solv-int/9708002.}
\Rref{AKS}{N.J.~Burroughs, Nonlinearity {\bf 6} (1993) 583; 
Nucl. Phys. {\bf B379} (1992) 340; \newline
L.~Feh\'er, {\sl KdV type systems and W-algebras in the
Drinfeld-Sokolov approach\/}, hep-th/9510001.}
\Rref{WILSON}{G.~Wilson, {\sl Ergod. Th. Dynam. Sys.} {\bf 1} (1981) 
361.}
\Rref{ANALS}{C.R.~Fern\'andez-Pousa, M.V.~Gallas, J.L.~Miramontes and
J.~S\'anchez Guill\'en, Ann. Phys. (N.Y.) {\bf 243} (1995) 372.}
\Rref{ABTOD}{M.A.~Olshanetsky and A.M.~Perelomov, Invent. Math. {\bf
54} (1979) 261; \newline
A.V.~Mikhailov, M.A.~Olshanetsky and A.M.~Perelomov, Commun. Math. Phys.
{\bf 79} (1981) 473.}
\Rref{LS}{A.N.~Leznov and M.V.~Saveliev, {\sl Group Theoretical Methods 
for  Integration of Non-Linear Dynamical Systems}, Progress in Physics
Series, v. 15, Birkha\"user-Verlag, Basel, 1992.}
\Rref{SAVGERV}{L.A.~Ferreira, J-L.~Gervais, J.~S\'anchez Guill\'en and 
M.V.~Saveliev,  Nucl. Phys. {\bf B 470} (1996) 236.} 
\Rref{NOS}{C.R.~Fern\'andez-Pousa, M.V.~Gallas, T.J.~Hollowood, and
J.L.~Miramontes, Nucl. Phys.~{\bf B 484} (1997) 609.}
\Rref{NOSQ}{C.R.~Fern\'andez-Pousa, M.V.~Gallas, T.J.~Hollowood, and
J.L.~Miramontes, Nucl. Phys.~{\bf B 499} (1997) 673.}
\Rref{NOSSOL}{C.R.~Fern\'andez-Pousa and J.L.~Miramontes, Nucl.
Phys.~{\bf B 518} (1998) 745.}
\Rref{FREEMAN}{M.~Freeman, Nucl. Phys. {\bf B 433} (1995) 657.}
\Rref{KBOOK}{V.G.~Kac, {\sl Infinite
dimensional Lie algebras\/} ($3^{rd}$ ed.), Cambridge
University Press (1990).}
\Rref{GO}{P.~Goddard and D.~Olive, Int. J.~Mod. Phys. {\bf A 1} (1986)
303; \newline
J.~Fuchs, {\sl Affine Lie Algebras and Quantum Groups\/}, Cambridge
University Press (1992).}
\Rref{KP}{V.G.~Kac and D.H.~Peterson, {\sl 112 constructions of
the basic representation of the loop group of $E_8$\/}, in {\it
`Symposium on Anomalies, Geometry and Topology'} (W.A.~Bardeen and
A.R.~White, eds.), World Scientific (1985) 276.}
\Rref{PK}{D.H.~Peterson and V.G.~Kac, Proc. Natl. Acad. Sci.
USA {\bf 80} (1983) 1778;\newline V.G.~Kac, {\sl
Constructing groups associated to infinite-dimensional Lie
algebras\/}, in {\it `Infinite dimensional groups with
applications'} (V.G.~Kac, ed.), Berkeley MSRI pub., Springer
Verlag (1985) 167.}
\Rref{TL}{K.~Ueno and K.~Takasaki, Adv. Stud. in Pure Maths. {\bf 4}
(1984) 1; \newline
D.~Olive and N.~Turok, Nucl. Phys. {\bf B 257} (1985) 277; Nucl.
Phys. {\bf B 265} (1986) 469; \newline 
T.~Takebe, Commun. Math. Phys. {\bf 129} (1990) 281.}
\Rref{LUIZKP}{H.~Aratyn, L.A.~Ferreira, J.F.~Gomes and A.H.~Zimmerman,
{\sl Solitons from Dressing in an Algebraic Approach to the
Constrained KP Hierarchy\/}, hep-th/9709004; {\sl Vertex Operators and
Solitons of Constrained KP Hierarchies\/}, in proceedings of
{\it `Integrable  Models and Supersymmetry'}.}
\Rref{NIEDER}{M.R.~Niedermaier, Commun. Math. Phys. {\bf 160} (1994)
391.}

\endref   

\ciao